\documentclass[fleqn,useAMS,usenatbib]{mnras}
\usepackage{graphicx}
\usepackage{amsmath,amssymb}
\usepackage{multirow}
\usepackage{enumerate}
\usepackage{bm}
\usepackage{enumitem}
\graphicspath{ {Figures/} }
\def\beq{\begin{equation}} \def\eeq{\end{equation}}
\def\etal{{\em et al.}}
\def\LCDM{$\Lambda$CDM}
\def\AJ#1{AJ, {#1}}
\def\MNRAS#1{MNRAS,\ {#1}}
\def\PRL#1{Phys.\ Rev.\ Lett.,\ {#1}} \def\PR#1#2{Phys.\ Rev.\ #1,\ {#2}}
\def\ApJ#1{ApJ,\ {#1}} \def\AaA#1{A\&A,\ {#1}}
\def\ApJs#1{ApJS,\ {#1}}
\def\CQG#1{Class.\ Quantum Grav.,\ {#1}}
\def\GRG#1{Gen.\ Relativ.\ Grav.,\ {#1}}
\def\lsim{\mathop{\hbox{${\lower3.8pt\hbox{$<$}}\atop{\raise0.2pt\hbox{$
\sim$}}$}}}
\def\gsim{\mathop{\hbox{${\lower3.8pt\hbox{$>$}}\atop{\raise0.2pt\hbox{$
\sim$}}$}}}\def\goesas{\mathop{\sim}\limits}
\def\Z#1{_{\lower2pt\hbox{$\scriptstyle#1$}}}
\def\X#1{_{\lower1pt\hbox{$\scriptscriptstyle#1$}}}
\def\ns#1{_{\rm #1}} \def\Ns#1{_{\lower2pt\hbox{$\scriptstyle\rm#1$}}}
\def\HCMB{H\Ns{CMB}} \def\HLG{H\Ns{LG}} \def\HX{H\Ns{X}}
 \def\OmM{\Omega\Z{M0}}
\def\OmL{\Omega\Z{\Lambda0}}
\def\ga{\gamma}\def\de{\delta}
\def\ph{\phi}\def\si{\sigma}
\def\ns#1{_{\rm #1}}\def\bs{{\bar\si}}\def\w#1{\,\hbox{#1}}
\def\h{\,h^{-1}}\def\hm{\h\hbox{Mpc}}\def\deg{^\circ}
\def\kms{\w{km}\;\w{sec}^{-1}}\def\kmsMpc{\kms\w{Mpc}^{-1}}
\def\ave#1{\left\langle#1\right\rangle}\def\mr{{\bar r}}
\def\Hm{H\Z0} \def\Ha{\bar H\Z0} \def\nus{\nu_s^{}}
\def\chN{\chi_{\rm a}^2}\def\chL{\chi_{\rm b}^2}\def\Nb{\nu_{\rm b}}
\def\frn#1#2{{\textstyle{#1\over#2}}}\def\vd{v\ns{der}}\def\vt{v\ns{true}}
\def\fb{f_p}\def\CS{{\textsc COMPOSITE}\ }
\catcode`@=11 \def\@volume{\bf 457}
\def\@pagerange{3285--3305 (2016); Erratum MNRAS {\bf463}, 3113}
\def\@pubyear{2016}\def\tod@y{}
\def\@printed{Compiled using MNRAS \LaTeX\ style file v\@version}
\catcode`@=12
\newcommand\T{\rule{0pt}{2.6ex}} % Top strut
\newcommand\B{\rule[-1.2ex]{0pt}{0pt}} % Bottom strut
\def\today{Submitted 13 March 2015; revised 6 January 2016; accepted 13 January 2016}
\def\latitudes{In all figures, the galactic longitudes $\ell=0\deg,180\deg,
360\deg$ are on the right edge, centre and left edge respectively.}
\title[Defining the frame of minimum nonlinear Hubble expansion variation]{Defining the frame of minimum nonlinear Hubble expansion variation}
\author[McKay \& Wiltshire]{James H. McKay\thanks{E-mail:
j.mckay14@imperial.ac.uk} \& David L. Wiltshire\thanks{E-mail:
david.wiltshire@canterbury.ac.nz}\\
Department of Physics \& Astronomy, University of Canterbury,
Private Bag 4800, Christchurch 8140, New Zealand\\
}
\begin{document}

\maketitle
\label{firstpage}
\begin{abstract}
We characterize a cosmic rest frame in which the monopole variation of the
spherically averaged nonlinear Hubble expansion is most uniform, under
arbitrary local Lorentz boosts of the central observer. Using the \CS sample of
4534 galaxies, we identify a degenerate set of candidate minimum nonlinear
variation frames, which includes the rest frame of the Local Group (LG) of
galaxies, but excludes the standard Cosmic Microwave Background (CMB) frame.
Candidate rest frames defined by a boost from the LG frame close to the plane
of the galaxy have a statistical likelihood similar to the LG frame. This may
result from a lack of constraining data in the Zone of Avoidance. We extend our
analysis to the \textit{Cosmicflows}-2 (CF2) sample of 8162 galaxies. While
the signature of a systematic boost offset between the CMB and LG frame
averages is still detected, the spherically averaged nonlinear expansion
variation in all rest frames is significantly larger in the CF2 sample than
would be reasonably expected. We trace this to the CF2 distances being
reported without a correction for inhomogeneous distribution Malmquist bias.
Systematic differences in the inclusion of the large SFI++ subsample into the
COMPOSITE and CF2 catalogues are analysed. Our results highlight the importance
of a careful treatment of Malmquist biases for future peculiar velocities
studies, including tests of the hypothesis of \citet{WSMW} [\PR D{88}, 083529]
that a significant fraction of the CMB temperature dipole may be nonkinematic
in origin.
\end{abstract}

\section{Introduction}

Although the Universe is spatially homogeneous in some statistical sense,
at the present epoch it exhibits a complex hierarchical structure, with galaxy
clusters forming knots, filaments and sheets that thread and surround voids,
in a complex cosmic web \citep{web1,web2,web3}. Deviations from homogeneity
are conventionally treated in the framework of peculiar velocities, by which
the mean redshift, $z$, and luminosity distance, $r$, of a galaxy cluster are
converted to a peculiar velocity at small redshifts according to
\beq
v\ns{pec}=cz-\Hm r\,\label{vpec}
\eeq
where $c$ is the speed of light and $\Hm=100\,h\kmsMpc$ is the Hubble constant.

While large galaxy surveys reveal that statistical homogeneity
emerges on scales $70$ -- $100\hm$ \citep{hogg05,Scrimgeour2012} as determined
by the two--point galaxy correlation function, in the standard
peculiar velocities framework relatively large bulk flows are seen on scales
larger than this relative to our own location. The amplitude of such flows,
and their consistency with the standard Lambda Cold Dark Matter (\LCDM)
cosmology, are a matter of ongoing debate
\citep*{WFH1,kash10,dn11,turn12,lah12,ms13,minrep,Ppec,HCT,ctlm}.
While it
may be possible to achieve convergence of such bulk flows to the CMB frame
within current uncertainties based on cosmic variance and $N$-body simulations
in the \LCDM\ model, given the deep mystery associated with the nature of dark
energy, it is important that alternative cosmological models are investigated.
Such alternatives \citep{buch00,buch08,clocks,sol,obs} may also
demand a treatment of cosmic expansion which includes effects of
inhomogeneities that fall outside the peculiar velocity framework, with
testable observational consequences \citep{ls08,WSMW}.

The peculiar velocity framework makes a strong geometrical assumption over and
above what is demanded by general relativity. In particular, the quantity
$v\ns{pec}$ defined by (\ref{vpec}) only has the physical characteristics
of a velocity if one implicitly assumes the spatial
geometry {\em on all scales} larger than those of bound systems is
exactly described by a homogeneous isotropic
Friedmann-Lema\^{\i}tre-Robertson-Walker (FLRW) model with a single
cosmic scale factor, $a(t)$, whose derivative defines a single global
Hubble constant, $\Hm=\left.\dot a/a\right|_{t_0}$.
Deviations from the uniform expansion are then ascribed to Lorentz
boosts of each galaxy cluster with respect to the spatial hypersurfaces of
average homogeneity, at each cluster location.

It is a consequence of general relativity, however, that inhomogeneous
matter distributions generally give rise to a differential expansion of space
that cannot be reduced to a single uniform expansion plus local boosts.
This is a feature of general exact solutions to the cosmological Einstein
equations, such as the Lema\^{\i}tre--Tolman--Bondi (LTB) \citep{L,T,B} and
\citet{Sz} models. Any definition of the expansion rate in such models depends
on the spatial scale relative to that of the inhomogeneities. Although one
can define scale dependent Hubble parameters for specific exact solutions
-- for example, given the spherical symmetry of the LTB model -- the actual
cosmic web is sufficiently complex that in reality one must deal with spatial
or null cone averages in general relativity.

In recent work \citet{WSMW} examined the variation\footnote{\citet{WSMW} used
the terminology ``Hubble flow variance'', where variance was understood in
the same loose sense as in the terminology ``cosmic variance''. Here we use
the more generic word ``variation'', to avoid any confusion with
the square of a standard deviation.} of
the Hubble expansion from a fresh perspective, by generalizing the earlier
approaches of \citet{ls08} and \citet{md07}. In particular, given that there
is a notion of statistical homogeneity on large ($\gg100\hm$) scales, then an
average expansion law characterized by a single asymptotic Hubble constant,
$\Ha$, is applicable on such scales. However, from the first principles
of general relativity one should make no geometrical assumptions about
cosmic expansion below the statistical homogeneity scale, where cosmic
expansion is nonlinear. One can nonetheless perform radial and angular averages
of the distance versus redshift of a large sample of galaxies in spherical
shells, and compare the results with the asymptotic Hubble constant in order
to quantify the nonlinear variation of the Hubble expansion.

\citet*{WSMW} conducted such an analysis on the \CS sample of 4534
cluster, group and galaxy distances \citep*{WFH1,WFH2}, with the following
results:
\begin{itemize}
\item A linear Hubble law with a spherically averaged Hubble constant, $H_s$,
which is statistically indistinguishable from the asymptotic Hubble constant,
$\Ha$, is found to emerge in independent radial shells with mean distances
in the range $\bar r_s>70\hm$.
\item On scales $r\lsim65\hm$ the spherically averaged value, $H_s$, in
independent shells is greater than the asymptotic value, $\Ha$. However,
the difference is significantly larger in the standard rest frame of the
CMB radiation when compared to either the
rest frame of the Local Group (LG) of galaxies or the Local Sheet (LS).
In other words, the spherically averaged Hubble expansion is more uniform
in the LG rest frame than in the CMB rest frame, with very strong Bayesian
evidence, $\ln B\gg5$. The uniformity of expansion in the LG and LS frames is
statistically indistinguishable.
\item By a variety of angular tests, the residual variation of the spherical
(monopole) Hubble expansion is found to be correlated with structures in the
range $32$ -- $62\hm$, which give a Hubble expansion dipole with a markedly
different character in the CMB and LG frames.
\item A skymap of angular variation of the Hubble expansion in the LG frame,
constructed by Gaussian window averaging \citep{md07}, has a very strong
dipole. The angular expansion skymap has a correlation coefficient of $-0.92$
with the residual CMB temperature dipole in the LG frame.
\item On scales $\gsim80\hm$ the magnitude of the dipole variation in the
Hubble expansion is less in the LG frame than in the CMB frame. While the
LG frame dipole is statistically consistent with zero in most individual outer
shells, by contrast the CMB frame dipole magnitude rises to a residual
level. Consequently the large bulk flow reported by \citet{WFH1}
may be partly due to a choice of rest frame.
\end{itemize}

\begin{figure}
\centering
\includegraphics[width=0.46\textwidth]{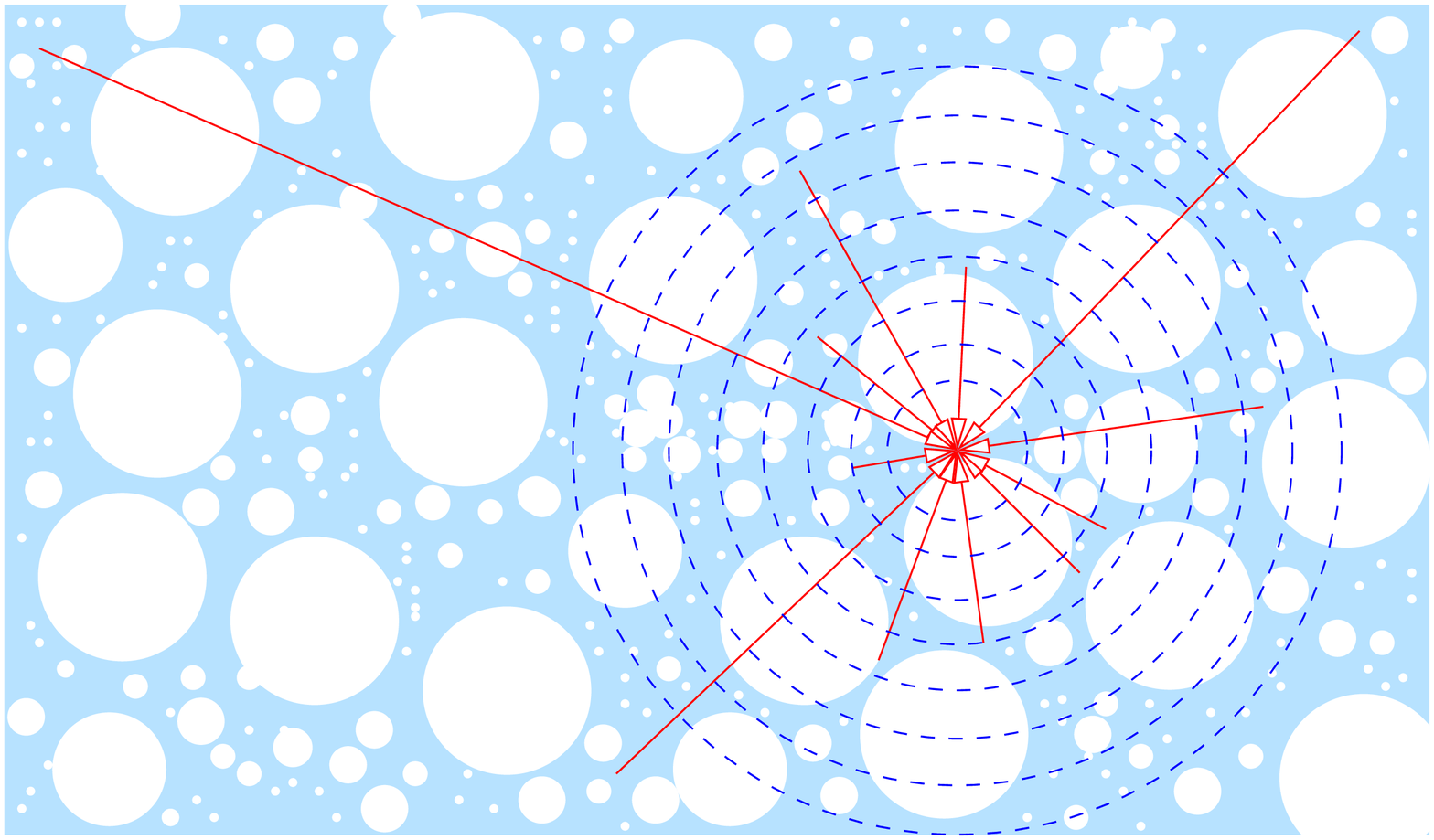}
\caption{Schematic diagram of spherical averaging (from \citet{WSMW}). The
universe is an ensemble of filaments, walls and voids, the largest
{\em typical} nonlinear structures being voids of diameter $\goesas30\hm$
\protect\citep{HV02,HV04,Pan2011}. If one averages $cz/r$ in spherical shells
(dotted lines) about any point then once the shells are a few times larger
than the typical nonlinear structures, an average Hubble law with small
statistical scatter is obtained (e.g., for the longest null geodesics depicted
by lines converging to the central observer). However, there are
considerable deviations for shells on scales comparable to the typical
nonlinear structures. This approach makes no assumptions about spatial
geometry of the universe on the scales in question, and thus is model
independent.}
\label{Hshells}
\end{figure}

The first of the results above is consistent with other observations which
find that a notion of statistical homogeneity emerges at scales of order $70$
-- $100\hm$ \citep{hogg05,Scrimgeour2012}. Furthermore, the fact that $H_s>\Ha$
on the $\lsim65\hm$ scales on which the Hubble expansion is nonlinear agrees
well with the observation that the largest typical\footnote{Larger structures
such as the $320\hm$ long Sloan Great Wall \citep{Gott2005} and the $350\hm$
long Large Quasar Group \citep{chr} are known, but it is arguable that these are
not typical.} structures in the late epoch Universe are voids of diameter $30
\hm$ \citep{HV02,HV04}. \citet{Pan2011} found that voids occupy 62\% of
the volume studied in the 7th release of the Sloan Digital Sky Survey
\citep{Sloan2009}. If one constructs averages in spherical shells, as
in Fig.~\ref{Hshells}, then once the shells are a
few times larger than the diameter of the largest typical nonlinear structures,
a well-defined asymptotic average, $\Ha$, is obtained which does not change
when shells are further enlarged. When shells are 1 -- 2 times the diameter of
the typical nonlinear voids, however, a variation in expansion rate is seen
and since the faster expanding voids dominate by volume then the average, $H_s$,
is increased relative to $\Ha$.

While the results just discussed may not be surprising to those familiar with
the statistics of the cosmic web, the remaining results of \citet{WSMW} listed
above are not at all expected in the conventional approach to peculiar
velocities in observational cosmology. In particular, the cosmic rest frame
should be the one in which the variation of cosmic expansion is a minimum in
some sense. The results of \citet{WSMW} show that such a frame of minimum
nonlinear Hubble expansion variation is not the conventional CMB frame.

One important consequence is that a substantial fraction of the CMB dipole
usually attributed to a local boost of the Local Group at $635\pm38\kms$
\citep{tsk08} in the direction $(\ell,b)=(276.4\deg,29.3\deg)$ would be due to
a nonkinematic differential expansion of space. It has been independently shown
that the radio galaxy number count dipole appears to be nonkinematic, at the
99.5\% confidence level, using the NRAO VLA Sky Survey \citep{rs13}.

\citet{WSMW} estimate that a 0.5\% anisotropy in the distance--redshift
relation on $\lsim65\hm$ scales would be required to achieve the observed
properties of the CMB dipole. Furthermore, using ray--tracing simulations on
an exact Szekeres solution of Einstein's equations which exhibits differential
expansion on such scales, it is found that it is possible to produce a Hubble
expansion dipole which is very close to that of the \CS sample, while
simultaneously generating CMB anisotropies that both account for the
residual temperature dipole in the LG frame and are consistent with the
amplitude of higher order multipoles \citep*{Bolejko}.

This may potentially provide an explanation for the angular scale dependence
observed in attempts to measure the boost to the CMB frame using the effects
of special relativistic aberration and modulation on the CMB anisotropy
spectrum \citep{Pboost}. Since subtraction of a kinematic dipole is a key
step in the map making process, this may also have a direct impact on various
large angle anomalies in the CMB anisotropy spectrum \citep{Piso}. Furthermore,
given that a transformation to the CMB frame is a step that is taken in many
observational procedures without question, the possibility that a substantial
fraction of the CMB dipole may be nonkinematic could have a large impact
in other areas of observational cosmology.

Given the potential importance of such a result, it is important to try to
characterize the frame of minimum nonlinear Hubble expansion variation in purely
observational terms. \citet{WSMW} compared the LG and LS frames with that
of the CMB, motivated by the starting point that a frame close to the LG frame
would be the natural standard of rest according to the ``Cosmological
Equivalence Principle'' \citep{cep} which underlies the approach of
\citet{clocks,sol,obs} to the averaging problem in inhomogeneous cosmology
\citep{fit1,buch00,buch08,dust,nw}.

This paper asks the question -- if we make no assumptions about our own
standard of rest, but make arbitrary boosts at our own location -- can we
determine in a model--independent fashion whether there is a rest frame
in which the nonlinear variation of the Hubble expansion is minimized?
We will consider all scales over which cosmic expansion is nonlinear and
transitions to linearity, and the question of how such a transition impacts
the characterization of a minimum variation frame.

While our principal results are determined from an analysis of the
\CS sample \citep{WFH1,WFH2}, we have also considered the recently
published {\em Cosmicflows}-2 (CF2) sample of \citet{Tully13}. We find that the
issue of the different treatments of Malmquist bias in the two datasets
at present prevents as detailed analysis as we perform for the \CS sample.
However, our investigation highlights how the implicit assumption of a FLRW
expansion law below the scale of statistical homogeneity via (\ref{vpec})
subtly influences the manner in which such biases are treated in practice,
and raises concerns about remaining distribution biases in the CF2 distances.

\section{Methodology}

In this paper we will use the methodology of \citet{WSMW}, who considered
both radial and angular averages of the Hubble expansion in the \CS sample,
adapting techniques that \citet{ls08} and \citet{md07} had previously applied
to radial and angular averages respectively in the much smaller
Hubble Telescope Key Project dataset \citep{freedman01}. We are primarily
interested in the radial spherical averages, as it was by this method that
\citet{WSMW} found the most decisive evidence that the Hubble expansion is
significantly more uniform in the LG or LS frames, as compared to the
standard CMB rest frame.

We begin by repeating the analysis of \citet{WSMW},
but without assuming that our own standard of rest is known. In particular,
we will make arbitrary local boosts of the central observer, and perform
numerical studies on the \CS sample to determine what cosmic rest frame(s)
are singled out by minimizing the variation in the spherically averaged
nonlinear Hubble expansion.

An arbitrary boost of the central observer has the effect of transforming the
inferred redshift of individual sources we observe from values, $z_i$, in
any one rest frame to new values given by
\beq
cz_i'=cz_i+v\cos\phi_i
\label{zp}\eeq
where $v$ is the boost magnitude at our origin, and $\phi_i$ is the angle
between each data point and the boost direction. We stress that in our
nonlinear framework the boost velocity $v$ of the central observer is a
distinct quantity from the peculiar velocities of individual
sources\footnote{In particular, our minimum nonlinear Hubble expansion
variation frame should not be confused with the terminology of
\citet{WFH1,WFH2,AFW12}, who use the term ``minimum variance'' with respect
to weights of peculiar velocities in bulk flow studies.}, $v\ns{pec}$, in
the linear peculiar velocities framework (\ref{vpec}).

Following \citet{WSMW}, we determine the best fit linear Hubble law -- even
in the regime in which the average expansion is nonlinear -- by standard
linear regression in independent radial shells, minimizing the quantity
\beq\label{Hs_chi}
\chi^2_s=\sum_{i=1}^{N_s}\left[\sigma_i^{-1}(r_i-cz_i/H)\right]^2,
\eeq
with respect to $H$, where $z_i$, $r_i$ and $\sigma_i$ are respectively the
redshift, distance and distance uncertainty of each object, and $N_s$ is the
number of data points in a given shell, $s$. This method, with distance as a
function of redshift, is chosen because all uncertainties in the \CS sample
have been included as distance uncertainties. Any corrections that would be
required due to noise arising from peculiar motion within galaxy clusters
have been accounted for by assigning a distance and uncertainty to the cluster
itself rather than individual galaxies \citep{WFH1}. The values of $z_i$ are
taken to be exact, whereas the distance uncertainties are large, approximately
15\% for most \CS sample objects. Fortunately, the large number of data points
within each shell nonetheless lead to statistically significant results.

The value of the Hubble constant in the $s$th shell is then determined as
\beq\label{Hsav}
H_s=\left(\sum_{i=1}^{N_s}\frac{(cz_i)^2}{\sigma_i^2}\right)\left(\sum_{i=1}
^{N_s}\frac{cz_ir_i}{\sigma_i^2}\right)^{-1}
\eeq
while its weighted average luminosity distance is
\beq\label{rsav}
\bar{r}_s=\left(\sum_{i=1}^{N_s}\frac{r_i}{\sigma_i^2}\right)\left(\sum_{i=1}
^{N_s}\frac{1}{\sigma_i^2}\right)^{-1}.
\eeq
The radial averages are computed in two different shell configurations of 11
shells, for luminosity distances with $r_s<r\le r_{s+1}$. Both configurations
use shells of width 12.5$\hm$ in most cases, but start from a different
innermost shell cutoff of either $2\hm$ or $6.25\hm$. The two shell
configurations are labelled using unprimed and primed integers respectively,
as given in Table~1 of \citet{WSMW}.
Given less data at large distances, the shells $10$ and $10'$ are taken to be
wider than the rest so as to include approximately the same number of data
points as the inner shells. They both have an outer cutoff at $156.25\hm$. The
outermost shell 11 has an outer bound given by the largest distance in the
sample, and is the same in both configurations,
to provide an anchor for the asymptotic value of the Hubble constant, $\Ha$.

The uncertainty in (\ref{Hsav}) is taken as
\beq\label{sigH}
\bar\sigma\Z s=\sqrt{\bar{\sigma}\Z{0s}^2+\bar{\sigma}\Z{1s}^2}
\eeq
where
\beq
\bar{\sigma}\Z{0s}=H_s\frac{\sigma_0}{\bar{r}_s}
\label{zero_point}\eeq and
\beq
\bar{\sigma}\Z{1s}=\left(\sum_{i=1}^{N_s}\frac{(cz_i)^2}{\sigma_i^2}\right)^{3/2}
\left(\sum_{i=1}^{N_s}\frac{cz_ir_i}{\sigma_i^2}\right)^{-2},\label{sig1}
\eeq
A zero point uncertainty (\ref{zero_point}) is added in quadrature to
the standard uncertainty (\ref{sig1}) determined through error propagation in
(\ref{Hsav}), to yield the total uncertainty (\ref{sigH}) for the Hubble
constant in each shell. The zero point uncertainty arises from the fact that a
linear Hubble law must necessarily pass through the
origin. However, there is an uncertainty in determining the origin, due to a
$20\kms$ uncertainty in the heliocentric peculiar velocity of the LG (and LS)
frames \citep{tsk08} and a 0.4\% uncertainty in the magnitude of the CMB dipole
\citep{Fixsen}, which combine to give $\sigma_0=0.201\hm$ \citep{WSMW}. For each
shell, the weighted zero point uncertainty (\ref{zero_point}) is the uncertainty
in the linear Hubble law for a shell with mean distance $\bar{r}_s$
due to the uncertainty in the origin alone. This uncertainty is more
significant for shells with a small mean distance compared to those at a larger
distance, due to the shorter lever arm of the linear Hubble law for shells
with smaller mean distances.

As a measure of the difference in the local Hubble expansion from its asymptotic
value, we use the quantity \citep{ls08,WSMW}
\beq\label{Delta_H}
\de H_s=\frac{\left(H_s-\Ha\right)}{\Ha}
\eeq
where $\Ha$ is the mean asymptotic value of the Hubble constant. In our case,
$\Ha$, and its uncertainty are calculated from the data points at distances
$r>156.25\hm$ (shell 11 in both unprimed and primed configurations).
\citet{WSMW} choose the distance scale for the outermost shell as one which
is statistically reliable for the \CS data, while including only data at
distances larger than the baryon acoustic oscillation scale, the largest scale
on which one might expect to see the effects of inhomogeneity on the local
Hubble expansion. They verify that a linear Hubble law with a very high goodness
of fit is found for $r>156.25\hm$, with a Hubble constant
$\Ha=(100.1\pm1.7)h\kmsMpc$ in the CMB frame or $\Ha=(101.0\pm1.7)h\kmsMpc$ in
the LG/LS frames, consistent with the normalization $H_0=100\kmsMpc$ used in
the \CS sample.

\section{Monopole expansion variation due to systematic boost offsets}%
\label{LGframe}

\citet{WSMW} propose that the much larger monopole variation of the linear
Hubble law observed in the CMB frame, as compared to the LG/LS frames,
may have a systematic origin. This arises from the nonlinear dependence of
$H_s$ in (\ref{Hsav}) on the individual $cz_i$ values, which change when
performing a local Lorentz boost to a different frame. By calculating the
result of an arbitrary boost on the individual $H_s$ values \citet{WSMW}
obtain an explicit form for this systematic variation.

Consider redshifts, $z_i$, observed in a frame of reference in which the
variation of the spherically averaged Hubble expansion is minimized. An
arbitrary local boost (\ref{zp}) of the central observer leads to inferred
redshifts, $z_i'$, in the new frame given by
\begin{align}
cz_i\to cz_i'&= c(\ga-1)+\ga\left[cz_i+v\cos\phi_i(1+z_i)\right]\nonumber\\
&\simeq cz_i+v\cos\phi_i(1+z_i)+{\rm O}\left(v^2\over c^2\right)\\
&\simeq cz_i+v\cos\phi_i+{\rm O}\left(v\over c\right)\,,
\label{approx}
\end{align}
where $\ga=\left[1-v^2/c^2\right]^{-1/2}$. We will adopt the widely used
Newtonian velocity addition approximation (\ref{approx}), which is also
assumed in (\ref{vpec}). This ignores the
O$(v/c)$ correction, which is at most 0.5\% for the boosts
considered here, at least one order of magnitude smaller than typical distance
uncertainties. This results in the changes
$(cz_i)^2\to(cz_i')^2=(cz_i)^2+2cz_iv\cos\phi_i+v^2\cos^2\phi_i$ in
the numerator of (\ref{Hsav}), and $cz_ir_i\to cz_ir_i+r_iv\cos\phi_i$ in the
denominator.
The linear contributions to the transformed quantities in the denominator and
numerator of (\ref{Hsav}) should be approximately self--cancelling for
spherical averages of data which is uniformly distributed on the celestial
sphere in each shell: on average each positive contribution from the term
$v\cos\ph_i$ will cancel with a negative contribution from a data point on the
opposite side of the sky\footnote{The lack of
data in the Zone of Avoidance does not pose a problem as this absence of data
is symmetrical on opposite sides of the sky. A rough self--cancellation of
the linear contributions would only be invalid when one side of the sky has a
significant lack of data as compared to the opposite side of the sky. The \CS
sample does indeed have sufficient sky coverage to satisfy this requirement,
with the exception of the first unprimed shell 1, with $2<r\le12.5\hm$
\citep{WSMW}, which is excluded in the data analysis.}.
With such a cancellation assumed we are left with the difference
\begin{align}
\Delta H_s=H'_s-H_s&\goesas\left(\sum_{i=1}^{N_s}{(v\cos\ph_i)^2\over\si_i^2}
\right)\left(\sum_{i=1}^{N_s}{cz_i r_i\over\si_i^2}\right)^{-1}\nonumber\\
&\approx\frac{v^2}{2\Ha\ave{r_i^2}_s}\label{eq1}
\end{align}
where
\beq
\ave{r_i^2}\equiv\left(\sum_{i=1}^{N_s}\frac{r_i^2}{\sigma_i^2}\right)
\left(\sum_{i=1}^{N_s}\frac{1}{\sigma_i^2}\right)^{-1}\label{rsqav}
\eeq
is a weighted average of the squared luminosity distance in each shell.
The second line of (\ref{eq1}) follows by assuming that:
(i) $\ave{(v\cos\ph_i)^2}=v^2\ave{\ph_i^2}\goesas\frac12v^2$ is roughly
constant from shell to shell; (ii) the leading order linear Hubble
approximation $cz_i\simeq\Ha r_i$ is made in the denominator. The uncertainty
in (\ref{rsqav}) is given by
\beq
\bar{\sigma}_{\ave{r_i^2}_s}=2\left(\sum_{i=1}^{N_s}\frac{r_i^2}{\sigma_i^2}
\right)^{1/2}\left(\sum_{i=1}^{N_s}\frac{1}{\sigma_i^2}\right)^{-1}.\label{sigr}
\eeq

Our first goal is to verify that the difference in the spherically averaged
Hubble expansion between the LG and CMB frames of reference is statistically
consistent with a systematic variation of the form (\ref{eq1}). To achieve
this we fit a power law to the observed data. As there is a correlated
uncertainty in both the independent and dependent variables, $\ave{r_i^2}_s$
and $\Delta H_s$ respectively, a standard least squares method is not
appropriate. Instead we use a total least squares fit, or ``error in
variables'' method with a model of the form
\beq\label{pl}
\Delta H_s=A\left(\ave{r_i^2}_s\right)^p\,.
\eeq
In comparison with (\ref{eq1}), we expect $p\approx-1$ and $A\approx v^2/
(2\Ha)$. The details of this method are presented in the Appendix.

Carrying out this analysis, we do indeed find a difference consistent with a
systematic boost offset between the LG and CMB frames of reference. Systematic
uncertainties arise in the choice of shell boundaries. Considering only primed
shells gives a value of $p=-1.01\pm0.27$. If we take the unprimed shells then
we obtain $p=-0.79\pm0.16$ if shell 1, with $2<r\le12.5\hm$, is
included and $p=-0.77\pm 0.35$ if this first shell -- which may have
insufficient sky coverage \citep{WSMW} -- is excluded. The data in the range
$6.25<r\le12.5\hm$ common to both the first primed and unprimed shells is
important in establishing the boost offset which is more pronounced at small
$r$. To account for systematic uncertainties, we have therefore applied a
continuous variation of the first shell boundary in the range $2$ -- $6.35\hm$,
while keeping the widths of the shells fixed. This leads to a value of
$p=-0.88\pm(0.25)\ns{stat}\pm(0.13)\ns{sys}$ where the first uncertainty is
the statistical and the second systematic. For the case of the primed shells
we also note the corresponding velocity calculated from the best fit value of
$A$ is $v=772^{+1024}_{-440}\kms$, which is indeed close to the actual boost
magnitude of $635\pm38\kms$, albeit with a very large uncertainty.

\begin{figure*}
\vbox{\centerline{\vbox{\halign{#\hfil\cr\scalebox{0.7}
{\includegraphics{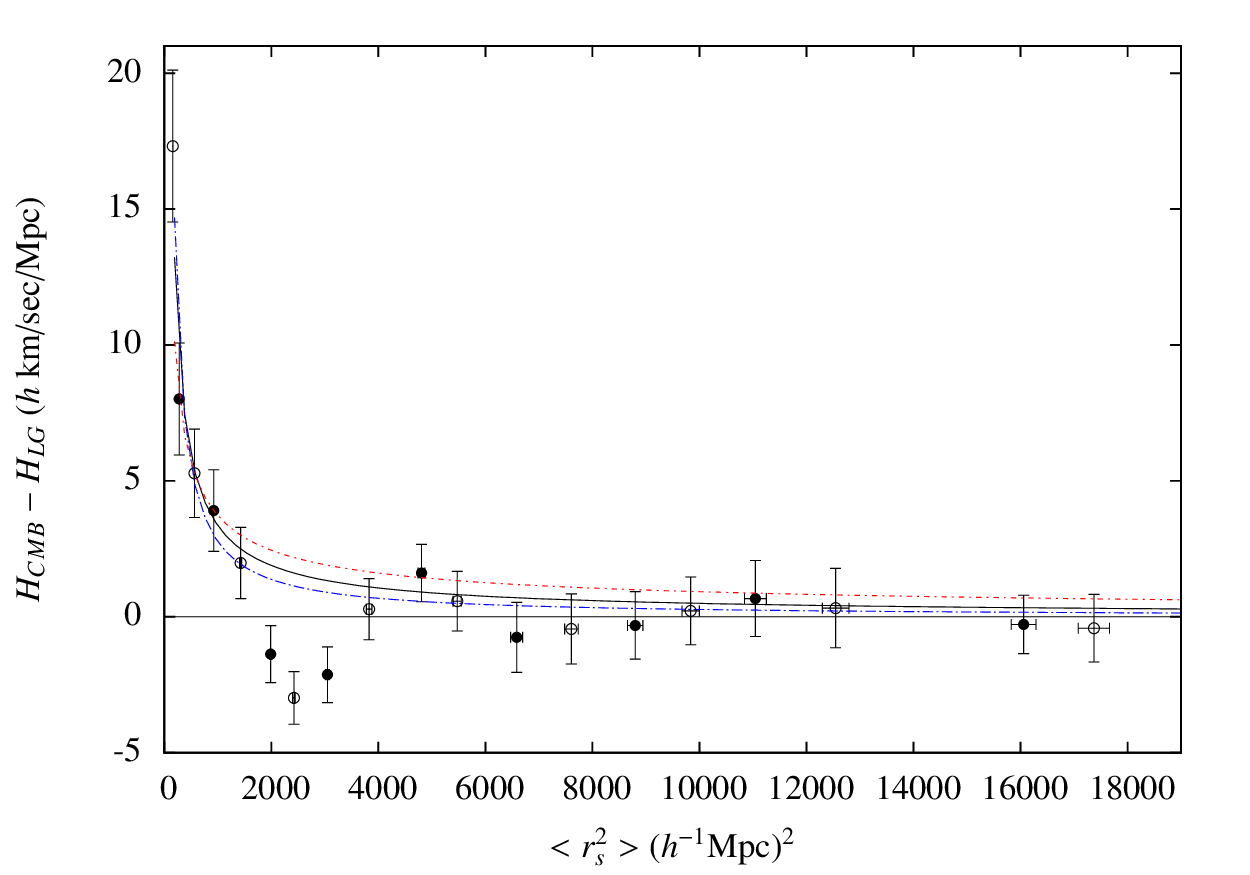}}\cr
\noalign{\vskip-5pt}\qquad{\bf(a)}\cr}}
\vbox{\halign{#\hfil\cr\scalebox{0.7}
{\includegraphics{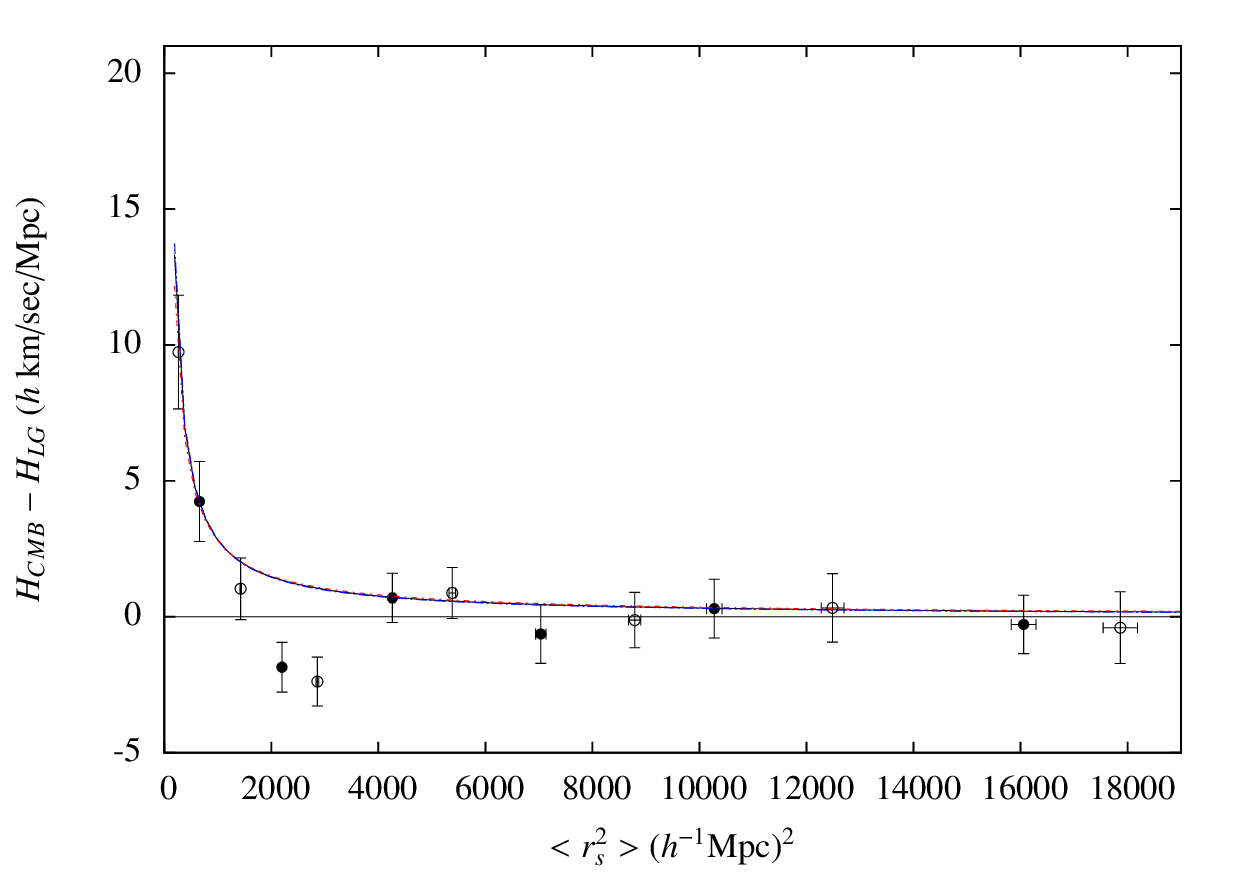}}\cr
\noalign{\vskip-5pt}\qquad{\bf(b)}\cr}}
}
\caption{Best fit power law to the radial variation in the spherically averaged
Hubble law in 2 configurations of: {\bf(a)} 11 shells; {\bf(b)} 8 shells. The
dashed blue (lower) curve indicates the best fit to primed shells only (empty
circles), the dotted red (upper) curve indicates the best fit to unprimed
shells only (filled circles). The solid black curve is the combined best fit
using both primed and unprimed shells.
The first data point -- corresponding to unprimed shell 1 -- is omitted in
each plot, as it is off the scale.
\label{Hdiff2}}}
\end{figure*}

We repeated the analysis using 8 shells rather than 11 to smooth out
variations that could interfere with the systematic boost offset. The second
configuration uses shells of width $18.75\hm$, starting from an inner cutoff of
$2\hm$ and $9.375\hm$ for unprimed and primed shells respectively. We find
$p=-0.89\pm 0.34$ for the primed shells and $p=-0.96\pm 0.26$ for the unprimed
shells. With a continuous variation of the inner shell boundary from $2$ --
$9.375\hm$, we arrive at a value of $p=-0.87\pm (0.33)\ns{stat}\pm(0.09)\ns{sys}
$. Fig.~\ref{Hdiff2}(b) shows the resultant best fit curves.

In Fig.~\ref{Hdiff2} we note a discrepancy between the best fit power law
and the negative values on shells in the range of $40$ -- $60\hm$ (or
$\ave{r_i^2}=1600$--$3600(\hm)^2$). This is understood to be the result of
structures in this particular range, which give rise to both a residual
monopole and dipole variation of the Hubble expansion in the LG frame. As
\citet{WSMW} show, there is evidence that the boost to the CMB frame somewhat
compensates for structures in this range. One finds that in the range
$40\lsim r\lsim 60\hm$ (and only in this range) the monopole
variation is less in the CMB frame, $(\de H_s)\Ns{CMB}<(\de H_s)\Ns{LG}$,
while the dipole magnitude is also less in the CMB frame,
becoming consistent with zero in the middle of the range. If the boost to the
CMB frame exactly compensated for structures in the range $40\lsim r\lsim
60\hm$ then the dipole magnitude should remain close to zero in shells at
larger distances. However, the magnitude of the CMB frame dipole increases to
a residual offset at large distances, while the LG frame dipole is consistent
with zero in most outer shells.

Thus it appears that the dipole almost -- but not entirely -- has the character
of a Lorentz boost dipole. Given that there are structures that give rise to a
residual nonlinear Hubble expansion in the range $40\lsim r\lsim60\hm$
we cannot expect a perfect power law fit (\ref{pl}) in Fig.~\ref{Hdiff2}.
However, the deviation from a power law is consistent with the observation that
$(H_s)\Ns{CMB}<(H_s)\Ns{LG}$ in the range over which the boost
almost compensates for nonlinear structures.

Now that we have verified the power law nature of the difference
$\HCMB-\HLG$ we must check whether this result is unique for the boost to
the LG frame. To investigate this we determine the Hubble constant in radial
shells for frames boosted arbitrarily with respect to the CMB, denoted by
frame ``X'', and then fit (\ref{pl}) to the resulting $\Delta H=
\HCMB-\HX$ curve. We vary the direction of the boost to frame X while
holding the magnitude constant, thus producing a sky map. We first choose a
magnitude of $635\kms$ corresponding to the boost from the CMB to LG
frame of reference.

\begin{figure*}
\centering\vskip-1.2\baselineskip
\scalebox{0.49}{\includegraphics{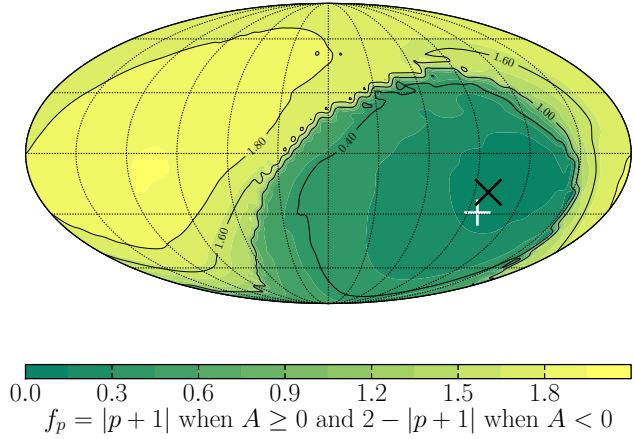}}
\caption{The best fit parameters to a systematic boost offset (\ref{pl})
for frames boosted from the CMB frame at $635\kms$. The black cross denotes the
boost to the LG frame and the white cross denotes the boost to the frame
with minimum variation in the spherically averaged Hubble law for a boost of
this magnitude, which will be discussed in \S \ref{chi_min}. \latitudes}
\label{boost_skymap}
\end{figure*}

To display these sky maps in a meaningful fashion we cannot simply plot
the value of $p$. Suppose that the CMB is boosted from a frame which has
$\Delta H_s=A(\ave{r_i^2}_s)^p$ with $p=-1$ and $A>0$, representing the best fit boost
offset. If one now boosts in the {\em opposite} direction by $635\kms$ then one
finds a best fit power law with $\Delta H_s=A(\ave{r_i^2}_s)^p$ with $p\approx-1$ but
$A<0$ since the CMB frame necessarily has the smaller value of $H_s$ on average.
In each case we must first of all determine whether (\ref{pl}) gives a
better overall fit with $A>0$ or $A<0$ -- given that {\em some} data points will
always be opposite to the overall trend. In Fig.~\ref{boost_skymap} we
plot\footnote{Both primed and unprimed shells are used (in the 11 shell case),
to produce a smoothed sky map without determining systematic uncertainties.}
\beq\label{fb}
\fb = \left\{
\begin{array}{lr}
|p+1|, &\ A\ge 0\\
2-|p+1|,& \ A<0
\end{array}
\right.
\eeq
which takes the value $\fb=0$ at the best fit with $A>0$ and $\fb=2$ at the
best fit with $A<0$. The
latter point turns out to be in the opposite direction, but not exactly
opposite the best fit direction, reflecting the uncertainties in the method.

Fig.~\ref{boost_skymap} shows that for a boost magnitude of $635\kms$ the
Local Group is indeed contained in a set of frames that display strong
evidence of being a minimum Hubble expansion variation frame as $p\approx -1$.
More precisely, the monopole Hubble expansion in the CMB frame compared
to frames boosted at $635\kms$ relative to the CMB has the mathematical
character of a systematic boost offset for directions close to that of the LG.
In Fig.~\ref{boost_skymap} we can see a distinctive difference between the
directions for which the boosted frame has the lesser variation ($A>0$, values
plotted closest to 0), and the directions for which the CMB frame has the
lesser variation ($A<0$, values plotted closest to 2).

We have verified that the additional monopole variation of the
Hubble expansion seen in the CMB frame does have the character of what is
expected by a boost from the LG frame, if the LG frame is close to the frame
in which the monopole variation is minimized. However, as yet we have not varied
the magnitude of the boost. Given the uncertainties we have already noted, it
is clear that any systematic boost offset will become very hard to confirm
statistically if the boost magnitude is small, since the other uncertainties
will then become dominant. The systematic boost offset method can only give a
rough indication of the boost direction for the large boosts rather than
defining a precise ``minimum Hubble expansion variation frame''. There are in
fact many degenerate frames.

\section{Minimizing the average spherical Hubble expansion variation}\label{min}

We will now investigate arbitrary variations of the boost magnitude and
direction to determine to what extent we can define a frame of reference in
which the monopole variation in the Hubble expansion is minimized.
\subsection{Variations in the ``nonlinear regime"}\label{chi_min}

Initially we will consider monopole variations with respect to a uniform
$\de H=0$ expectation below the scale of statistical homogeneity
($\lsim100\hm$). This is quantified by summing the mean square differences
of (\ref{Delta_H}), to give a statistic
\beq\label{chi_sq}
\chN(n_f,n_i)=\sum_{i=n_i}^{n_f} \frac{\Ha^4\de H_i^2}{\Ha^2\sigma\Z{H_i}^2
+H_i^2\sigma\Z{\bar H\X0}^2},
\eeq
where $n_f$ and $n_i$ define the upper and lower shells included in the range
of the calculation respectively. We will take the primed shells with, $n_i=1'$
and $n_f=8'$, covering the range $6.25<r\le106.25\hm$. This includes all
data potentially in the regime of nonlinear Hubble expansion, while excluding
the innermost unprimed shell which may have incomplete sky coverage.

In identifying a frame of ``minimum nonlinear Hubble variation'' we must also estimate an uncertainty about this minimum. Since we do not expect an exact fit to a linear Hubble expansion in {\em any} frame it is inappropriate to use the ordinary confidence intervals of the $\chN$ distribution, as this would not give a fair measure of the improvement in uniformity between two reference frames, instead rendering all reference frames at least\footnote{At the global minimum we find a $\chN$ per degree of freedom of 2.0.} $1\,\si$ from the ``expectation''. Instead we follow \citet{WSMW} in using Bayesian statistics to give a measure of the relative probability of uniformity of Hubble expansion in different reference frames. In particular, using the complementary incomplete gamma function
for the $\chN$ distribution we directly calculate the probabilities
$P\Ns{U}$, $P\Ns{V}$ of the spherically averaged Hubble law
giving a value that coincides with $\Ha$, using the shells $1'$ to $8'$,
for the pair of reference frames $U$ and $V$.
A Bayes factor $B=P\Ns{U}/P\Ns{V}$ is then computed. We use the bounds $\ln B=1,3,5$
to denote confidence regions, where the range $1<\ln B\le3$ represents
{\em positive} Bayesian evidence, $3<\ln B\le5$ represents {\em strong} Bayesian evidence
and $\ln B>5$ represents {\em very strong} evidence \citep{KR}. Thus if we
take frame $U$ to be the frame with the minimum $\chN$ within a specified
set of reference frames, any boosted frame $V$ with Bayes factor $\ln B\le1$
relative to $U$ can be considered statistically equivalent to
$U$ in the sense that the difference in probability is {\em``not worth more
than a bare mention''} \citep{KR}.

The frame of reference with the minimum monopole variation is found using a
downhill optimization with (\ref{chi_sq}). This reveals a global minimum
variation (MV) frame for a boost in the direction $(\ell,b)=(59.3\deg,16.6\deg)$
with magnitude $740.6^{+258.4}_{-487.0}\kms$ with respect to the LG frame. (The
uncertainty is taken from the $\ln B\le1$ bound.) Even though the boost is
large, the uncertainties are also large. The Bayes factor $\ln\left(P\Ns{MV}/
P\Ns{LG}\right)=2.77$ constitutes positive but not strong evidence for the
MV frame relative to the LG frame. However,
the corresponding boost from the CMB reference frame to the global minimum
frame has a magnitude $1203_{-458}^{+375}\kms$, which is even larger and now
very significantly nonzero, with $\ln\left(P\Ns{MV}/P\Ns{CMB}\right)=33.3$,
which is extremely strong evidence against the CMB frame.

\begin{figure*}
\vskip-1.2\baselineskip
\vbox{\centerline{\vbox to 230pt{\halign{#\hfil\cr\quad\scalebox{0.49}
{\includegraphics{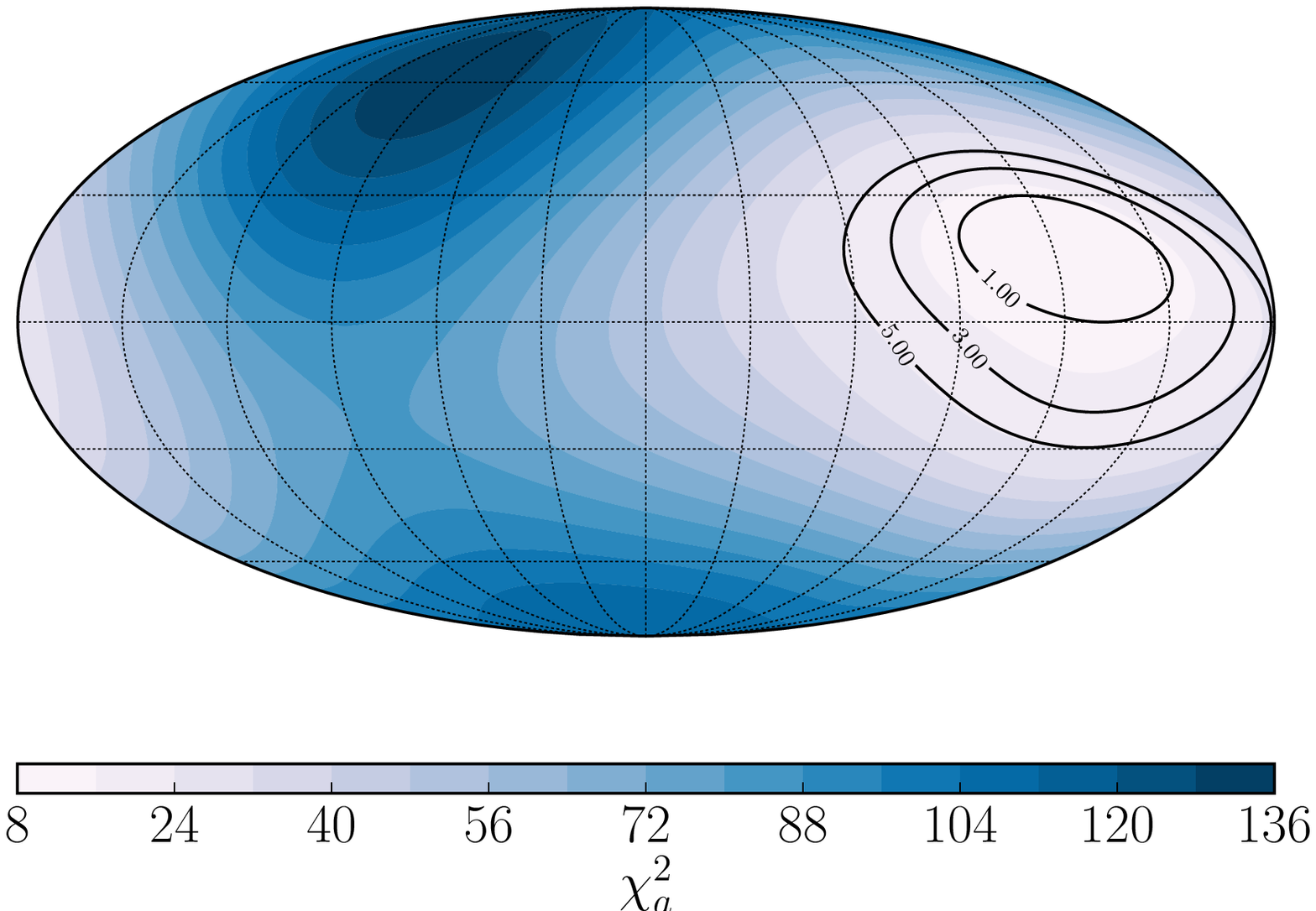}}\cr
\noalign{\vskip-50pt}\qquad\qquad{\bf(a)}\cr}\vfil}\hskip-20pt
\vbox to 230pt{\halign{#\hfil\cr\scalebox{0.49%0.6
}{\includegraphics{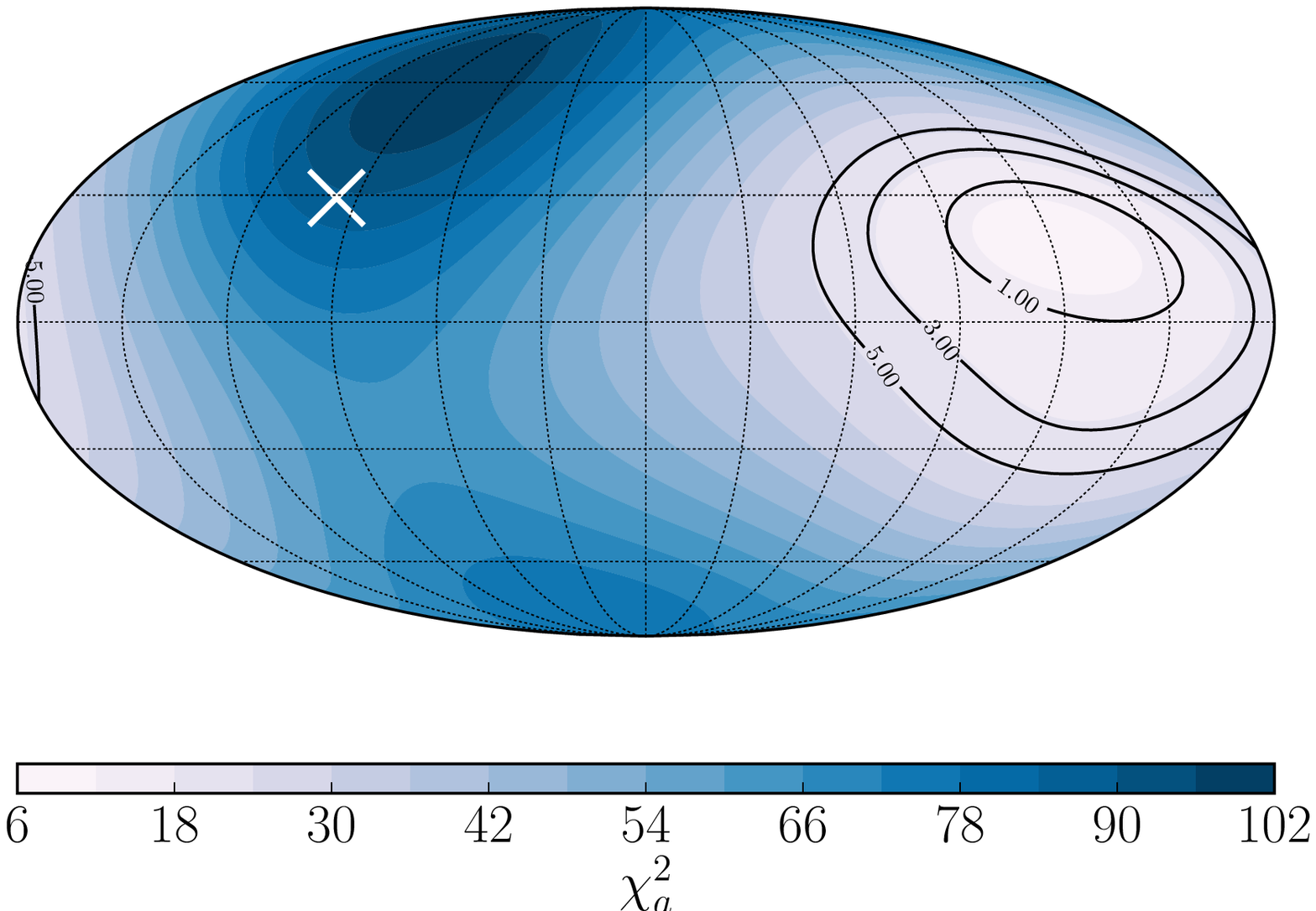}}\cr
\noalign{\vskip-50pt}\quad\qquad{\bf(b)}\cr}}}}
\caption{Contour maps of angular variation of $\chN$ for two choices of
boost magnitude with respect to the Local Group: {\bf(a)} 740$\kms$ {\bf(b)}
635$\kms$. The solid contours show the levels of Bayesian evidence
$\ln B=\ln(P\Ns{min}/P_{(\ell,b)})$ for each direction $(\ell,b)$ with respect to the frame of minimum $\chi^2_a$ in each case.
The white cross on (b) shows the direction of the boost to the CMB frame
(also of magnitude $\approx635\kms$). \latitudes\label{conf}}
\end{figure*}
\begin{figure*}
\centering\scalebox{0.45}{\includegraphics{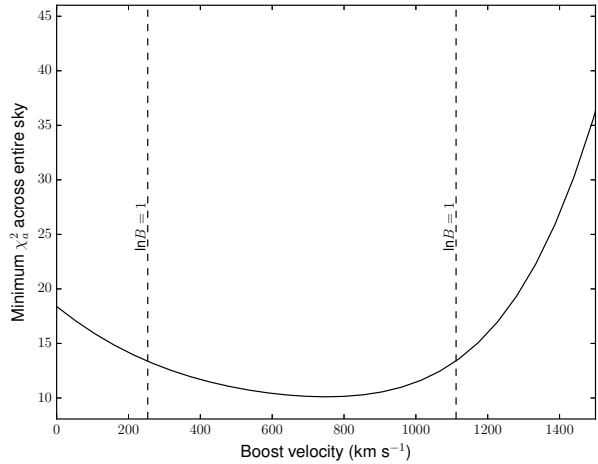}}
\caption{Variation of the minimum $\chN$ with fixed boost velocity, $v$.
The locus of $(\ell,b)$ values for which this minimum is found lies within
$\approx 20\deg$ of the galactic plane for cases in which
$\ln B=\ln(P\Ns{MV}/P\Ns{V})\le1$ as indicated by the dashed line (where MV is
the global minimum and V any frame along the locus).}
\label{va}
\end{figure*}

To visualize the distribution of $\chN$ in the 3-dimensional parameter
space $\{v,\ell,b\}$ we show two angular slices at fixed values of $v$ in
Fig.~\ref{conf}, and a slice along the locus of $(\ell,b)$ values for which
$\chN$ is minimized for fixed $v$ in Fig.~\ref{va}. The angular distribution
in Fig.~\ref{conf} is similar for both velocities. However, the
Bayesian confidence regions grow large as $v$ is decreased, eventually
taking up the whole sky for very small velocities. Only for large boosts is
there a well-defined boost direction that reduces monopole variation.

In Fig.~\ref{va} for each boost magnitude we locate the direction of
minimum $\chN$ and plot the corresponding value. The angular coordinates
$(\ell,b)$ of the minimum are different in each case. The distribution of
$\chN$ relative to the $\ln B\le1$ confidence interval (dashed vertical lines
on Fig.~\ref{va}(a)) reveals the main problem in constraining a
``minimum variation frame'' by this technique. The near flat distribution
of $\chN$ values within the $\ln B\le1$ bound of the global minimum are found
to lie on a locus of boost directions $(\ell,b)$ which all lie close to the
galactic plane. This is of course the Zone of Avoidance region, where the \CS
sample lacks data, due to the Milky Way obscuring distant galaxies.
Evidently, we are free to perform large boosts in the plane of the galaxy,
as the data is not constrained there. This hypothesis could be checked by
simulating data with the same characteristics as the \CS sample, using exact
solutions of Einstein's equations \citep{Bolejko}. Such an investigation is
beyond the scope of this paper. For now, with the available data we can only
conclude that the LG frame is not ruled out as the standard of rest by
this criterion, since the Bayesian distinction between the two frames
with $|\ln B|=2.77$ is not strong.

\begin{table*}
\begin{center}
\begin{minipage}[t]{\linewidth}
\caption{Hubble expansion variation in radial shells in minimum Hubble
expansion variation (MV) and LG frames. Spherical averages (\ref{Hsav}) are
computed for two different choices of shells,
$r_s<r\le r_{s+1}$, the second choice being labeled by primes. In each case we
tabulate the inner shell radius, $r_s$; the weighted mean distance, $\mr_s$;
the shell Hubble constants, $(H_s)\Ns{LG}$ and $(H_s)\Ns{MV}$ in the LG and
MV frames, and their uncertainties determined by linear regression within each
shell, together with its ``goodness of fit'' probability $Q_s$ and reduced
$\chi^2$ (for $\nus=N_s-1$); $\ln B$ where $B$ is the Bayes factor for the
relative probability that the MV frame has more uniform $\de H_s=0$ than the
LG frame when $\chi^2$ is summed in all shells with $r>r_s$.
$H_s$ and $\bs_s$ are given in units \protect{$h\kmsMpc$.}
\label{shell}}
\centering
\begin{tabular}{|lrrrrrrrrrrr|}
\hline\hline
Shell $s$&1&2&3&4&5&6&7&8&9&10&11\B\T\\
$N_s$& 92&505&514&731&819&562&414&304&222&280& 91\B\T\\
$r_s$ ($\hm$)& 2.00& 12.50& 25.00& 37.50& 50.00& 62.50& 75.00& 87.50&100.00&112.50&156.25\B\T\\
$\bar{r}_s$ ($\hm$)& 5.43& 16.33& 30.18& 44.48& 55.12& 69.24& 81.06& 93.75&105.04&126.27&182.59\B\T\\
$(H_s)_{LG}$&117.9&103.1&106.5&105.5&104.8&102.1&102.8&103.2&103.7&102.4&101.0\B\T\\
$(\bar{\sigma}_s)_{LG}$&4.6&1.4&1.0&0.7&0.7&0.7&0.9&0.9&1.0&0.8&1.7\B\T\\
$(Q_s)_{LG}$&0.000&0.000&0.000&0.000&0.998&0.940&1.000&1.000&1.000&0.993&0.999\B\T\\
$(\chi_s^2/\nus)_{LG}$&23.656&7.767&2.185&1.419&0.864&0.909&0.594&0.542&0.622&0.803&0.590\B\T\\
$(H_s)_{MV}$&118.5&102.9&106.7&104.5&104.8&102.9&102.6&103.9&104.9&102.7&102.0\B\T\\
$(\bar{\sigma}_s)_{MV}$&4.6&1.4&1.1&0.7&0.7&0.7&0.9&0.9&1.0&0.8&1.7\B\T\\
$(Q_s)_{MV}$&0.000&0.000&0.000&0.000&0.330&0.887&1.000&1.000&1.000&0.964&0.999\B\T\\
$(\chi_s^2/\nus)_{MV}$&29.130&12.320&3.037&2.005&1.021&0.928&0.682&0.600&0.667&0.854&0.603\B\T\\
$\ln B$ ($r\ge r_s$)& 3.53& 2.85& 2.79& 1.99& 0.85& 0.33& 0.36& 0.14& 0.07& 0.45&\B\T\\
\B\T\\
Shell $s$&1$'$&2$'$&3$'$&4$'$&5$'$&6$'$&7$'$&8$'$&9$'$&10$'$&11\B\T\\
$N_s$&321&513&553&893&681&485&343&273&164&206& 91\B\T\\
$r_s$ ($\hm$)& 6.25& 18.75& 31.25& 43.75& 56.25& 68.75& 81.25& 93.75&106.25&118.75&156.2\B\T5\\
$\bar{r}_s$ ($\hm$)& 12.26& 23.46& 37.61& 49.11& 61.74& 73.92& 87.15& 99.12&111.95&131.49&182.59\B\T\\
$(H_s)_{LG}$&103.5&103.5&103.9&106.6&103.9&102.0&103.2&103.6&101.6&102.7&101.0\B\T\\
$(\bar{\sigma}_s)_{LG}$&1.8&1.1&0.9&0.7&0.8&0.8&0.9&0.9&1.0&0.9&1.7\B\T\\
$(Q_s)_{LG}$&0.000&0.000&0.000&0.031&0.960&1.000&1.000&1.000&0.996&0.999&0.999\B\T\\
$(\chi_s^2/\nus)_{LG}$&11.427&3.246&1.792&1.090&0.907&0.701&0.592&0.608&0.728&0.711&0.590\B\T\\
$(H_s)_{MV}$&102.7&104.3&103.3&106.1&104.2&102.9&102.9&104.7&102.7&102.9&102.0\B\T\\
$(\bar{\sigma}_s)_{MV}$&1.8&1.1&0.9&0.7&0.8&0.8&0.9&0.9&1.0&0.9&1.7\B\T\\
$(Q_s)_{MV}$&0.000&0.000&0.000&0.000&0.481&1.000&1.000&1.000&0.967&0.997&0.999\B\T\\
$(\chi_s^2/\nus)_{MV}$&18.547&4.940&2.429&1.428&1.002&0.734&0.704&0.613&0.807&0.752&0.603\B\T\\
$\ln B$ ($r\ge r_s$)& 2.38& 2.32& 2.34& 1.93& 0.55& 0.33& 0.41& 0.14& 0.24& 0.50& \B\T\\
\hline
\end{tabular}
\end{minipage}\hfill
\end{center}
\end{table*}

We have found a set of degenerate frames of reference which might
be taken as the minimum average monopole variation frame. Although large
uncertainties still exist, we are able to see that the boost to the CMB frame,
shown by a white cross in Fig.~\ref{conf}(a), is far from our degenerate set of
possible boosts to the minimum variation frame. In Table \ref{shell} we make
a comparison similar to Table~1 of \citet{WSMW} between the LG frame and the
global minimum $\chN$ (MV) frame. In particular, we directly calculate the
probabilities $P\Ns{LG}$, $P\Ns{MV}$ of the spherically averaged Hubble law
giving a value that coincides with $\Ha$, when all shells larger
than a given shell are included. If we exclude the innermost unprimed
shell 1, then the Bayesian evidence for the MV frame having a more uniform
average Hubble expansion than the LG frame, lies at most in the range
$1<\ln B\le 3$, which represents positive but not strong Bayesian
evidence \citep{KR}. By contrast, the Bayesian evidence that the LG
frame has a more uniform spherically average Hubble expansion than the
CMB frame is very strong with $\ln B>5$ \citep{WSMW}.

\subsection{Variations in the ``linear regime"}

The goodness of fit, $Q_s$, of a linear Hubble law in the innermost shells of
Table~\ref{shell} is poor, as we would expect since these shells are in the
{\em nonlinear} regime. Beyond approximately $75\hm$ we expect to pass into
the linear regime \citep{Scrimgeour2012}, and this is seen with the decreasing
values of $\chi_s^2/\nus$, where $\chi_s^2$ is given by (\ref{Hs_chi}) for
each shell.

However, contrary to expectation, the MV frame determined from (\ref{chi_sq})
has an overall poorer goodness of fit $Q_s$ than the LG frame to a linear
Hubble law in shells with $s\ge4'$, including in particular in shells $9'$
and $10$ which should be in the linear regime. We note that the asymptotic
value, $\Ha$, is 1\% higher in the MV frame as compared to the LG frame,
and this may contribute to $\de H_s$ being smaller, even though the goodness
of fit in individual shells is poorer in some cases\footnote{The value of
$\Ha$ is also 1\% larger in the LG frame as compared to the CMB frame, but
by contrast the goodness of fit in the linear regime shells is better in the
LG frame than in the CMB frame \citep{WSMW}.}.

\begin{figure*}
\centering\vskip-1.2\baselineskip
\scalebox{0.45}{\includegraphics{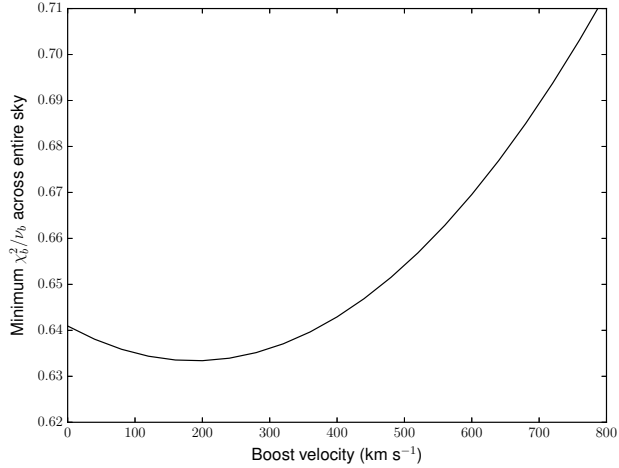}}
\caption{Variation of the minimum $\chL/\Nb$ with fixed boost velocity, $v$.}
\label{vb}
\end{figure*}
\begin{figure*}
\vbox{\centerline{\vbox{\halign{#\hfil\cr\scalebox{0.45}
{\includegraphics{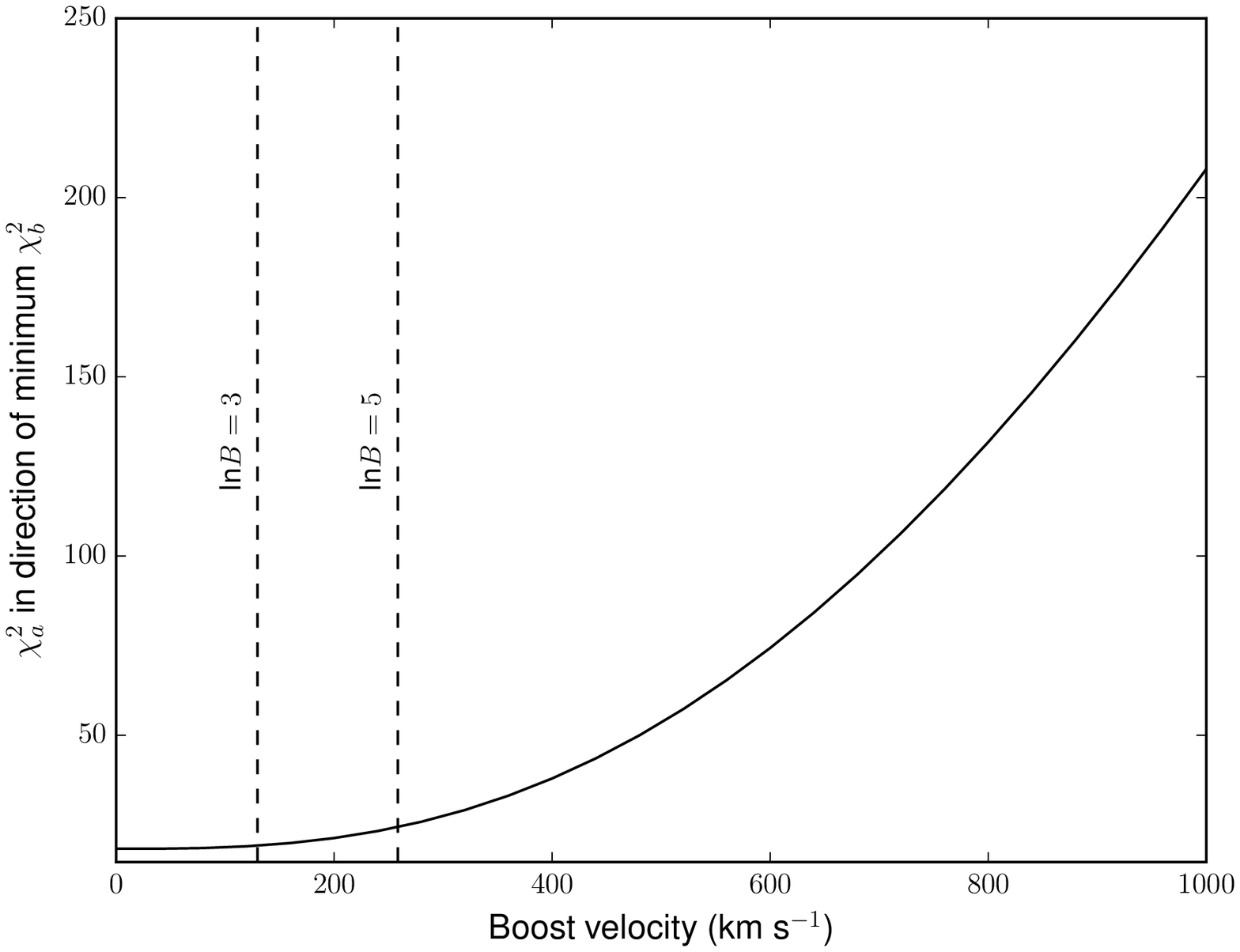}}\cr
\noalign{\vskip-5pt}\qquad{\bf(a)}\cr}}
\vbox{\halign{#\hfil\cr\scalebox{0.45}
{\includegraphics{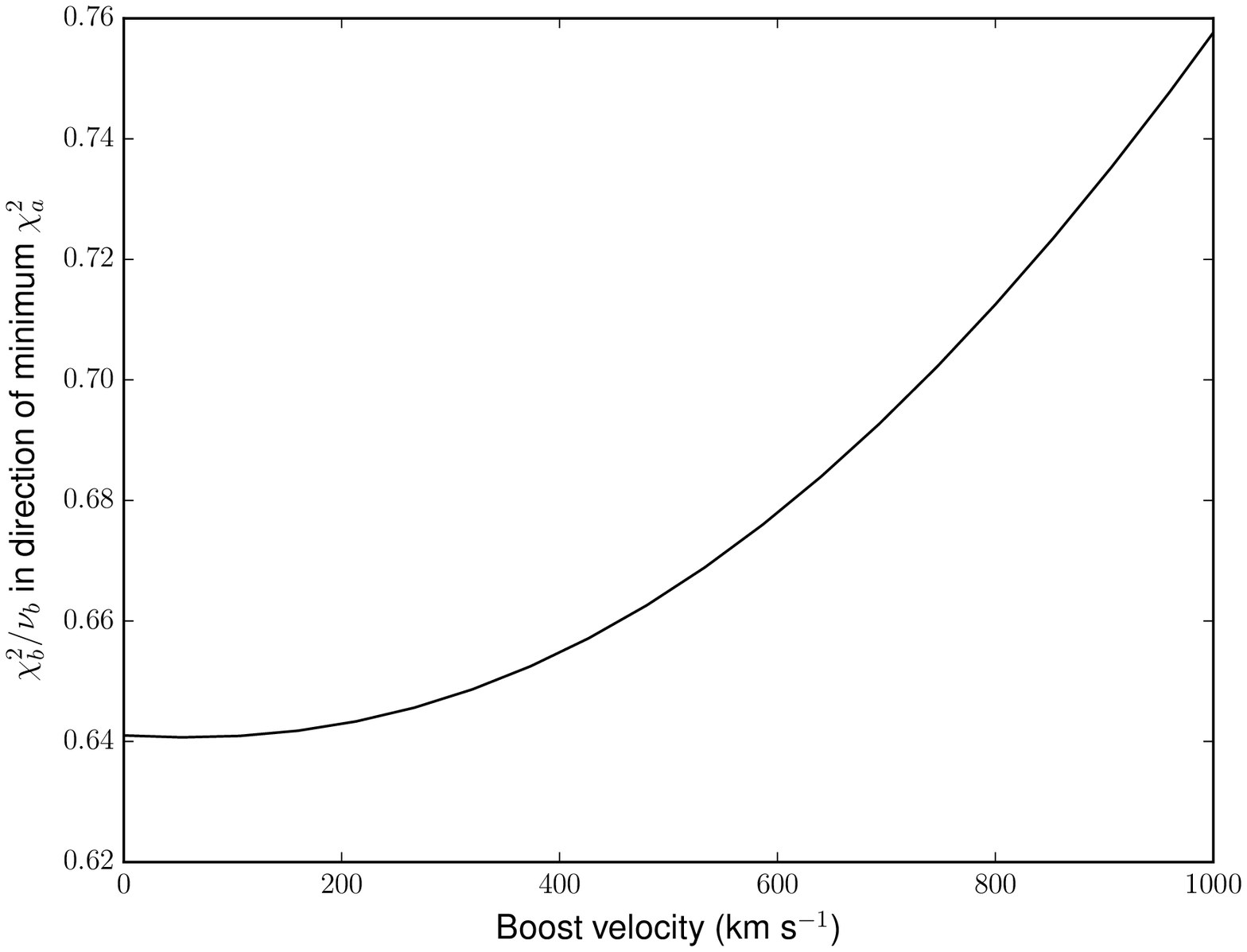}}\cr
\noalign{\vskip-5pt}\qquad{\bf(b)}\cr}}
}
\caption{{\bf(a)} Variation of $\chN$ along a locus of $(\ell,b)$ values
for which $\chL$ is minimized at each fixed boost velocity $v$. {\bf(b)}
Variation of $\chL/\Nb$ along a locus of $(\ell,b)$ values for which $\chN$
is minimized at each fixed boost velocity $v$.\label{vspace2}}}
\end{figure*}
Any true candidate for a minimum Hubble expansion variation frame should also
clearly demonstrate the emergence of a linear Hubble law consistent with the
existence of a statistical homogeneity scale. The $\chN$ statistic
(\ref{chi_sq}) involves minimizing the variation $\de H_s=(H_s-\Ha)/\Ha$
relative to the asymptotic Hubble constant. But a boost can also alter $\Ha$
in a way which makes for a worse goodness of fit to a linear Hubble law.
Therefore it is not a completely suitable candidate statistic.

In order to quantify the emergence of a linear Hubble law we
will therefore alternatively minimize the quantity
\beq\label{xi}
\chL=\sum_{s=7}^{11}\chi_s^2
\eeq
where $\chi_s^2$ is given by (\ref{Hs_chi}) in the $s$th shell. This sum is
performed over the unprimed configuration of shells as shell 7 has an inner
cutoff near the boundary of the nonlinear and linear regimes. Thus (\ref{xi})
gives a measure of the goodness of fit to a linear Hubble law averaged over
the outer 5 shells, without normalizing the asymptotic Hubble constant.
Defining the total degrees of freedom $\Nb=\left(\sum_{s=7}^{11} N_s\right)-1$,
we find $\chL/\Nb=0.631$ for the LG frame, $\chL/\Nb=0.692$ for the MV (minimum
$\chN$) frame, and $\chL/\Nb=0.653$ for the CMB frame. Thus even the CMB
frame shows a clearer emerging linear Hubble law than the minimum $\chN$
frame.

To determine whether there is any frame with a more definitive emerging linear
Hubble law than the LG frame, we determine the distribution of (\ref{xi})
upon making arbitrary boosts with respect to the LG frame.

In Fig.~\ref{vb} we locate the direction of minimum $\chL$ at each boost
magnitude and plot the corresponding value, analogously to Fig.~\ref{va}. We
find a best fit boost of $222.3\kms$ in the direction $(\ell,b)=(241.84\deg,
70.53\deg)$ with respect to the LG frame, with a value of $\chL/\Nb=0.621$.
This is $45\deg$ from the direction of the residual CMB temperature dipole in
the LG frame, and so does not appear related.

In Fig.~\ref{vspace2} for each boost magnitude we calculate the value of
$\chN$ in an $(\ell,b)$ direction determined by minimizing with respect to
$\chL$, and vice versa. Thus, we compute the locus of $(\ell,b)$ values in
the $\{v,\ell,b\}$ parameter space that minimize $\chL$ for each fixed $v$,
and then compute $\chN$ at these parameter values, and vice versa. It is
apparent that making boosts of the order $\gsim100\kms$ along the locus of
$(\ell,b)$ values which minimize $\chL$ results in an increase in
$\chN$ to values which are disfavoured by Bayesian evidence.
This is because making improvements in $\chL$ requires boosts in $(\ell,b)$
directions away from our minimum $\chN$.

\begin{figure*}
\vskip-1.2\baselineskip
\vbox{\centerline{\vbox to 230pt{\halign{#\hfil\cr\quad\scalebox{0.49}
{\includegraphics{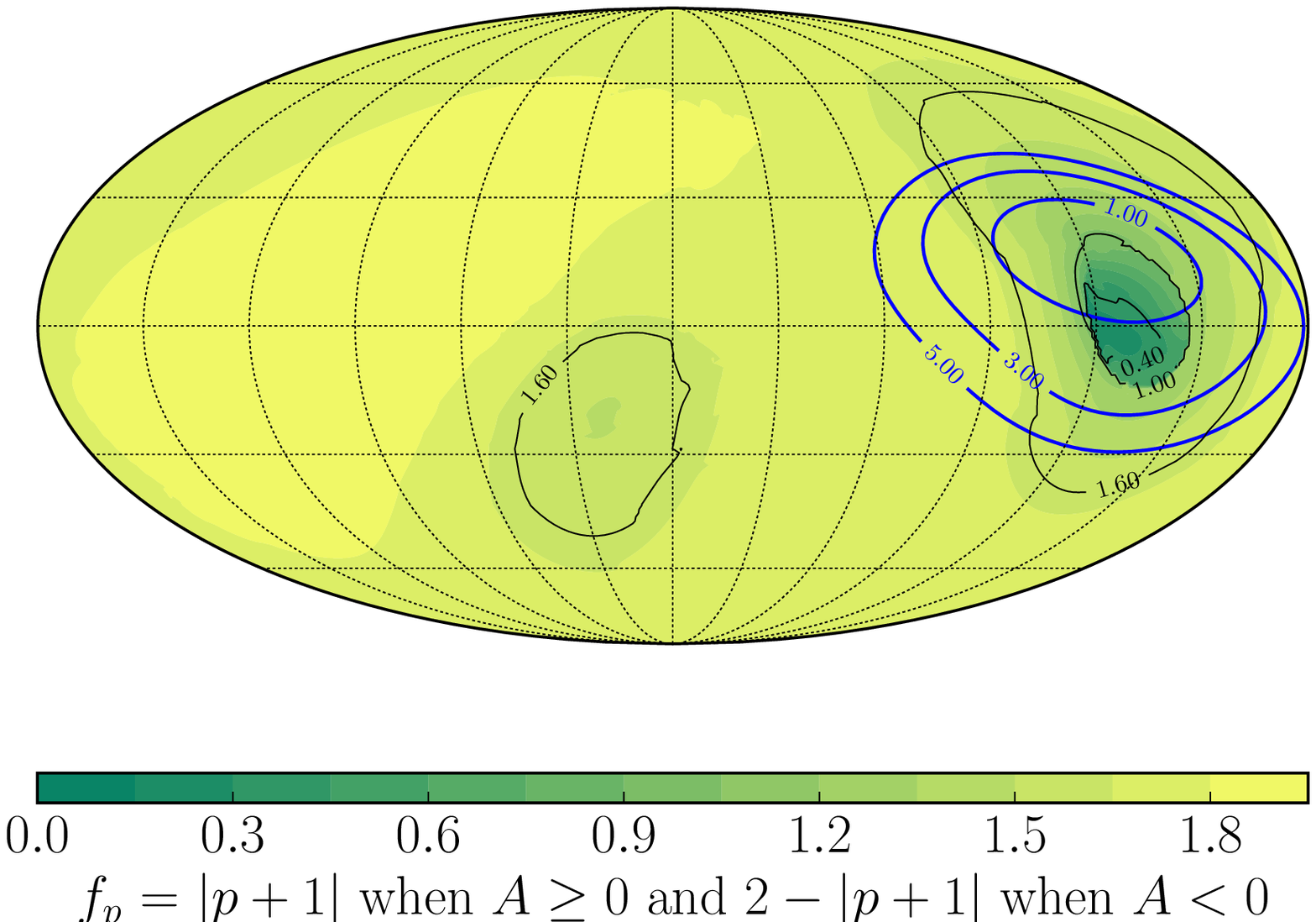}}\cr
\noalign{\vskip-50pt}\qquad\qquad{\bf(a)}\cr}\vfil}\hskip-20pt
\vbox to 230pt{\halign{#\hfil\cr\scalebox{0.49%0.6
}{\includegraphics{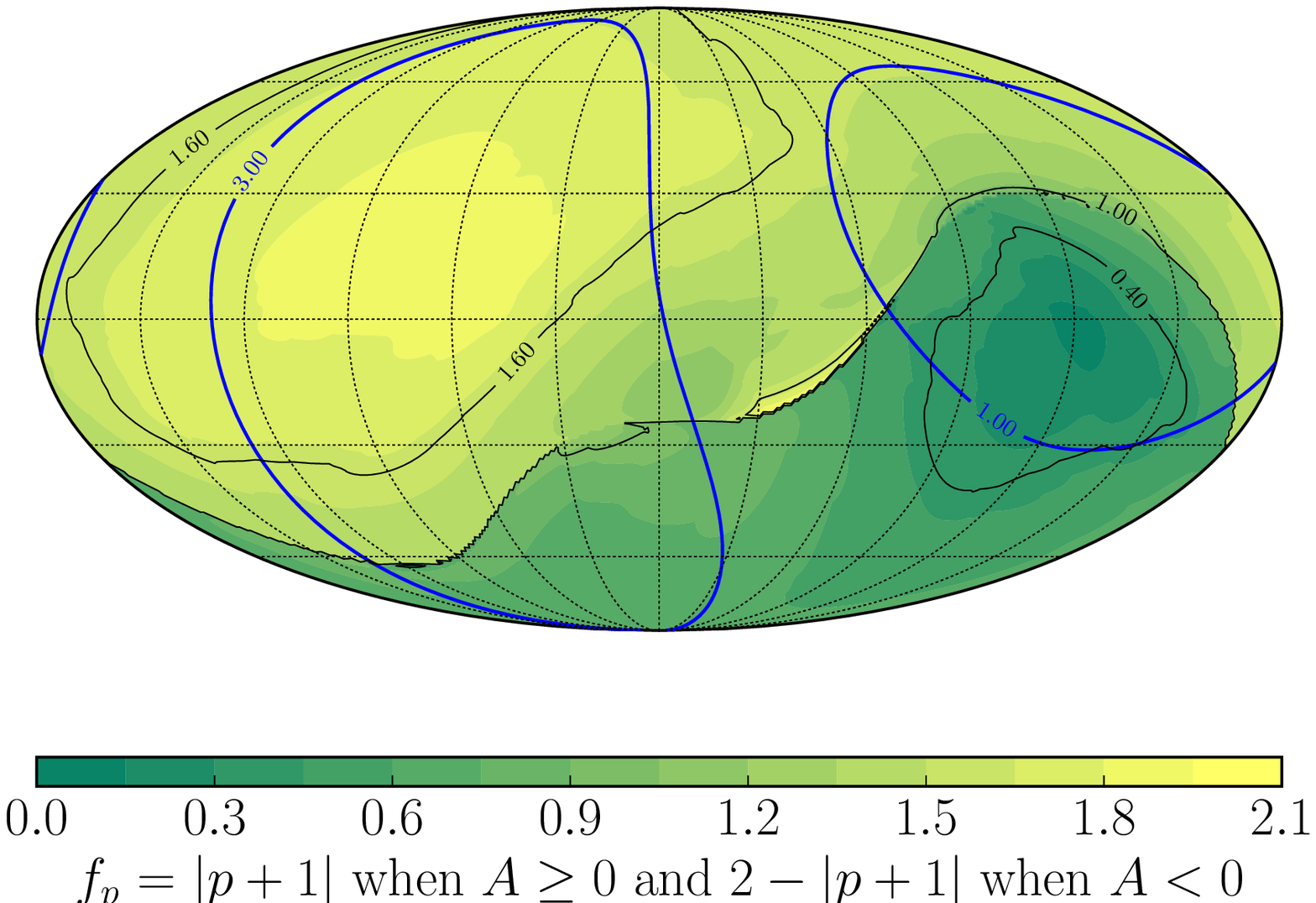}}\cr
\noalign{\vskip-50pt}\quad\qquad{\bf(b)}\cr}}}}
\caption{The best fit parameters to a systematic boost offset for boosts from
the Local Group of magnitude: {\bf(a)} 740$\kms$; {\bf(b)} 200$\kms$. The
thick blue contours denote the corresponding $\chN$ distribution.
The solid contours show the levels of Bayesian evidence $\ln B=\ln(P\Ns{min}/
P\Ns{(\ell,b)})=1,3,5$ for each direction $(\ell,b)$ with respect
to the frame of minimum $\chi^2_a$ at each boost magnitude.
In (a) $\ln B=1,3,5$ contours are visible while in (b) only
the $\ln B=1,3$ contours are visible, as no directions have $\ln B>5$
in that case. \latitudes}
\label{boost_skymaps}
\end{figure*}

However, we note that: (i) all values of $\chL$ shown in Fig.~\ref{vb} and
\ref{vspace2} are consistent with very probable fits, given the low values of
$\chL/\Nb$; (ii) there are less data points per se in those outer
shells used in $\chL$ as compared to the inner shells used in $\chN$; and
(iii) local boosts have a relatively small effect on the values of $cz_i$ in
the outer shell. These facts mean we should also exercise caution about drawing
strong conclusion solely from the minimization of $\chL$. The primary utility
of the $\chL$ statistic is not as a sole discriminator, but as a check against
the potential bias in the $\chN$ statistic due to its anchoring to the value
$\Ha$ in the outermost shell.

Finding a joint minimum for $\chN$ and $\chL$ would be feasible if
the $(\ell,b)$ values of each global minimum were close on the sky. However,
this is not the case -- the global minimum for $\chN$ is at an angle of
$93\deg$ from the global minimum of $\chL$, i.e., they are roughly
orthogonal. As there is no unbiased way to weight these two statistics we
cannot set out to determine a weighted minimum as any result would be highly
sensitive to our choice of weighting.

\subsection{Systematic boost offsets from the Local Group}\label{boosts}

Neither statistic $\chN$ nor $\chL$ appears entirely satisfactory for
establishing a global minimum expansion variation frame. The $\chN$
statistic is the better measure of Hubble expansion variation in the nonlinear
regime but is also affected by potential bias in the anchoring of $\Ha$. The
most we can say is that there is a freedom to perform large boosts in the
plane of the galaxy, given the lack of data in the Zone of Avoidance.

If $\chN$ is taken as the better statistic, then a criterion for breaking
the boost degeneracy may be possible by returning to the systematic boost
offset analysis of \S \ref{LGframe}. Any true best fit frame should show a
clear signal of a boost offset (\ref{eq1}) with respect to the Local Group.
The ``best'' boost offset can be characterized in 3 ways, each with its own
challenges:

\begin{enumerate}[label={(\arabic*)}]
\item Determine the boost for which $p=-1$. This is hindered by
the fact that there are many boosts that satisfy this criterion, at almost every
magnitude from the LG.
\item From the value of $A$ in (\ref{pl}) determine a derived boost
velocity, $\vd$. Any boost offset should have $\vd$ consistent with
the true boost magnitude $v_{true}$ within uncertainties. However, this is
difficult due to the large uncertainties associated with the value of $A$.
\item Determine a measure of variation in the fit of the boost offset, given
by (\ref{mindist}) in the Appendix. This is also problematic since all fits
are extremely good due to the large uncertainties in the $H_s$.
\end{enumerate}

\begin{figure*}
\vbox{\centerline{\vbox{\halign{#\hfil\cr\scalebox{0.45}
{\includegraphics{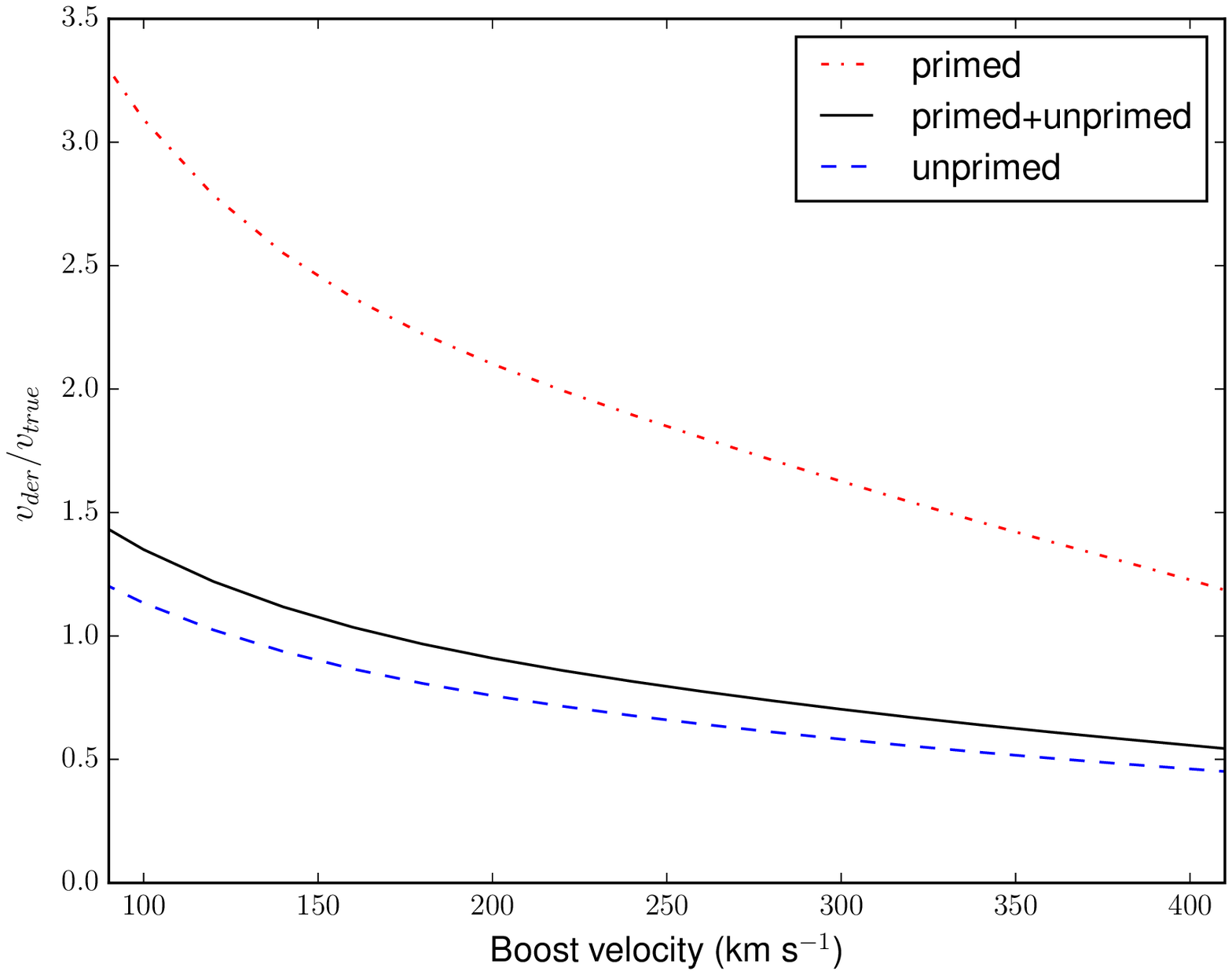}}\cr
\noalign{\vskip-5pt}\qquad{\bf(a)}\cr}}
\vbox{\halign{#\hfil\cr\scalebox{0.45}
{\includegraphics{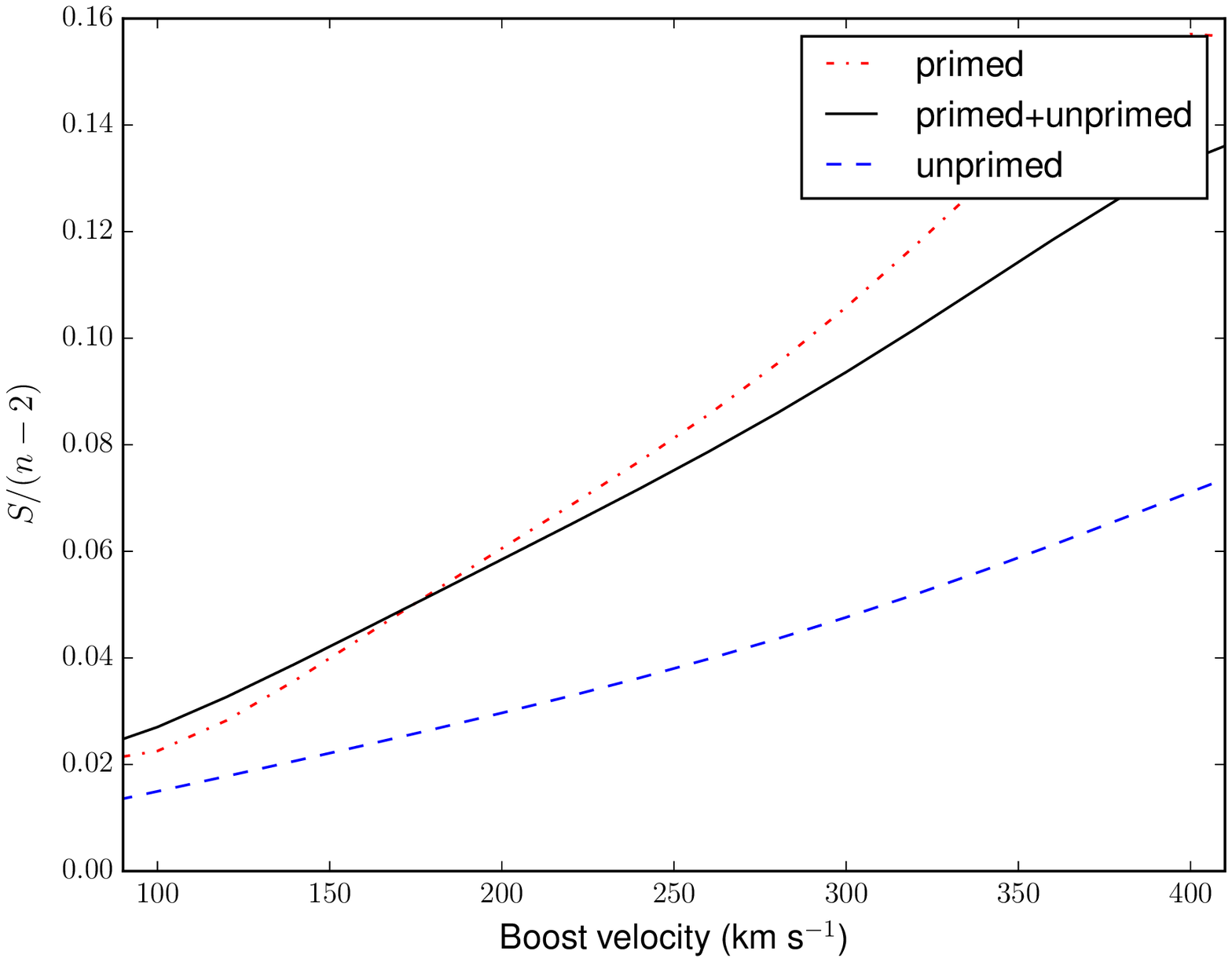}}\cr
\noalign{\vskip-5pt}\qquad{\bf(b)}\cr}}
}
\caption{{\bf(a)} The ratio of the derived velocity from the best fit power
law and the true boost velocity, at each $v$ the direction is determined by
the best fit for which $p$ is closest to $-1$. {\bf(b)} The statistical
variation in the same direction as determined by $S/(n-2)$ from (\ref{mindist}),
where $n$ is the number of data points being fit.\label{vder}}}
\end{figure*}

We will therefore use method (1) to determine the direction of the boost on
the sky, and then consider (2) and (3) to constrain the magnitude\footnote{The
choice of shell boundaries introduces systematic uncertainties in this analysis.
In the production of the sky maps in Fig.~\ref{boost_skymaps} we calculate both
primed and unprimed shells and fit the power law (\ref{pl}) to all points.
For Fig.~\ref{vder} we consider fits to primed and unprimed only, and to
both.}.

First we check for a systematic boost offset for the global $\chN$ minimum
frame determined in \S \ref{chi_min}.
A sky map of $\fb$ values as given by (\ref{fb}) for
boosts of magnitude $740\kms$ is given in Fig.~\ref{boost_skymaps}(a), with the
$\ln B\le 1$ and $1<\ln B \le 3$ confidence regions for $\chN$ displayed.
We note that there are in fact boosts with values of $p\approx-1$ consistent
with the $\chN$ minimum. However, these directions are far more constrained
and do not align with the exact minimum. In addition, the $(\ell,b)$ direction
at this magnitude with the best fit to (\ref{pl}) has an inconsistent
value of the derived velocity. Thus we do not see a clear systematic boost
offset between the Local Group and the frame corresponding to the global
minimum $\chN$, further ruling this out as a potential candidate for the
standard of rest we are looking for.

The next step is a global search for the best systematic boost offset from the
LG. In order to further understand the angular distribution of $\fb$ values
(\ref{fb}) for boosts from the LG frame, we arbitrarily choose a boost
magnitude of $200\kms$ and plot $f_p$ with respect to $(\ell,b)$ in
Fig.~\ref{boost_skymaps}(b). We have found that for all interpolating velocities
between the $200\kms$ and $740\kms$ cases displayed in Figure
\ref{boost_skymaps} there is a region of $(\ell,b)$ values for which
$\fb\approx 0$. Thus in order to use this method to find a realistic systematic
boost offset we must use an additional criterion.

In Fig.~\ref{vder} we plot the values of $\vd/\vt$ and $S/(n-2)$ (for $n$
data points) from (\ref{mindist}) with respect to the boost magnitude, where
for each magnitude the $(\ell,b)$ direction is that for which $p$ is closest to
$-1$ and $A>0$ (i.e. $\fb\approx0$). Thus, we can use these additional
quantities to constrain a systematic boost offset along the locus of $(\ell,b)$
directions in the 3-dimensional $\{v,\ell,b\}$ parameter space. The expected
value of $S$ has a $\chi^2$ distribution for $(n-2)$ degrees of freedom, and
thus $S/(n-2)$ has an expectation value of unity \citep{York04}. Clearly,
the values of $S/(n-2)$ in Fig.~\ref{vder}(b) are consistent with a very good
fit to (\ref{pl}) for all boosts. Our inability to tightly constrain
the boost magnitude is no doubt due to the lack of data in the Zone of
Avoidance and large uncertainties in the values of $\HLG-\HX$. Although
(\ref{mindist}) it is not useful for constraining the boost
magnitude, we nonetheless see that the ratio of derived and true
velocities in Fig.~\ref{vder} does show a meaningful difference on
this interval.

\begin{figure*}
\vbox{\centerline{\vbox{\halign{#\hfil\cr\scalebox{0.45}
{\includegraphics{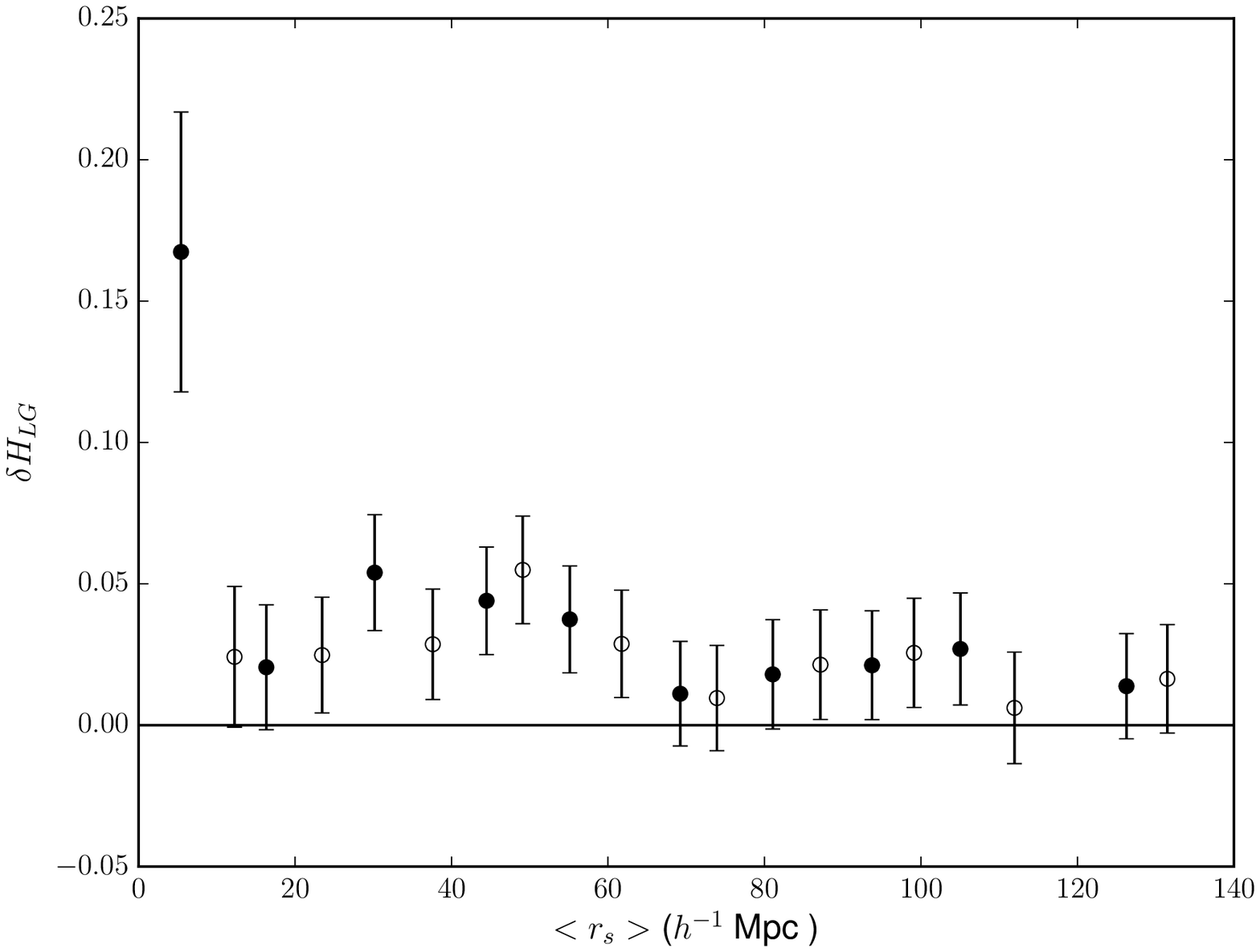}}\cr
\noalign{\vskip-5pt}\qquad{\bf(a)}\cr}}
\vbox{\halign{#\hfil\cr\scalebox{0.45}
{\includegraphics{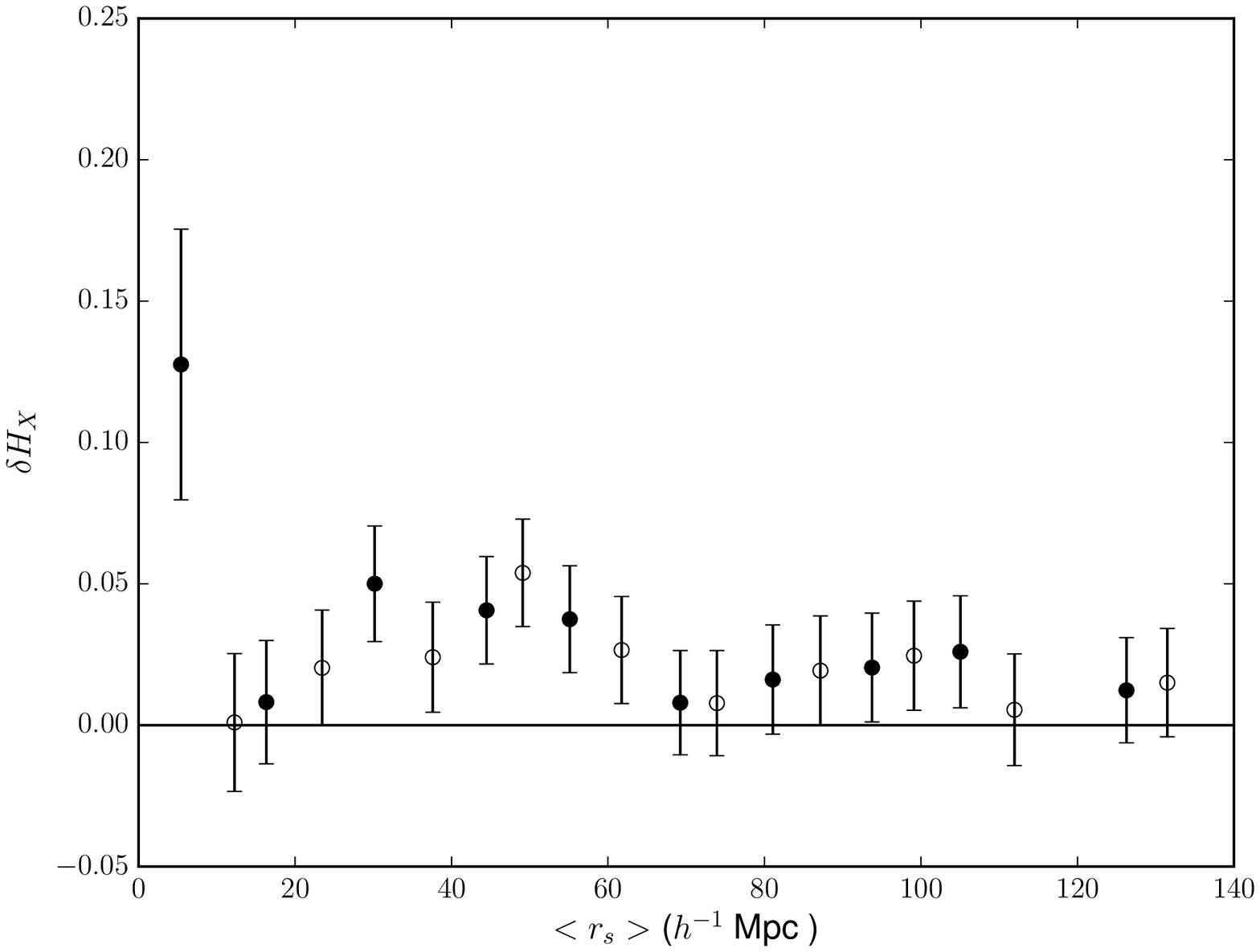}}\cr
\noalign{\vskip-5pt}\qquad{\bf(b)}\cr}}
}
\centerline{\vbox{\halign{#\hfil\cr
{\scalebox{0.45}{\includegraphics{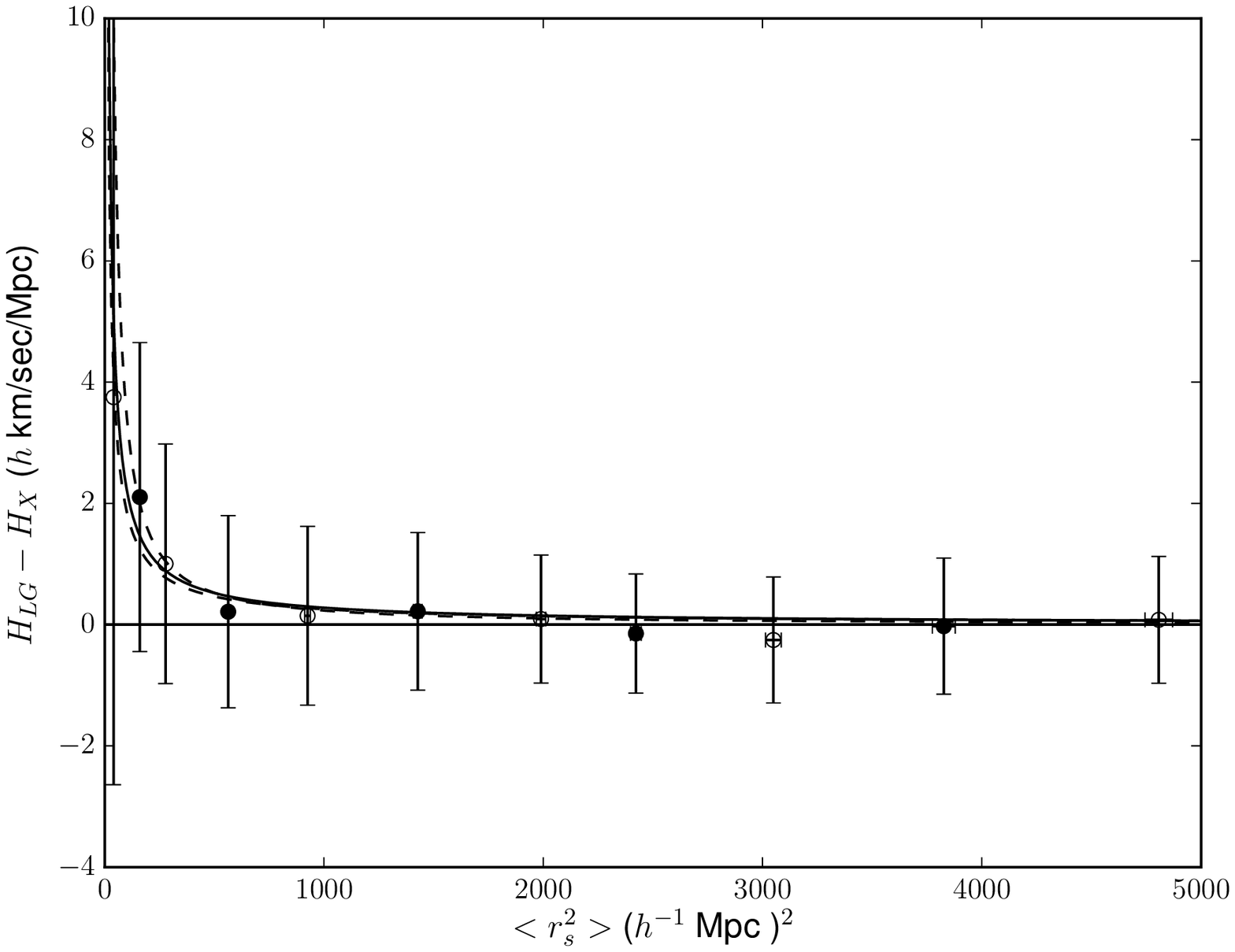}}}\cr
\noalign{\vskip-15pt}\qquad{\bf(c)}\cr}}}
\caption{The variation (\ref{Delta_H}) in the spherically averaged Hubble law:
{\bf(a)} $\de H_s$ in the LG frame; {\bf(b)} $\de H_s$ in frame X;
{\bf(c)} the systematic boost offset between the LG and frame $X$.
(Note: the first unprimed shell 1 is shown.)\label{boost_X}}}
\end{figure*}

Using both primed and unprimed shells in our calculation we find that the
$\vd/\vt=1$ at $\vt=171\kms$ in a direction $(\ell,b)=(59\deg,-4\deg)\pm(5\deg,4\deg)$.
For this boost, we find the result $p=-1.18\pm(0.95)\ns{stat}\pm(0.32)\ns{sys}$,
where the systematic uncertainty is determined as in \S \ref{LGframe},
consistent with $p=-1$. The difference between the spherically averaged
Hubble expansion in this frame, which we denote by $X$ and the LG is shown in
Fig.~\ref{boost_X}. One can see that a systematic boost offset is apparent
here. However, due to the large uncertainties that arise when taking
differences of the $H_s$ the result is also statistically consistent with
zero. (Thus the question of whether the first unprimed shell 1 should be
included in the analysis, due to its incomplete sky cover, is immaterial. Indeed, if we take only primed shells then we obtain $p=-1.17\pm(1.2)\ns{stat}$.)
This frame $X$ is only weakly disfavoured compared to the global $\chN$
minimum, $\ln(P\Ns{MV}/P\Ns{X})= 1.7$, and is within $1\,\si$ of the global
minimum of $\chL$; similar to the LG frame on both counts.

Our choice of frame X above is based on taking $\vd/\vt=1$, a condition which
may only be approximately matched in reality, given our huge uncertainties in
the coefficient $A$. We again have a degeneracy in the choice of minimum
Hubble expansion variation frame that satisfies the two conditions $p=-1$ and
$\vd=\vt$.

The results of this section confirm the finding of \S \ref{chi_min} that
our determination of a suitable cosmic rest frame is limited in the \CS sample
by a degeneracy under boosts close to the plane of the galaxy. The consistency
between the methods of this section and \S \ref{chi_min} may be
less significant, as they are not completely independent. In the LG frame the
primary source of the monopole variation is the increased value of
$H_s$ in the innermost shells, while the more distant shells closer to the
linear regime show closer to asymptotic values. Thus boosting to a frame with
a reduced $H_s$ in the innermost shells will give the most significant
improvement to $\chN$, relative to which small changes in the more
distant shells are negligible. This is precisely the type of difference we
model with a power law of the form (\ref{pl}) with $p\approx-1$.
Consequently, if our hypothesis concerning (\ref{pl}) is correct
then it is not surprising that we see the consistency in the angular directions
that minimize $\chN$ on one hand, and which give values of $p\approx-1$
with $A>0$ on the other.

\subsection{Angular Hubble expansion variation}\label{angvar}

\citet{WSMW} also explore the extent to which angular averages of the
Hubble expansion offer an independent characterization of a minimum Hubble
variation frame of reference. When one takes angular Gaussian window averages
of the Hubble expansion a dipole becomes apparent. \citet{WSMW} show that
this dipole is strongly correlated with the residual CMB temperature dipole
when both are referred to the LG (or LS) rest frame. If we are to define a
new cosmic standard of rest, within which we still observe a residual CMB
temperature dipole, then \citet{WSMW} argue that such a dipole must have a
nonkinematic origin. The correlation of the residual CMB temperature dipole and
Hubble expansion dipole supports the proposal that structures in the nonlinear
regime of expansion are simultaneously responsible for these effects. Thus we
are interested in finding a frame of reference in which this correlation is
maximized.

We have investigated this question, and find that correlation of the
residual CMB temperature and Hubble variation dipoles under arbitrary boosts
does not offer a viable characterization of the minimum variation frame
\citep{thesis}. By making boosts in the appropriate direction from the LG
we are able to artificially increase both the magnitude of the Hubble
expansion dipole \textit{and} the CMB residual temperature dipole
simultaneously, at the expense of also increasing the monopole expansion
variation. By calculating the
Hubble expansion dipole and higher multipole coefficients using
HEALPIX\footnote{http://healpix.jpl.nasa.gov/\citep{gorski05}} in the
\textit{frame of maximum correlation} for a given boost magnitude from the LG
we can observe this artificial increase in the strength of the dipole relative
to the higher multipoles. Boosting in directions which make the dipoles more
pronounced will naturally increase the correlation, but this is irrelevant
if the monopole variation is also increased. Therefore, as this artificially
induced increase in the correlation cannot be distinguished from a physically
meaningful increase, this method does not offer a viable characterization of
the minimum variation frame and was abandoned as a line of investigation
\citep{thesis}.

\section{Hubble expansion variation in the \textit{Cosmicflows}-2 catalogue}%
\label{cf2}

Thus far our investigation has been based entirely on the \CS catalogue of
distances and redshifts. In this section we aim to repeat the monopole Hubble
expansion variation analysis on the recently released \textit{Cosmicflows}-2
(CF2) catalogue. Systematic differences become apparent in this analysis which
we will investigate in \S \ref{determination_of_distance}.

\begin{figure*}
\vbox{\centerline{\vbox{\halign{#\hfil\cr\scalebox{0.7}
{\includegraphics{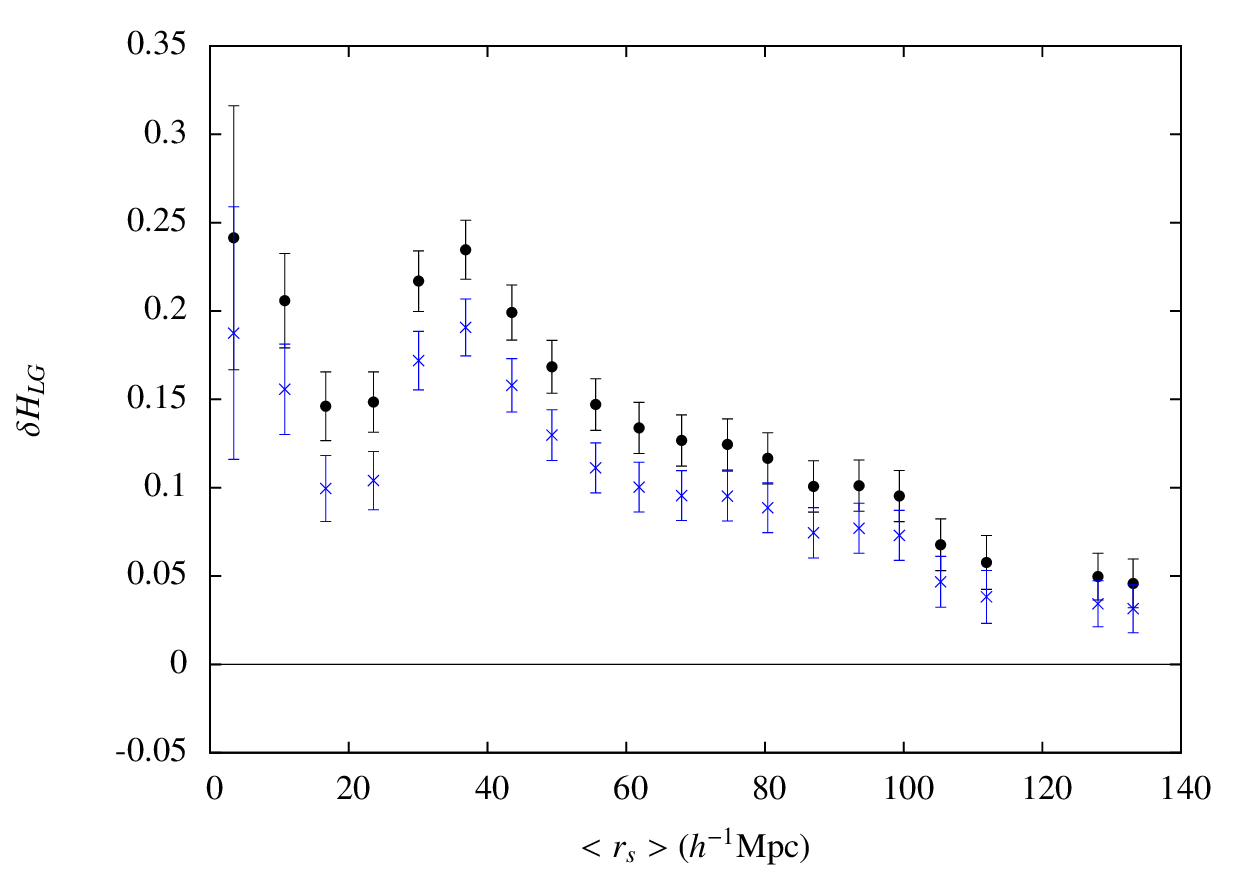}}\cr
\noalign{\vskip-5pt}\qquad{\bf(a)}\cr}}
\vbox{\halign{#\hfil\cr\scalebox{0.7}
{\includegraphics{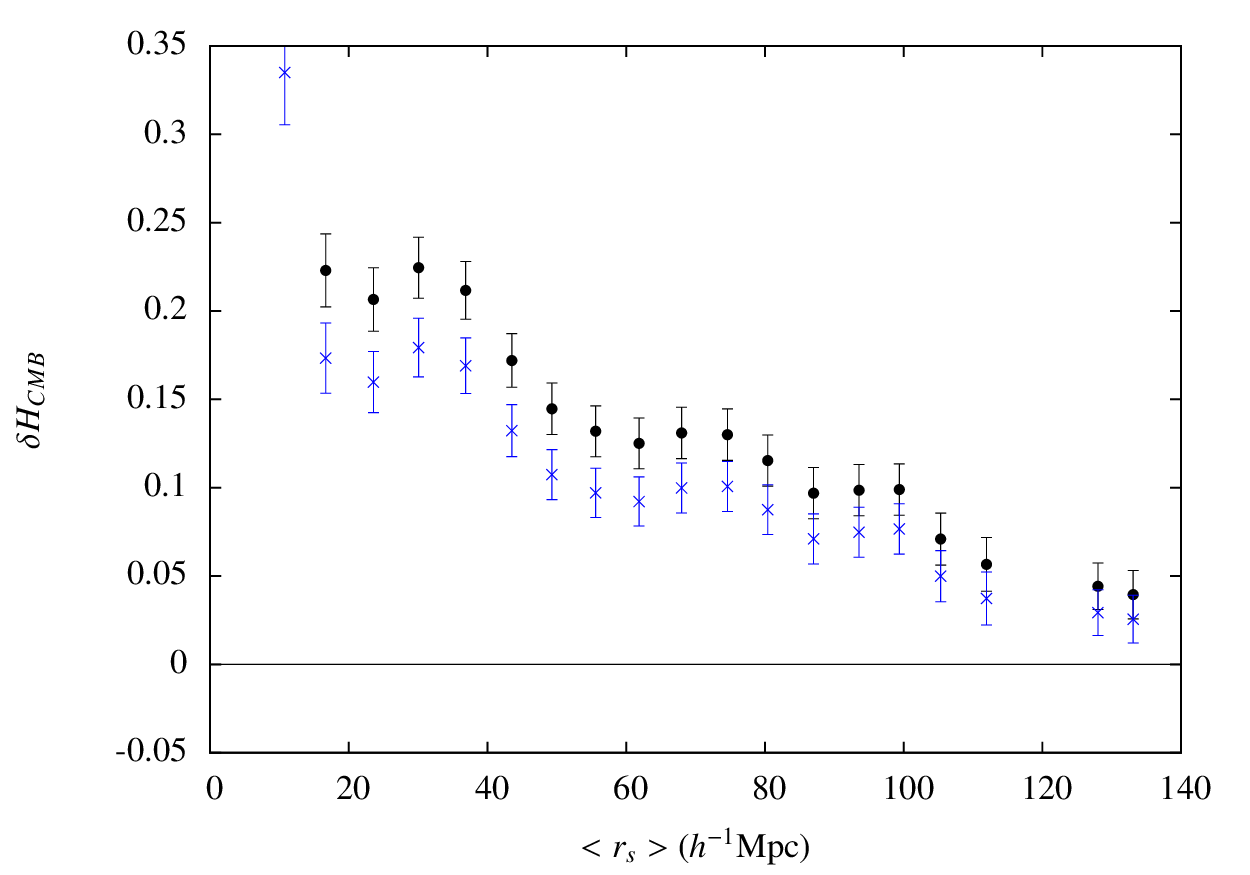}}\cr
\noalign{\vskip-5pt}\qquad{\bf(b)}\cr}}
}
\caption{The monopole Hubble expansion variation for the CF2 sample without
the FLRW ``correction'' (\ref{vmod}) (\textit{black} filled circles) and with
the FLRW ``correction'' (\ref{vmod}) (\textit{blue} crosses) in the: {\bf(a)} Local
Group frame of reference; {\bf(b)} CMB frame of reference.\label{compare_vmod}}}
\end{figure*}
CF2 is a compilation of distances and redshifts from both new and existing
sources of observational data. The entire catalogue consists of over 8000
galaxies both locally and extending beyond the scale of statistical homogeneity.
This compilation has a large subset of galaxies and galaxy clusters in common
with the \CS sample, including the large SFI++ sample of \citet{springob07}.
In total, the distances are determined by six different methods and compiled
together by \citet{Tully13}. The CF2 data are presented in two sets, one with
all individual galaxies included, and one condensed into galaxy groups,
including groups consisting of one galaxy. We will use the entire data set of
8162 galaxy redshifts and distances, freely available from the extragalactic
distance database\footnote{\textit{Cosmicflows}-2 distances retrieved
14 October 2014 from http://edd.ifa.hawaii.edu/.}.

\subsection{Spherically averaged Hubble expansion and treatment of biases}

We repeat our earlier analysis to calculate $\delta H_s$ (\ref{Delta_H}) in
the same two configurations of 11 spherical shells. The CF2 data are presented
with a modified ``recession velocity'', $cz$, and a raw observed redshift.
Given the prevalence of the use of such modifications, particularly in bulk
flow studies, it is worthwhile to briefly investigate the effect such
modifications can have on monopole Hubble expansion variation. \citet{Tully13}
define an adjustment {\em assuming a FLRW cosmology with $\OmM=0.27$ and
$\OmL=0.73$}. This adjustment is given by a Taylor expansion to
$\mathcal{O}(z^3)$ of a homogeneous isotropic expansion law,
\beq\label{vmod}
cz_{\text{mod}}=cz\left[1+\frn12(1-q_0)z-\frn16(1-q\Z0-3q\Z0^2+1)z^2\right],
\eeq
where $q\Z0=0.5(\OmM-2\OmL)$ and $z$ is the redshift in the CMB frame.

Since we wish to deal with cosmological model--independent quantities, this is
not the type of adjustment that should be made. In particular, a homogeneous
isotropic expansion law cannot be assumed below the scale of statistical
homogeneity if the conclusions of \citet{WSMW} are correct. Nor should such
an expansion law be assumed in the CMB rest frame. However, for completeness
we consider the adjustment (\ref{vmod}) in order to rule it out as the cause
of much larger systematic differences we will discuss shortly.

Fig.~\ref{compare_vmod} shows $\delta H$ using the CF2 sample. It is
immediately evident that these plots are very different to those found with the
\CS sample, as given in Fig.~3 of \citet{WSMW}. Most notably, the uniformity
of Hubble expansion in spherical shells is considerably worse in CF2 as
compared to the \CS sample, in both rest frames. While there is a small shift
introduced by the redshift modification, it is only makes a modest impact on
the very large differences from a uniform expansion seen in the unadjusted
redshifts.

It should be noted that by (\ref{Delta_H}) $\de H_s$ depends on the
asymptotically normalized Hubble constant, $\Ha$. Thus the vertical shift
seen in Fig.~\ref{compare_vmod} when the FLRW ``correction'' is applied
is primarily due to a change in the asymptotic value, $\Ha$, as data at
smaller values of $cz$ are barely affected by the correction in (\ref{vmod}).
Table \ref{shell2} gives the numerical values of the Hubble constant with
adjusted and raw redshifts, along with the number of objects in each shell
and the mean shell radii, making it transparent how the shift in Figure
\ref{compare_vmod} arises.

\begin{table*}
\begin{center}
\begin{minipage}[t]{\linewidth}
\caption{Hubble expansion variation in radial shells in CMB and LG frames for
the CF2 data. Spherical averages (\ref{Hsav}) are computed for two different
choices of shells, $r_s<r\le r_{s+1}$, the second choice being labeled by
primes. In each case we tabulate the number of data points per shell, the
weighted mean distance, $\mr_s$; the shell Hubble constants, $(H_s)\Ns{LG}$
and $(H_s)\Ns{CMB}$ and their associated uncertainties in the LG and
CMB frames for both the raw redshifts and those adjusted with (\ref{vmod}).
\label{shell2}}
\centering
\begin{tabular}{|lrrrrrrrrrrr|}
\hline\hline
Shell $s$& 1& 2& 3& 4& 5& 6& 7& 8& 9&10&11\B\T\\
$N_s$&579&946&834&936&959&794&739&670&497&825&333\B\T\\
$r_s$ ($\hm$)& 2.00& 12.50& 25.00& 37.50& 50.00& 62.50& 75.00& 87.50&100.00&112.50&156.25\B\T\\
$\ave{r}_s$ ($\hm$)& 3.41& 16.67& 30.07& 43.49& 55.59& 67.99& 80.40& 93.57&105.34&128.00&186.90\B\T\\
$(H_s)\Z{\text{CMB}}$&177.3&110.6&110.8&106.0&102.4&102.3&100.9& 99.4& 96.9& 94.5& 90.5\B\T\\
$(\sigma_s)\Z{\text{CMB}}$&10.5&1.5&1.0&0.8&0.7&0.8&0.8&0.8&0.9&0.7&0.9\B\T\\
$(H_s)\Z{\text{CMB,adjusted}}$&177.7&111.3&111.9&107.4&104.1&104.3&103.2&102.0& 99.6& 97.7& 94.9\B\T\\
$(\bs_s)\Z{\text{CMB,adjusted}}$&10.5&1.5&1.1&0.8&0.8&0.8&0.8&0.8&0.9&0.7&1.0\B\T\\
$(H_s)\Z{\text{LG}}$&112.2&103.6&110.0&108.4&103.7&101.8&100.9& 99.5& 96.5& 94.9& 90.4\B\T\\
$(\sigma_s)\Z{\text{LG}}$&6.6&1.4&1.0&0.8&0.8&0.8&0.8&0.8&0.9&0.7&0.9\B\T\\
$(H_s)\Z{\text{LG,adjusted}}$&112.6&104.2&111.1&109.8&105.3&103.8&103.2&102.1& 99.2& 98.0& 94.8\B\T\\
$(\sigma_s)\Z{\text{LG,adjusted}}$&6.7&1.4&1.1&0.8&0.8&0.8&0.8&0.8&0.9&0.7&1.0\B\T\\
\B\T\\
Shell $s$& $1'$& $2'$& $3'$& $4'$& $5'$& $6'$& $7'$& $8'$& $9'$&$10'$&$11'$\B\T\\
$N_s$&869&867&846&989&889&777&643&648&412&625&333\B\T\\
$r_s$ ($\hm$)& 6.25& 18.75& 31.25& 43.75& 56.25& 68.75& 81.25& 93.75&106.25&118.75&156.25\B\T\\
$\ave{r}_s$ ($\hm$)& 10.76& 23.54& 36.85& 49.29& 61.86& 74.59& 87.01& 99.37&111.95&133.10&186.90\B\T\\
$(H_s)\Z{\text{CMB}}$&126.1&109.2&109.6&103.6&101.8&102.2& 99.2& 99.4& 95.6& 94.0& 90.5\B\T\\
$(\sigma_s)\Z{\text{CMB}}$&2.5&1.2&0.9&0.8&0.7&0.8&0.8&0.8&0.9&0.8&0.9\B\T\\
$(H_s)\Z{\text{CMB,adjusted}}$&126.7&110.0&110.9&105.1&103.6&104.4&101.6&102.2& 98.4& 97.3& 94.9\B\T\\
$(\sigma_s)\Z{\text{CMB,adjusted}}$&2.5&1.2&0.9&0.8&0.8&0.8&0.8&0.8&1.0&0.8&1.0\B\T\\
$(H_s)\Z{\text{LG}}$&109.0&103.8&111.6&105.6&102.5&101.6& 99.5& 99.0& 95.6& 94.5& 90.4\B\T\\
$(\sigma_s)\Z{\text{LG}}$&2.1&1.1&1.0&0.8&0.7&0.8&0.8&0.8&0.9&0.8&0.9\B\T\\
$(H_s)\Z{\text{LG,adjusted}}$&109.5&104.6&112.9&107.1&104.3&103.8&101.8&101.7& 98.4& 97.8& 94.8\B\T\\
$(\sigma_s)\Z{\text{LG,adjusted}}$&2.1&1.1&1.0&0.8&0.8&0.8&0.8&0.8&1.0&0.8&1.0\\
\hline
\end{tabular}
\end{minipage}\hfill
\end{center}
\end{table*}

As we will see in \S \ref{determination_of_distance}, the treatment of Malmquist
biases is the most likely cause of the systematic difference we see in Figure
\ref{compare_vmod} as compared to Fig.~3 of \citet{WSMW} for the \CS sample.
To study the nature of this systematic difference we make use of the SFI++
sample, which is a subset of both the \CS and CF2 samples. SFI++
\citep{springob07} consists of \citet*{tfr} Relation (TFR) derived distances for
4861 field and cluster galaxies. Since \citet{springob07} supply the SFI++
sample with and without corrections for Malmquist biases, it makes it an ideal
candidate with which to study the effects that the treatment of such biases
has on the spherically averaged Hubble expansion.

\begin{figure*}
\vbox{\centerline{\vbox{\halign{#\hfil\cr\scalebox{0.7}
{\includegraphics{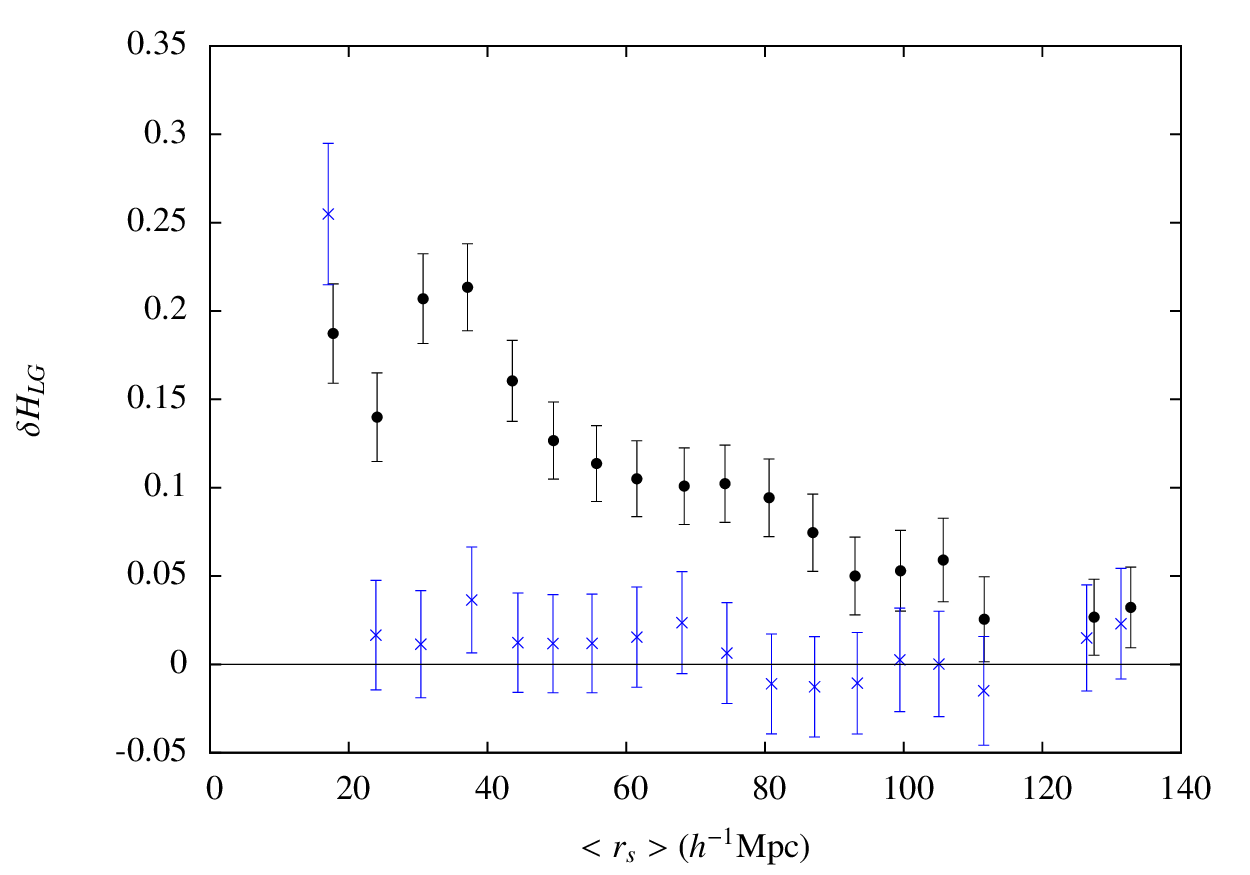}}\cr
\noalign{\vskip-5pt}\qquad{\bf(a)}\cr}}
\vbox{\halign{#\hfil\cr\scalebox{0.7}
{\includegraphics{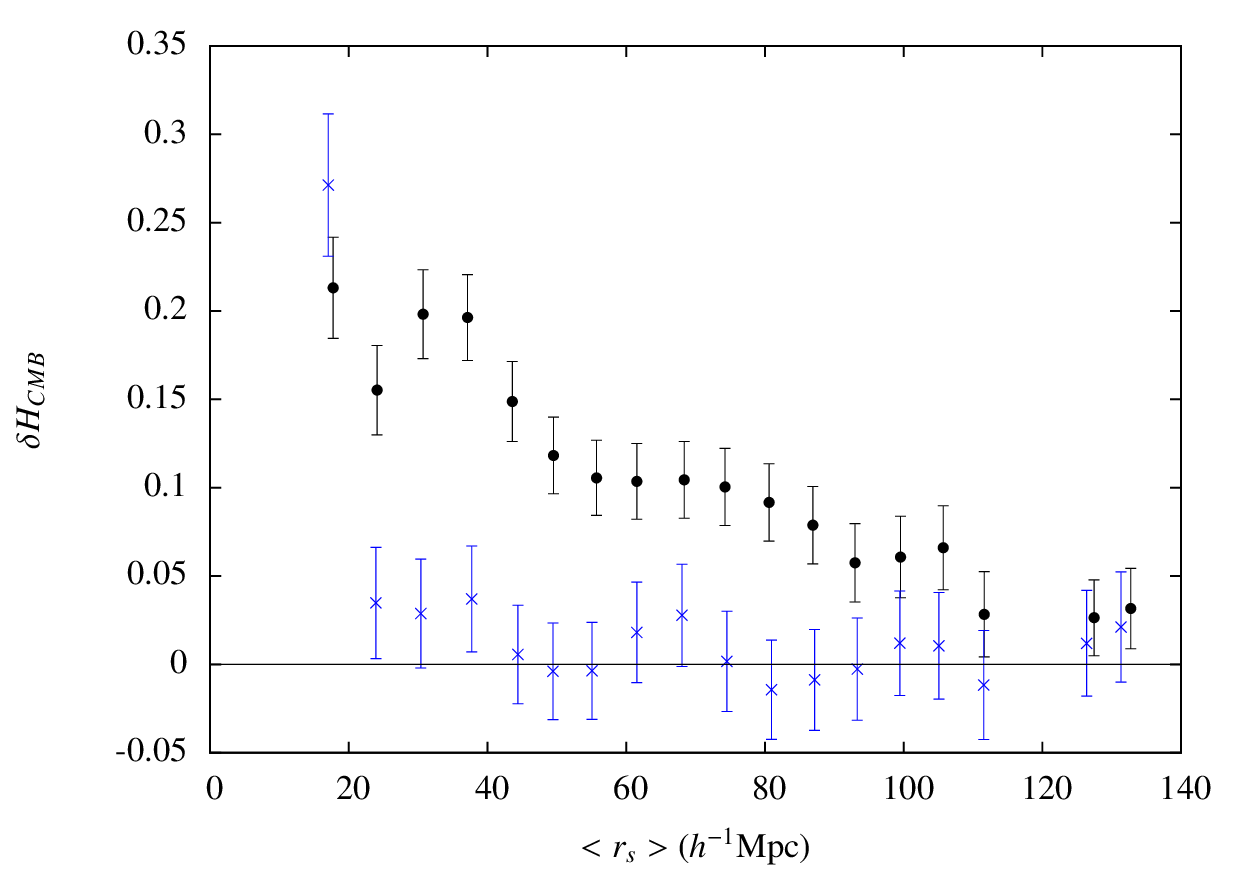}}\cr
\noalign{\vskip-5pt}\qquad{\bf(b)}\cr}}
}
\caption{The monopole Hubble expansion variation for the SFI++ sample without
corrections for Malmquist bias (\textit{black} filled circles) and with
corrections (\textit{blue} crosses) in: {\bf(a)} the Local Group frame of
reference; {\bf(b)} the CMB frame of reference.\label{SFI_malmquist}}}
\end{figure*}

To understand the effect of the Malmquist bias corrections applied by
\citet{springob07}, we calculate the monopole variation of the Hubble
expansion for the SFI++ sample with and without corrections. Figure
\ref{SFI_malmquist} shows the significant difference in $\de H_s$ between
these two treatments.

\begin{figure*}
\vbox{\centerline{\vbox{\halign{#\hfil\cr\scalebox{0.7}
{\includegraphics{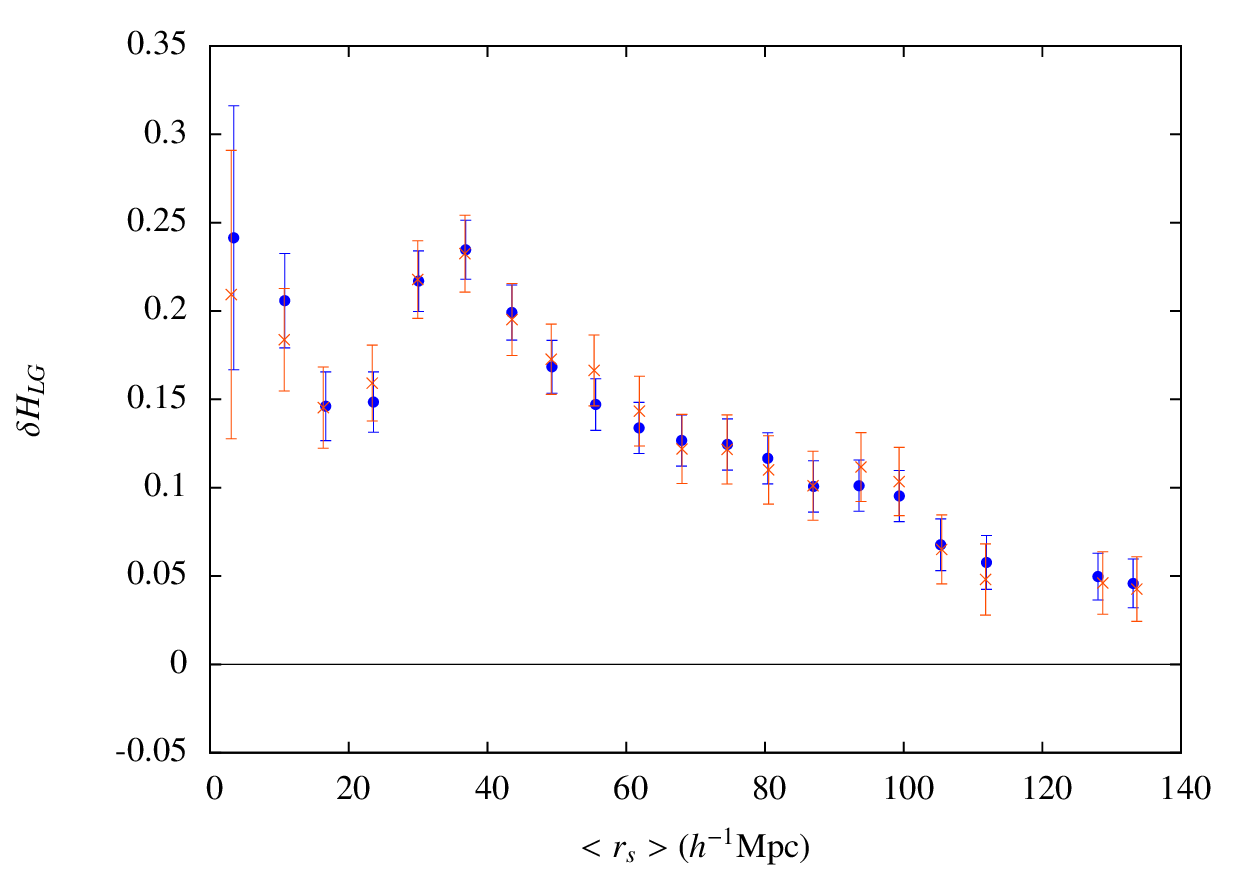}}\cr
\noalign{\vskip-5pt}\qquad{\bf(a)}\cr}}
\vbox{\halign{#\hfil\cr\scalebox{0.7}
{\includegraphics{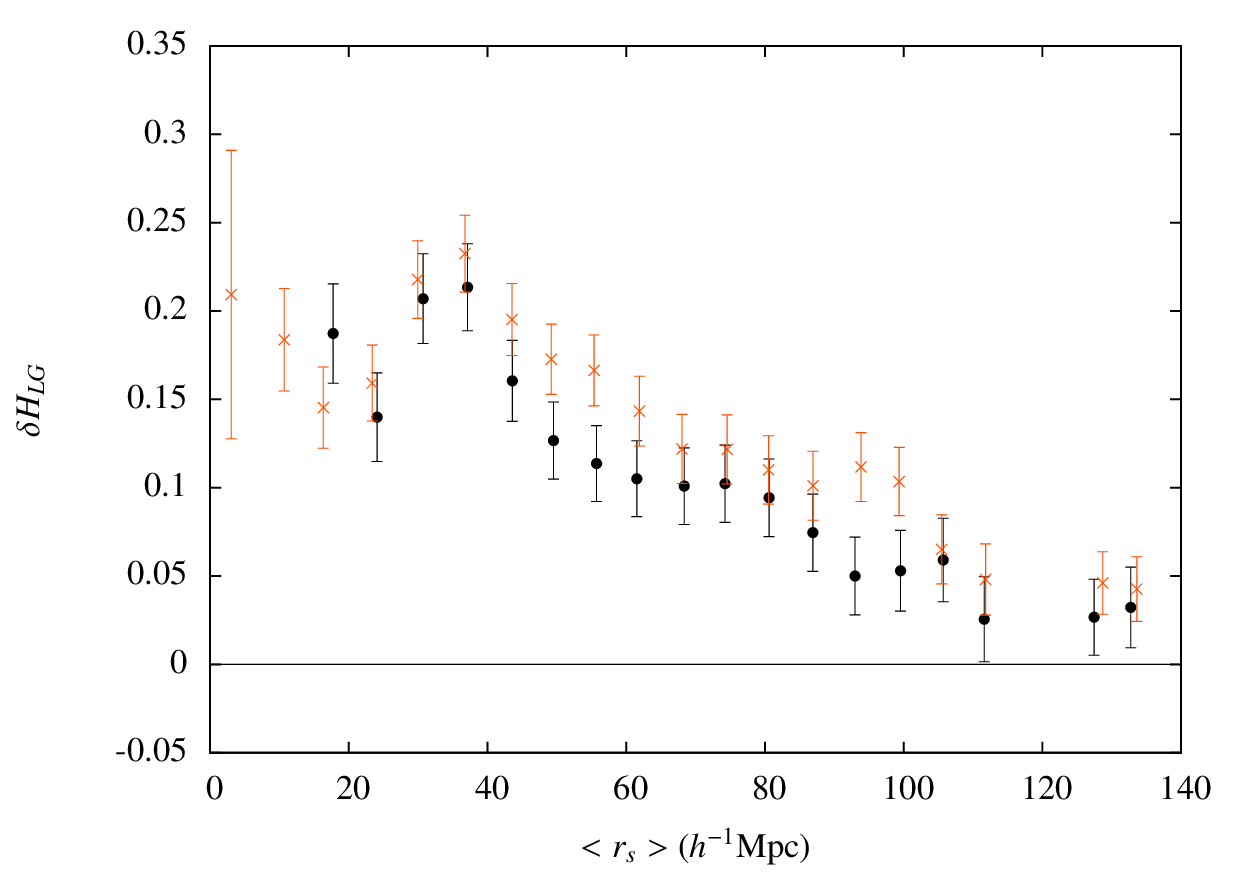}}\cr
\noalign{\vskip-5pt}\qquad{\bf(b)}\cr}}
}
\caption{The monopole Hubble expansion variation in the LG frame for: {\bf(a)}
all CF2 data (\textit{blue} filled circles) and CF2 data without SFI++ data
(\textit{red} crosses); {\bf(b)} CF2 data without SFI++ data (\textit{red}
crosses) and SFI++ only without Malmquist corrections (\textit{black} filled
circles). For the uncorrected SFI++ data the first unprimed and primed values
are not shown, these are 2.7 and 0.65 respectively.\label{CF2_noSFI}}}
\end{figure*}

Since uncorrected SFI++ data points are included in the CF2 catalogue, we
can determine whether there is any systematic difference between this
subsample and the remainder of the CF2 catalogue. If we take the CF2 catalogue
and remove the 3625 points in common with our SFI++ sample, and repeat the
analysis, we arrive at Fig.~\ref{CF2_noSFI}. We find that $\de H_s$ does not
change to any statistically significant extent, by removal of this potentially
biased data. This is an indication that the systematic bias present in the
SFI++ raw distances -- uncorrected for Malmquist bias -- is likely to also be
present in the rest of the CF2 data. Thus we are confident that the discrepancy
seen between the monopole variation of Fig.~\ref{compare_vmod} for CF2
and Fig.~3 of \citet{WSMW} for the \CS sample is a systematic issue, arising
from the treatment of Malmquist bias in the CF2 catalogue, as we discuss
further in \S\ref{determination_of_distance}.

\subsection{The systematic boost offset revisited}\label{CF2boost}

One may ask whether, despite the obvious problems with a systematic bias
in the CF2 data, any of the analyses applied to the \CS sample can nonetheless
yield meaningful results.

We applied the analysis of \S\ref{chi_min} in the CF2 catalogue, but found no
reference frame in which $\chN$ approaches unity, or even within the same
order of magnitude. While some decrease in the variation of Hubble expansion
was found for boosts in the galactic plane, the uncertainties were too large
to give any statistically significant results. Such investigations must be
abandoned until the bias problems in the CF2 catalogue are dealt with.

By contrast, we found that in spite of the bias problem, the signature of a
systematic boost offset studied in \S\ref{LGframe} is nonetheless evident in
CF2, as this involves the {\em difference} of the $H_s$ values in
the LG and CMB frames from Table \ref{shell2}, as plotted in panels (a) and
(b) of Fig.~\ref{compare_vmod}.

\begin{figure*}
\centering%\vskip-1.2\baselineskip
\scalebox{0.7}{\includegraphics{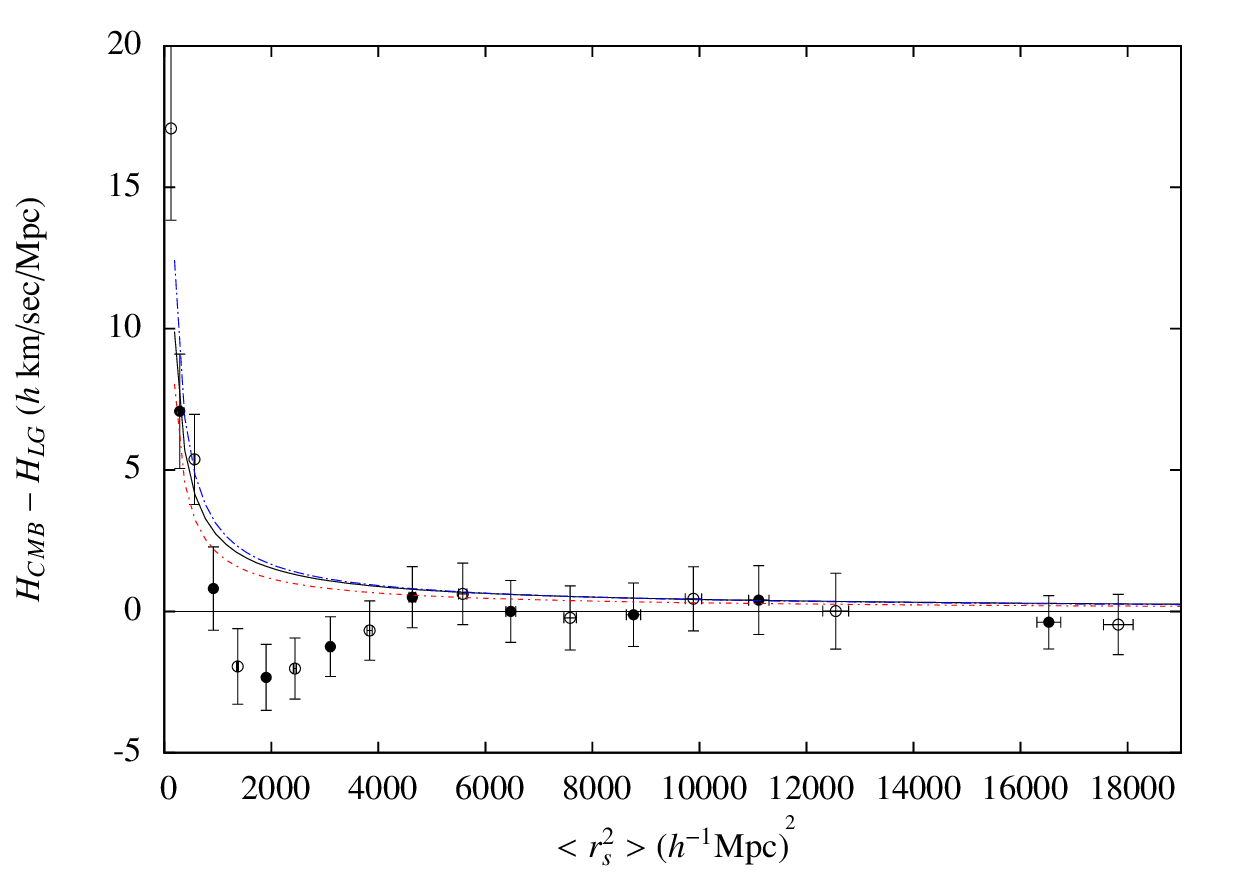}}
\caption{Best fit power law to the radial variation in the spherically
averaged Hubble law in 11 shells for the CF2 galaxies data. The dashed blue
curve is the best fit to primed shells only (empty circles), the dotted red
curve is the best fit to unprimed shells only (filled circles) and the
solid black curve is the best fit to all data points.
The first data point -- corresponding to unprimed shell 1 -- is omitted,
as it is off the scale.} \label{hdiff_CF2_gal}
\end{figure*}

Fig.~\ref{hdiff_CF2_gal} shows the results of repeating the analysis used in
\S \ref{LGframe}, using the unadjusted CF2 distances. The best fit value for
$p$ in (\ref{pl}) is found to be
$p=-0.83\pm 0.17$ for unprimed and $p=-0.86 \pm 0.26$ for primed shells.
Varying the shell boundaries as in \S \ref{LGframe} we find a value of
$p=-0.84\pm (0.21)\ns{stat} \pm (0.06)\ns{sys}$. However, if we compare
Fig.~\ref{hdiff_CF2_gal} with Fig.~\ref{Hdiff2}, we see that there are
more data points with $(H_s)\Ns{CMB}<(H_s)\Ns{LG}$, which do not conform
to the power law (\ref{pl}). However, the range of distances of the
shells for which this is true coincides in Fig.~\ref{Hdiff2} and Figure
\ref{hdiff_CF2_gal}, being $40\h\lsim r\lsim60\hm$ in the \CS sample and
$30\h\lsim r\lsim67\hm$ in the CF2 sample. This is consistent with the hypothesis that aside
from the systematic boost offset, there are structures responsible for
nonlinear deviations in the monopole Hubble expansion in the range identified
in the \CS sample, but untreated biases in the CF2 catalogue have broadened the
range of distances attributed to the same structures\footnote{These data
points are necessarily disregarded when we perform the logarithmic
transformation to fit the power law (\ref{pl}), and so do not
contribute to the stated uncertainties.}.

\begin{figure*}
\centering\vskip-2.5\baselineskip
\scalebox{0.49}{\includegraphics{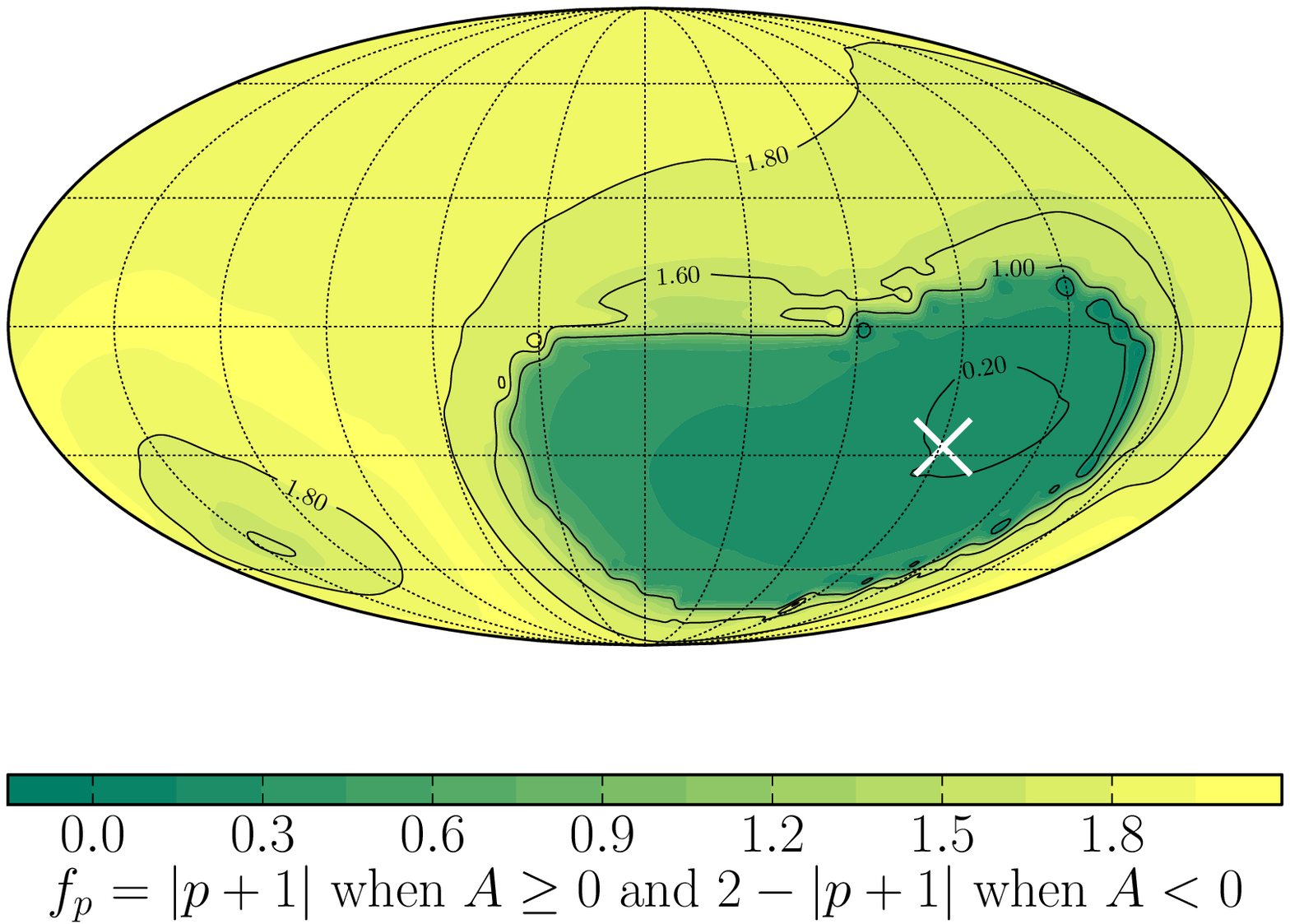}}
\caption{Best fit power law parameters to (\ref{pl}) across entire sky
for boosts of $635\kms$ from the CMB frame. The cross indicates the direction
of the boost to the LG, which is also of magnitude 635$\kms$. \latitudes}
\label{boost_from_CMB_634_CF2}
\end{figure*}

The systematic boost offset is still evident in the innermost shells of the
CF2 sample. However, the fit is somewhat worse than in the \CS sample due to
more data points lying in the increased range which deviates from
(\ref{pl}). Nonetheless, we can still check if the boost offset signature is
unique to the angular direction of the residual CMB dipole in the LG frame.
Fig.~\ref{boost_from_CMB_634_CF2} shows the value of $\fb$ from (\ref{fb})
which represent the best fit parameters for a systematic boost offset for
boosts of $635\kms$ across the entire sky. The results are consistent with
those found for the \CS sample, providing further evidence that this is indeed
not a random statistical outcome but is consistent with our hypothesis.

\begin{figure*}
\vskip-1.2\baselineskip
\vbox{\centerline{\vbox to 230pt{\halign{#\hfil\cr\quad\scalebox{0.49}
{\includegraphics{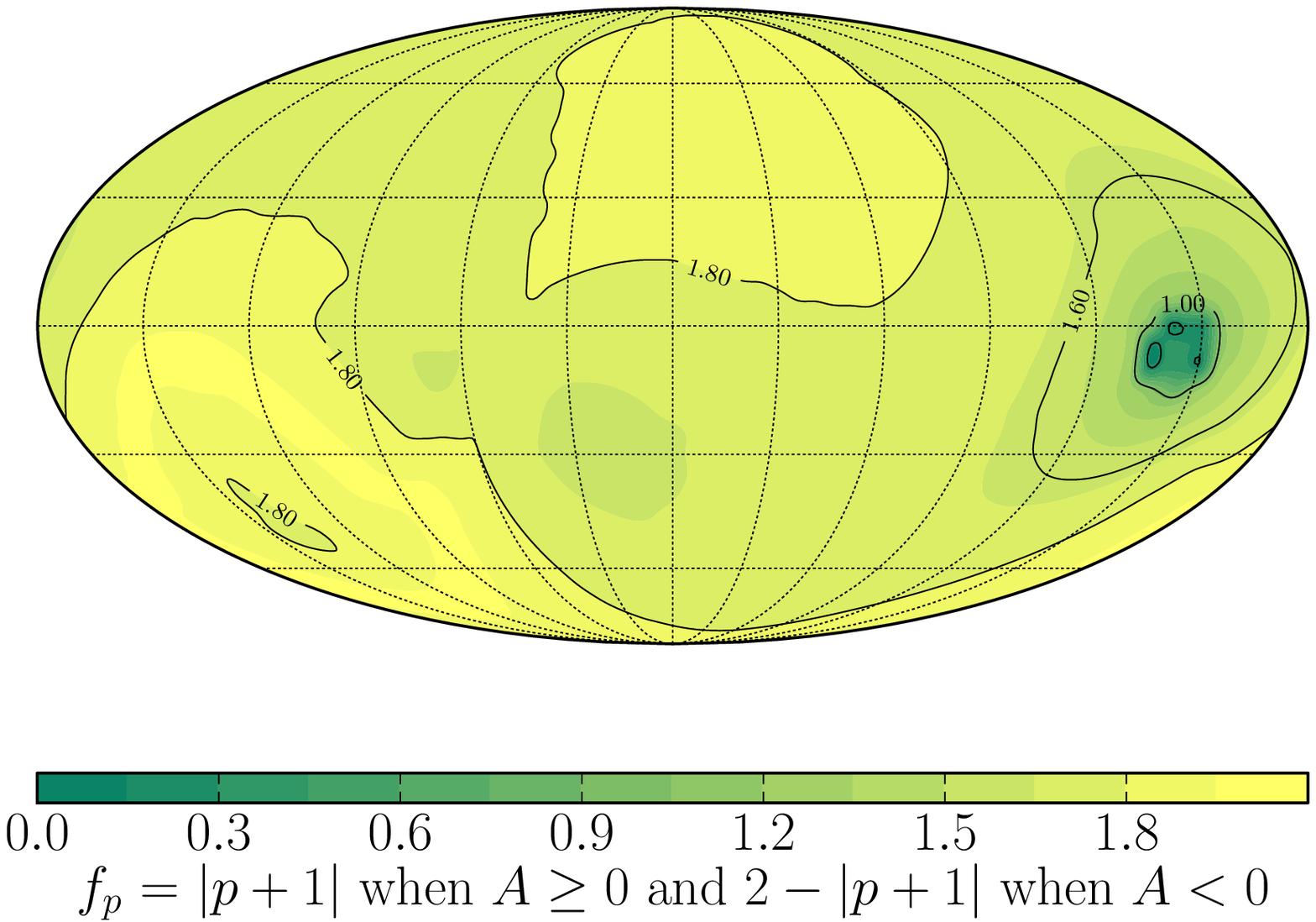}}\cr
\noalign{\vskip-50pt}\qquad\qquad{\bf(a)}\cr}\vfil}\hskip-20pt
\vbox to 230pt{\halign{#\hfil\cr\scalebox{0.49%0.6
}{\includegraphics{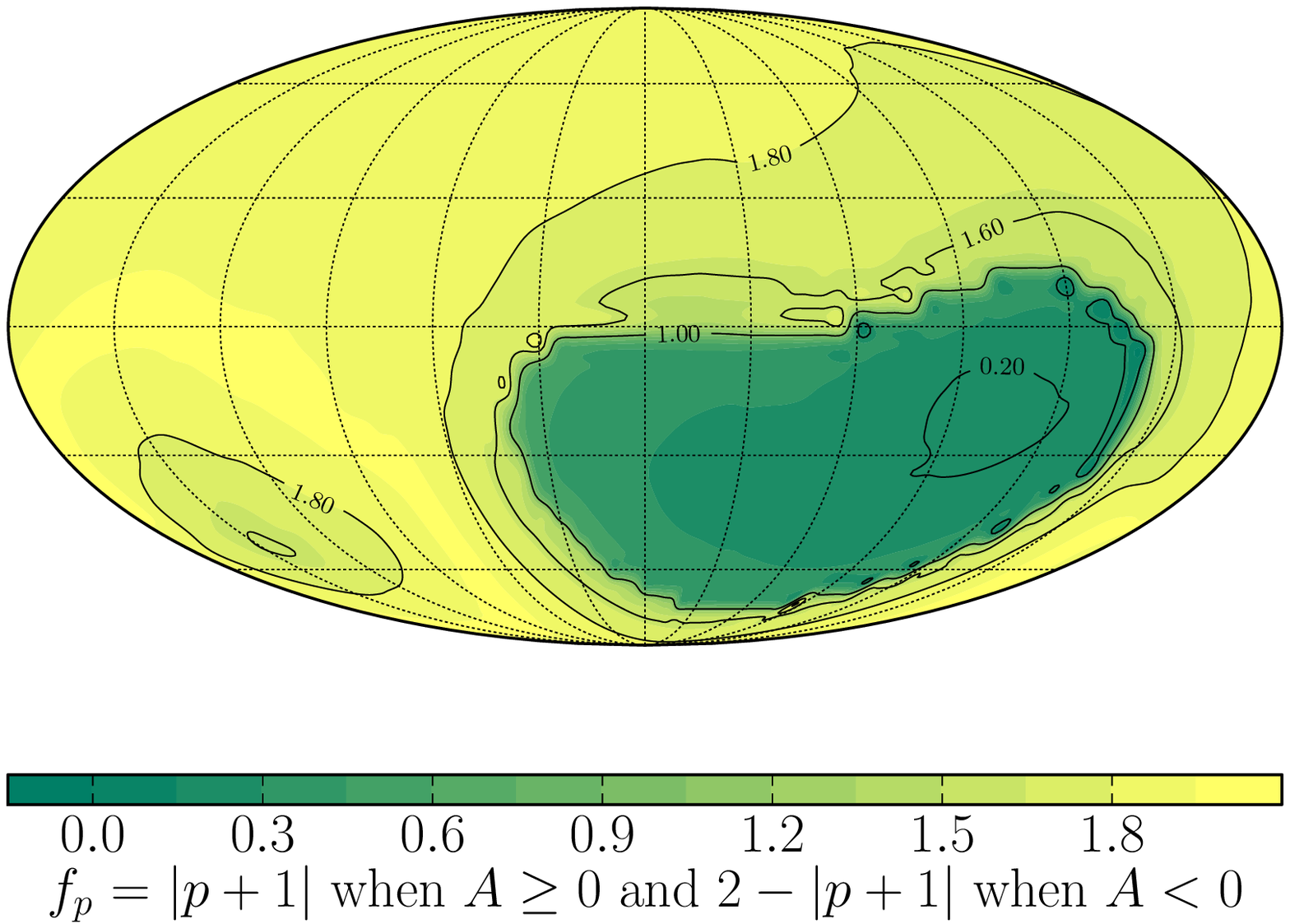}}\cr
\noalign{\vskip-50pt}\quad\qquad{\bf(b)}\cr}}}}
\caption{Best fit parameters to (\ref{pl}) using unadjusted CF2 data
for boosts from the LG of: {\bf(a)} $450\kms$; {\bf(b)} $200\kms$.
\latitudes \label{CF2_boost_LG}
}
\end{figure*}

We have also repeated the analysis of \S\ref{boosts} for the best fit to a
systematic boost offset from the LG frame, as shown in Figure
\ref{CF2_boost_LG}.
The angular directions found are consistent with our results for the \CS sample.
For example, on the $200\kms$ sky map the best fit value is in the direction
$(\ell,b)=(55\deg,-5\deg)$, close to that found earlier. However, the value
of $p=-0.92\pm0.75$ has a far greater statistical uncertainty, which may well
be due to the untreated biases.

In conclusion, the greater number of data points in the CF2 catalogue
may potentially yield statistically more accurate results than the \CS sample.
However, at present this is prevented on account of untreated biases, to which
we now turn.

\section{\textit{Cosmicflows}-2 Malmquist treatment}\label{determination_of_distance}

The \textit{Cosmicflows}-2 (CF2) and \CS catalogues deal with Malmquist
distance biases in different ways. The results of the last section indicate
that the treatment of such biases is crucial in establishing the actual
nature of the variation of cosmic expansion below the statistical
homogeneity scale. Therefore we will perform further analyses to better
understand these differences.

Since the treatment of Malmquist bias is complex, we will first briefly
remind the reader that in its current usage this term refers to at least
three distinct biases that affect the average derived distance of a galaxy
cluster:
\begin{enumerate}
\item {\em Selection bias}\quad This is the simple homogeneous systematic error
that results from using objects in a magnitude limited sample \citep{Malmquist}.
If the galaxies in a
cluster have a true average flux $\ave{F}_0$, since we can only observe the
brightest of these, we measure a biased average flux $\ave{F}_b>\ave{F}_0$,
with the effect of bias increasing with distance. Thus the luminosity distance
${d_L}^2=\ave{L}_0/(4\pi\ave{F})$ will be less when using a biased average flux,
where $\ave{L}_0$ is the average luminosity as determined from standard
candles calibrated using nearby objects.
\item {\em Homogeneous distribution bias}\quad This is a systematic average
error for objects around the same derived distance which can be understood in
terms of statistical scatter. If we assume a standard Gaussian scatter, $\si$,
in the derived distances about an estimated mean, then -- since the radial
number density grows as $N(r)\propto r^3$ -- there are more objects with true
distances larger than the estimated distance, than smaller. Thus at a given
derived distance, more galaxies will have been scattered by the
errors down from larger true distances than up from smaller ones
\citep{Lynden_Bell1988,Hanski1999}. This
is equivalent to giving more statistical weight to more distant values and thus
the probability distribution for the true distance is no longer Gaussian
along the line of sight, centred on the measured distance, instead being
skewed towards greater distances. For the special case of a constant density
distribution, for example, one arrives at Eddington's formula
\citep{Eddington1914}
\beq\label{Eddington}
E(\mu_{\text{true}}|\mu_{\text{der}})=\mu_{\text{derived}}+1.382\sigma^2,
\eeq
where $\mu\equiv 5\log r+25$ is the standard distance modulus when $r$ is
given in Mpc.
\item {\em Inhomogeneous distribution bias}\quad The inhomogeneous bias is
analogous to the homogeneous one in that it involves a systematic error in the
statistical scatter of objects due to their distribution in 3--dimensional
space. However, in this case it arises as number counts are higher in regions of
greater density, resulting in a systematic scatter of measurements out of higher
into lower density regions \citep{Strauss1995}. Thus the inhomogeneous effect is
very sensitive to large variations in large-scale structure along the line of
sight. Failure to account for this type of bias can give spurious infall
signatures on to high density regions. This bias is of course far more difficult
to account for, requiring accurate density fields for structure along the
line of sight for each observation.
\end{enumerate}

The selection and homogeneous distribution biases typically lead to
underestimates of distances that increase as the distance grows, so
that a plot of $\mu\ns{true}-\mu\ns{der}$ versus redshift has a positive slope.
However, the inhomogeneous distribution bias can lead to the opposite
effect. For example, \citet{Feast} shows that in applying the TFR method
when the spatial density of objects at a given 21cm line width is constant,
then the required Malmquist correction is the classical one given by
\citet{Eddington1914}. However, when this is not the case it is possible to
obtain overestimated distances \citep{Feast}.

Moreover, regardless of the details of any inhomogeneous matter distribution,
one just needs the spherically averaged $N(r)$ to decrease sufficiently
quickly for the direction of scatter in the standard homogeneous distribution
bias to be reversed, giving overestimated distances.

Some biases can be dealt with in the data reduction. In particular, by applying
the inverse TFR method rather than the direct method one can in principle
effectively eliminate the selection bias \citep{Schechter1980}, leaving only a
considerably smaller bias \citep{Willick1994,Willick1995}.

In the CF2 catalogue \citet{Tully13} use an inverse TFR procedure to reduce the
selection bias only, stating that only a small subsequent correction for
residual bias is required. In particular, they ``make no adjustments for the
distribution Malmquist effects'' in their reported CF2 distances
\citep{Tully13}. Their calibration carried out for this relation
follows the procedures of \citet{Pierce2000}, \citet{Courtois2012} and
\citet{Sorce2013}.

On the other hand, the SFI++ catalogue \citep{springob07} which forms the
major part of the \CS sample includes corrections to account for
all homogeneous and inhomogeneous biases in their data. In their view
their own treatment of the homogeneous and inhomogeneous distribution
biases was ``straightforward'', given access to a reconstruction
of the local density field \citep{el06}. However, they stated that
their treatment of the selection bias was {\em ad hoc} because the selection
criteria used are designed to mimic the observational properties of the
survey as closely as possible, and so are very inhomogeneous. They therefore
provided both the raw and corrected distances, should other researchers
adopt alternative methods for dealing with selection bias.

The \CS sample incorporates the SFI++ distances with the Malmquist bias
corrections of \citet{springob07}, whereas CF2 uses a subset of uncorrected
SFI++ distances. While we do not have the data to independently repeat
any of the bias corrections, we are able to test the consistency of
assumptions made by \citet{Tully13}.

\subsection{The SFI++ subsamples of CF2}

\begin{figure*}
\vbox{\centerline{\vbox{\halign{#\hfil\cr\scalebox{0.7}
{\includegraphics{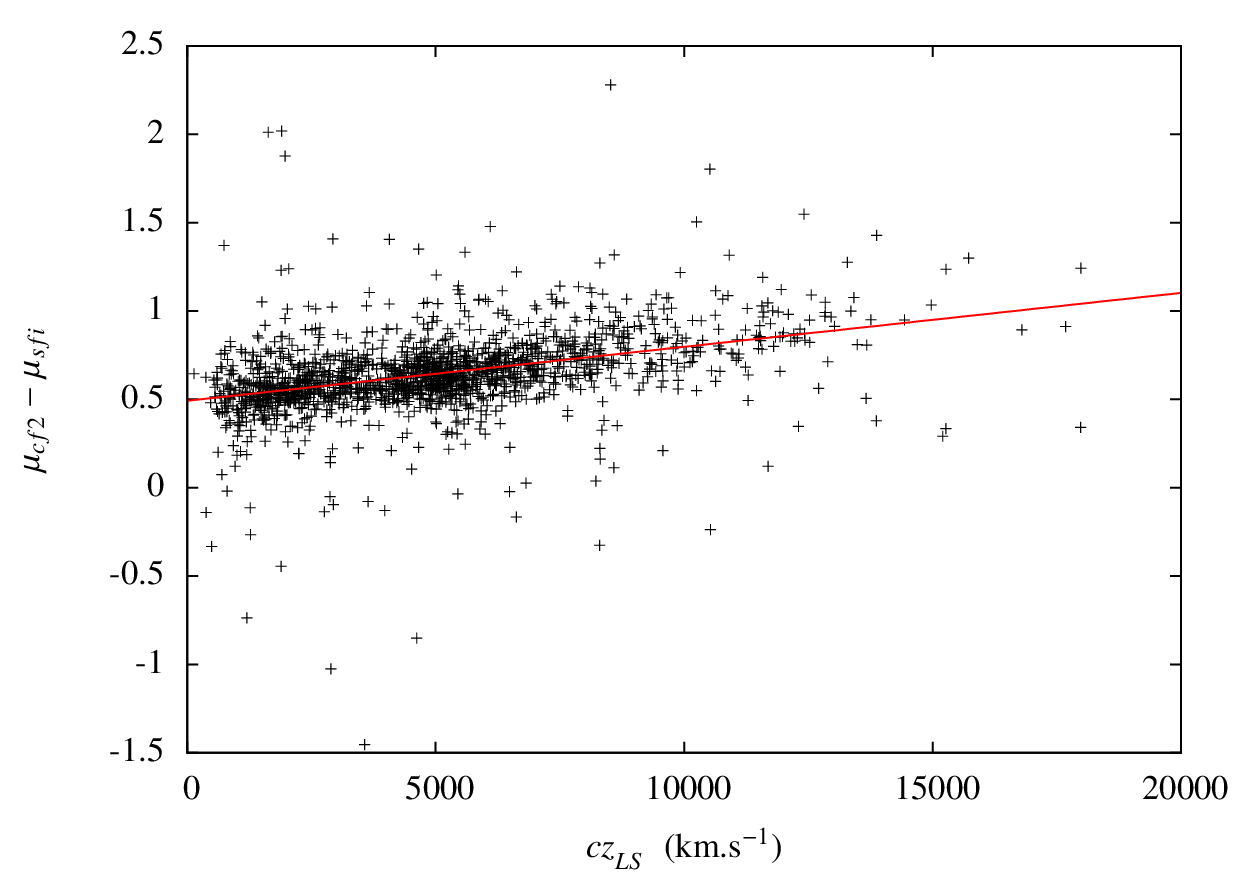}}\cr
\noalign{\vskip-5pt}\qquad{\bf(a)}\cr}}
\vbox{\halign{#\hfil\cr\scalebox{0.7}
{\includegraphics{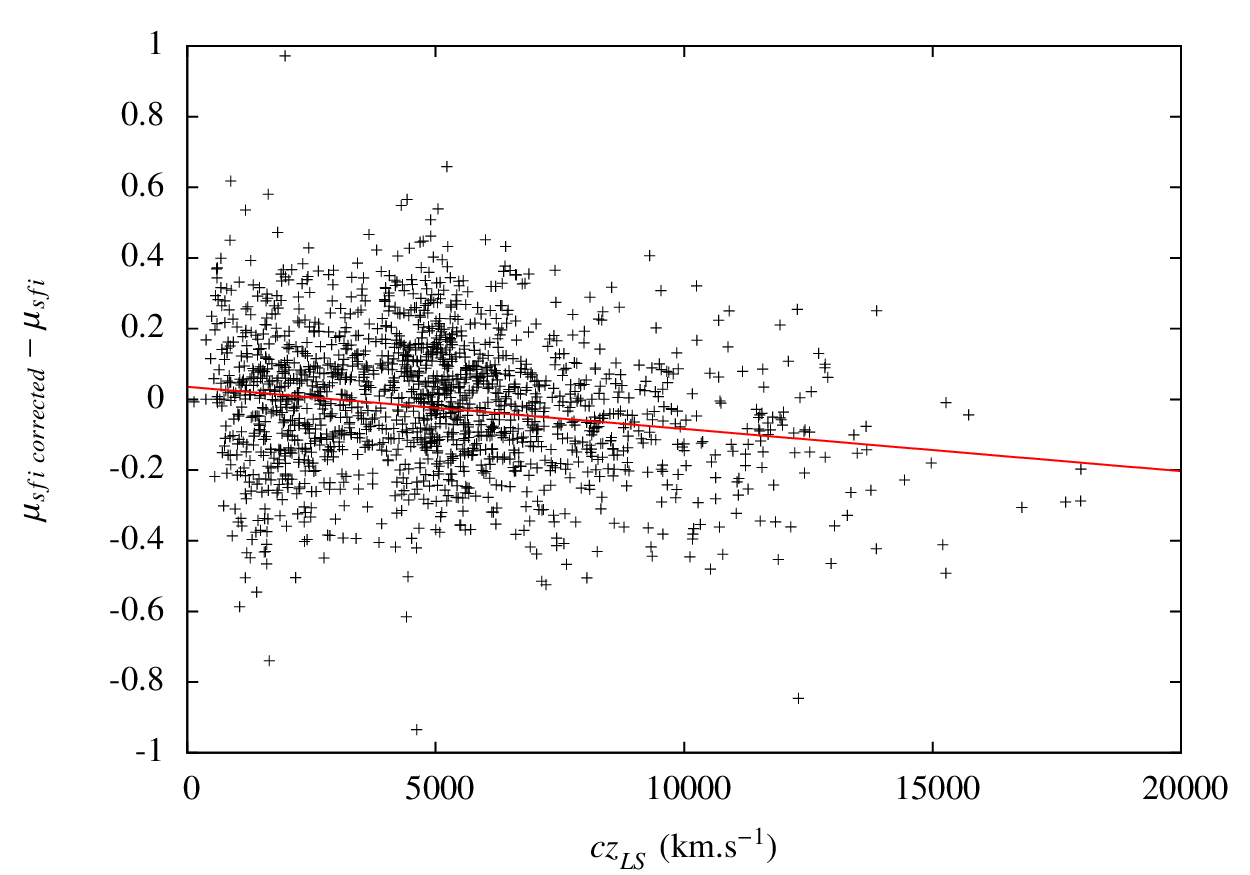}}\cr
\noalign{\vskip-5pt}\qquad{\bf(b)}\cr}}
}
\caption{The difference in distance modulus between: {\bf(a)} the CF2 and
raw SFI++ distances for the points common to the two surveys, as
presented in \citet{Tully13}. {\bf(b)} the SFI++ raw and corrected catalogues
for the same points.\label{malm_compare}\label{CF2_malm_correction}}}
\end{figure*}

\citet{Tully13} find that for 2071 common points between their own survey
and the SFI++ survey (excluding 5 points judged to be
``bad'') there was a ``correction'' of the form
\beq\label{Malm_correction_CF2}
\Delta\mu_1=0.492(\pm0.011)+0.000031(\pm0.000002)cz\Ns{LS}
\eeq
where $\Delta\mu_1\equiv\mu_{cf2}-\mu_{sfi}$, $\mu_{cf2}$ is the CF2 distance
modulus with the zero point established by \citet{Courtois2012}, and
$\mu_{sfi}$ is the \citet{springob07} unadjusted modulus with a nominal zero
point consistent with $H_0=100h\kmsMpc$, and $z\Ns{LS}$ is the raw\footnote{The
FLRW adjustment (\ref{vmod}) is not applied.} redshift in the rest frame of
the Local Sheet, which is close to the Local Group frame \citep{tsk08}. We
independently confirm the slope in (\ref{Malm_correction_CF2}) using the
appropriate zero point\footnote{The intercept in (\ref{Malm_correction_CF2})
is determined by a scaling of the data so we are not interested in independently
confirming this for our investigation.}, and plot this in Figure
\ref{CF2_malm_correction}. Note that this comparison is made for a subset of
the SFI++ sample, henceforth SFI++A, consisting only of objects that are
common between the SFI++ survey and the independently obtained CF2 distances.
In their final analysis \citet{Tully13} use averages of the CF2 and SFI++
distances with double weight given to the CF2 distances.

The intercept in Fig.~\ref{malm_compare}(a) simply reflects the different
normalizations of $H_0$ in the two samples. The positive slope of the linear
relation in Fig.~\ref{malm_compare}(a) is consistent with the CF2 inverse
TFR distances having a correction that accounts for the raw SFI++ distances
being increasingly underestimated due to the Malmquist selection bias.

Let us now compare (\ref{malm_compare}) to the Malmquist correction used by
\citet{springob07} in the SFI++ sample, and subsequently adopted in the \CS
sample. We repeat the analysis of Fig.~\ref{CF2_malm_correction} on the SFI++A
subsample but now using the distances as corrected by \citet{springob07}, which
include both selection and distribution bias corrections. We find a linear
relationship of the form
\beq\label{Malm_correction_SFI}
\Delta\mu_2=0.0356(\pm0.0063)-0.000012(\pm0.000001)cz\Z{\text{LS}}
\eeq
where $\Delta\mu_2\equiv \mu_{sfi,corrected}-\mu_{sfi}$ is the difference
in distance moduli between the corrected and raw distances.
The data and best fit line are displayed in Fig.~\ref{malm_compare}.

We can immediately see from (\ref{Malm_correction_CF2}) and
(\ref{Malm_correction_SFI}) that there is a significant difference between the
corrections. For small redshifts the \citet{springob07} correction is positive
indicating raw distances are underestimated, while for large redshifts the
correction is negative indicating raw distances are overestimated.
Adjusting the intercept of Fig.~\ref{malm_compare}(a) to zero we then
find a hierarchy $\mu_{sfi,corrected}<\mu_{sfi}<\mu_{cf2}$ in the limit of large
redshifts. This is consistent with the observation of
\citet{Watkins14} that: ``the distances are systematically larger in the
\textit{Cosmicflows}-2 catalogue [than in the \CS catalogue] due to a
different approach to bias correction".

The fact that Fig.~\ref{Malm_correction_SFI}(b) has a negative slope means
that the dominant correction cannot arise solely from selection and
homogeneous distribution biases, since as noted above both of these effects
underestimate true distances. The difference therefore must be due to
the treatment of the inhomogeneous distribution bias, which \citet{springob07}
have included but \citet{Tully13} have not.

\begin{figure*}
\vbox{\centerline{\vbox{\halign{#\hfil\cr\scalebox{0.7}
{\includegraphics{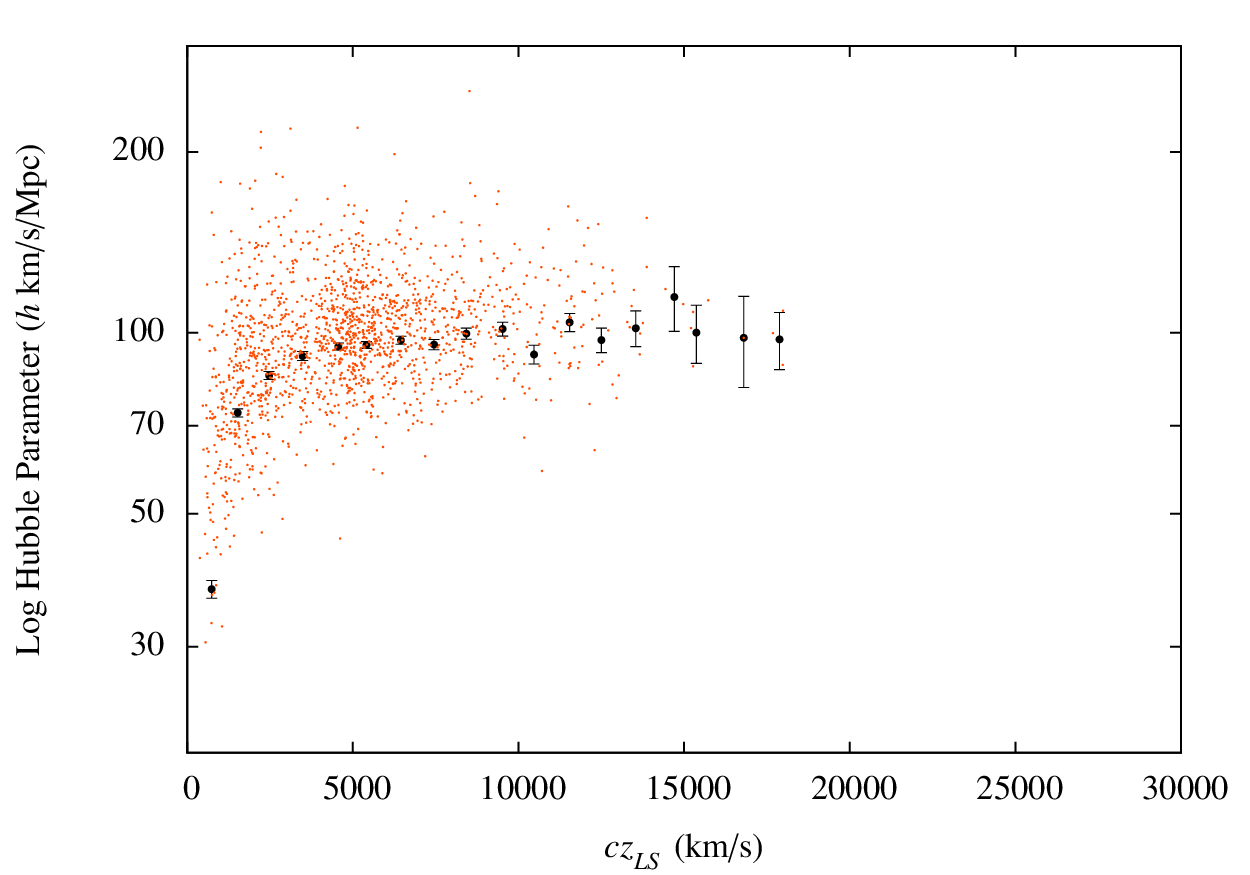}}\cr
\noalign{\vskip-5pt}\qquad{\bf(a)}\cr}}
\vbox{\halign{#\hfil\cr\scalebox{0.7}
{\includegraphics{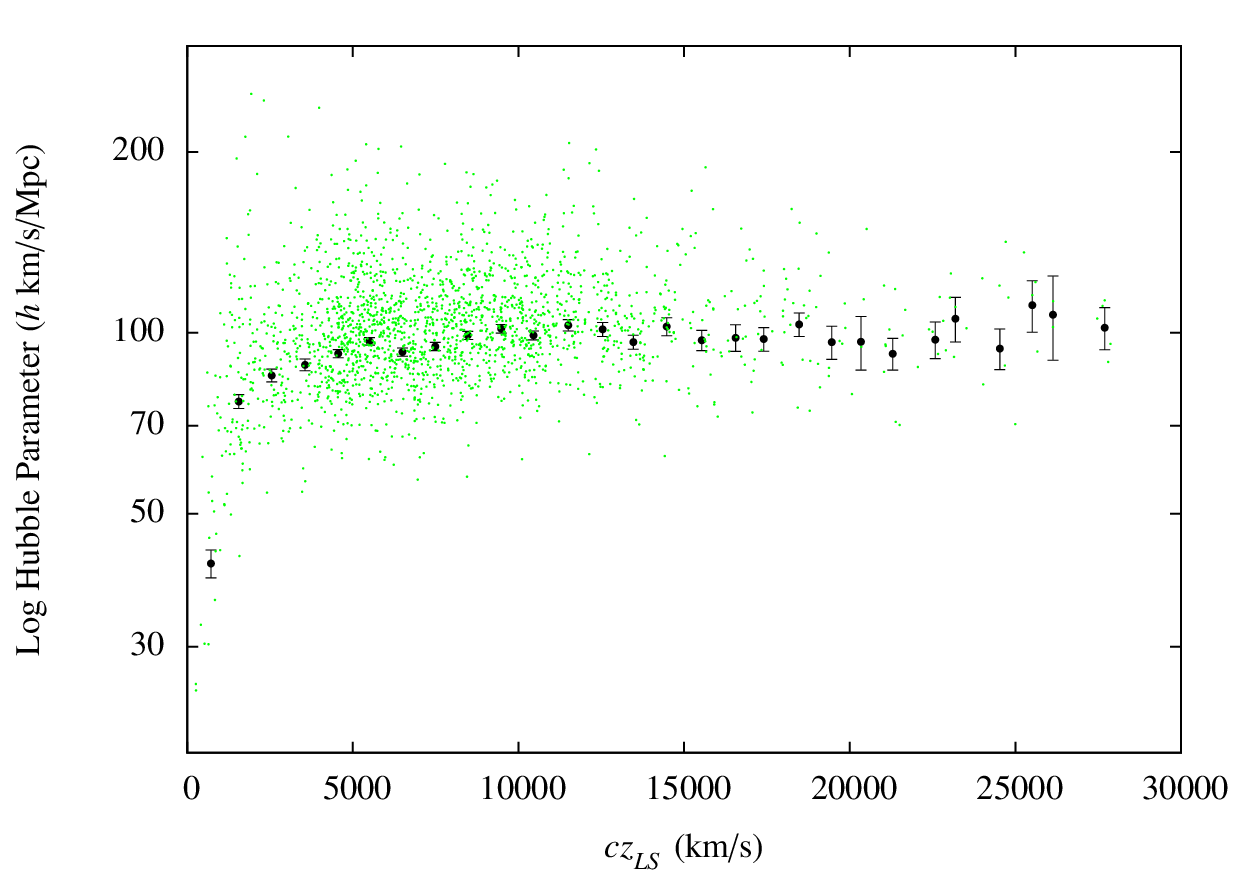}}\cr
\noalign{\vskip-5pt}\qquad{\bf(b)}\cr}}
}
\centerline{\vbox{\halign{#\hfil\cr
{\scalebox{1}{\includegraphics[width=0.5\textwidth]{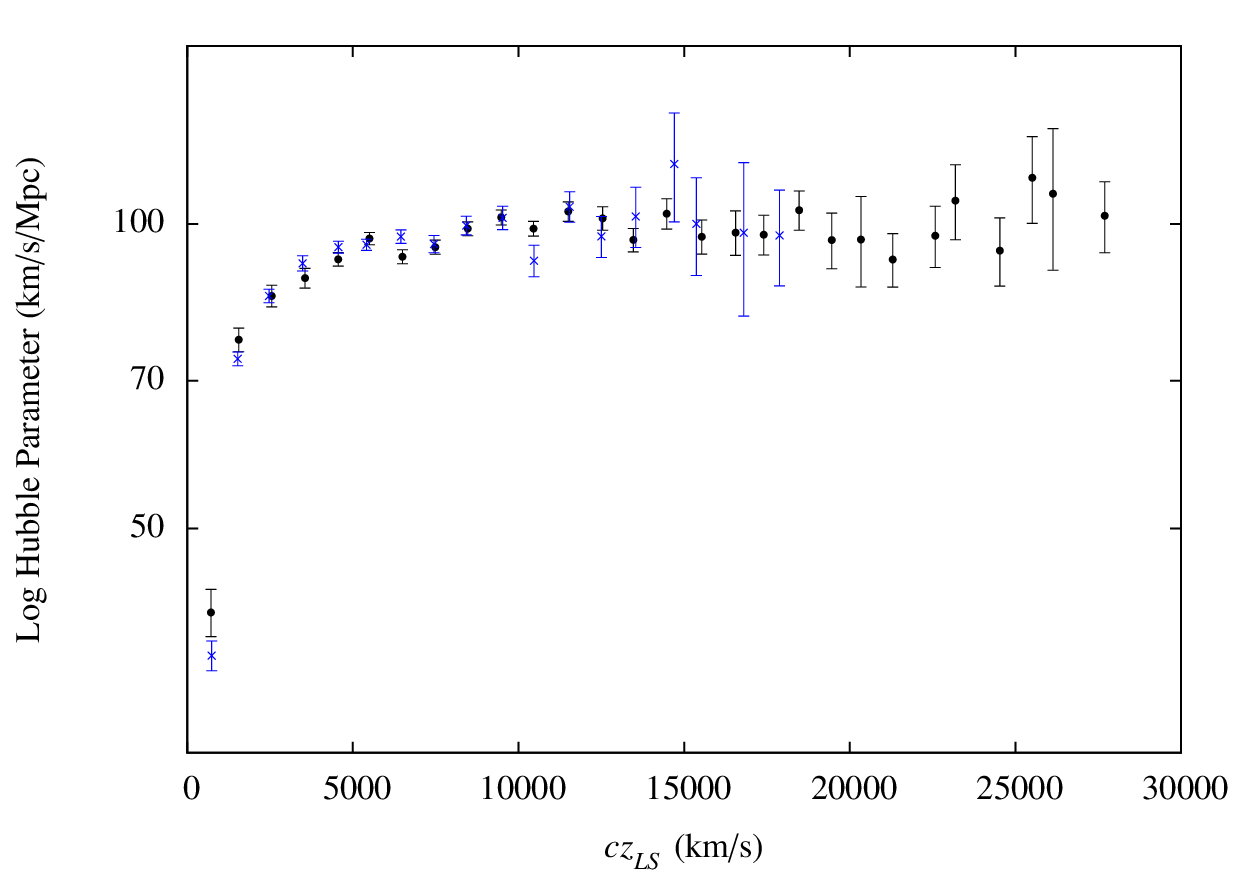}}}\cr
\noalign{\vskip-15pt}\qquad{\bf(c)}\cr}}}
\caption{The Hubble parameter, $H_i=cz_i/r_i$, computed for each individual
data point (coloured points) and from averaging in 1000$\kms$ bins (black
points) using the: {\bf(a)} SFI++A subsample; {\bf(b)} SFI++B
subsample. {\bf(c)} Comparison of the averaged points in (a) and (b) with blue
crosses being from SFI++A and black filled circles from SFI++B.\label{H_bins}}}
\end{figure*}

\begin{figure}
\centering
\includegraphics[width=0.5\textwidth]{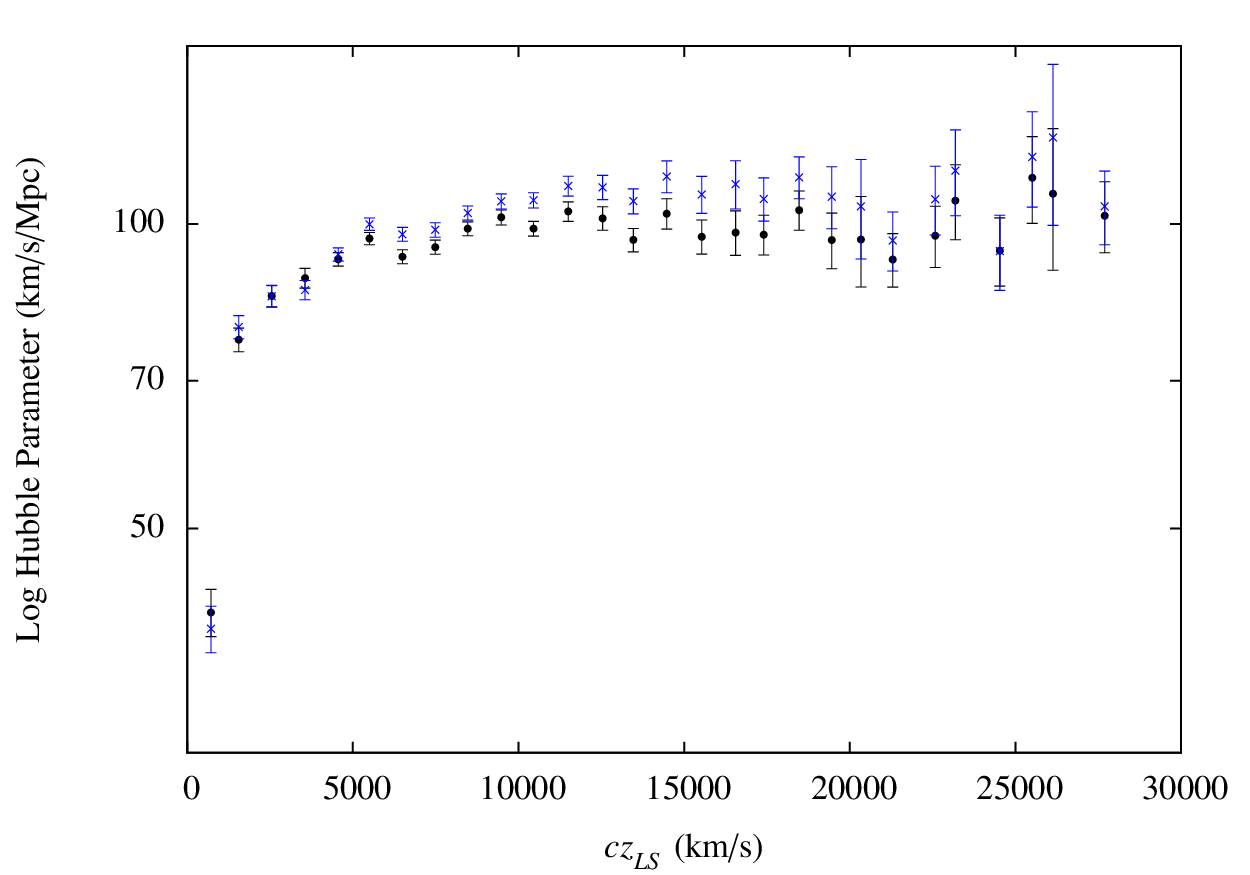}
\caption{The Hubble parameter, $H_i=cz_i/r_i$, computed from averaging in
1000$\kms$ bins using the SFI++B subsample with corrections for Malmquist
bias (blue crosses) and without corrections (black filled circles).}
\label{H_bins2}
\end{figure}

The 1970 points in the SFI++ sample that are not also contained in the
original CF2 survey, henceforth SFI++B, have been incorporated into CF2 without
using their correction (\ref{Malm_correction_CF2}). This data covers a larger
range, up to redshifts of almost $z=0.1$, whereas SFI++A only covers up to
$z=0.06$. \citet{Tully13} state that these points, if corrected using
(\ref{Malm_correction_CF2}) cause a ``highly significant decrease in the
Hubble parameter with increasing velocity''. We independently verified this
result\footnote{A decreasing Hubble constant below the scale of statistical
homogeneity is, to a limited extent, what is expected from the analysis of
\citet{WSMW}. Thus trends which appear anomalous as compared
to a standard FLRW expectation should not automatically be regarded as a signal
of unaccounted observational bias. However, there are also systematic
differences that occur when binning in redshift, as in Figs.~\ref{H_bins} and
\ref{H_bins2}, as opposed to binning in distance with (\ref{Hsav}), so careful
analysis is required to make sense of the different approaches. The direct
calculation of the Hubble parameters in each bin is also different to that
described in (\ref{Hsav}), as it is not clear which method \citet{Tully13} use.
We apply a simple weighted average of $cz_i/r_i$ values and obtain consistent
results (although as no values are tabulated by \citet{Tully13} we can
only verify by inspection).}. Thus \citet{Tully13} do not adjust these
distances, instead claiming that they are of a different nature altogether,
the main difference being that these consist of cluster samples from a
different survey \citep{Dale99a,Dale99b}. In these samples rotation
information for the galaxies was obtained from optical spectroscopy rather
than the standard 21cm Hydrogen line widths. However, Tully et al. state
that ``it is not clear to us why this component of SFI++ does not manifest
the selection Malmquist bias''.

Since it appears two halves of the SFI++ have been incorporated into CF2 in
different ways we determine whether they do actually show different
characteristics. Considering $\Delta\mu_2$ for SFI++B we find a correction
\beq
\Delta\mu_2=0.0417(\pm0.0061)-0.000012(\pm0.000001)cz\Z{\text{LS}}
\eeq
which has an intercept consistent within uncertainties and identical slope to
(\ref{Malm_correction_SFI}). Since there is no apparent difference using this
test we repeat a similar analysis to that performed by \citet[Fig.~10]{Tully13}
to test for bias. This analysis is based on the
fact that selection bias is manifest by an increase in the Hubble parameter
with redshift \citep{Teerikorpi93}, for data binned by redshift. In
Figs.~\ref{H_bins} and \ref{H_bins2} we repeat the analysis of
the Hubble constant in redshift bins performed by \citet{Tully13} for the
subsets of interest, and compare the results.

In Fig.~\ref{H_bins} we produce plots equivalent to \citet[Fig.~10]{Tully13}
for the SFI++A and SFI++B subsamples, both using raw distances.
We subsequently find that the difference in the Hubble constant in individual
redshift bins for the SFI++A and SFI++B ranges from $0.03\sigma$ to
$1.8\sigma$ in individual bins. The weighted mean of these differences is
$0.84\sigma$, and thus we do not see a significant difference between SFI++A
and SFI++B.

We note that two mutually consistent halves of the SFI++ sample have been
incorporated into CF2 in different ways, but are uncertain as to what impact
this may have on the full CF2 catalogue.

On the other hand the \CS sample \citep{WFH2} uses a much larger subset of
SFI++ distances corrected for Malmquist biases (after rejection of outliers).
While it appears that there may be inconsistencies in the inclusion of
the raw SFI++ distances into the CF2 catalogue, it is possible that the
SFI++ corrected distances are subject to systematic error also. It is
for this reason that \citet{springob07} included both corrected and raw
distances, to allow others to take on the challenging task of Malmquist bias
corrections. However, the CF2 catalogue is only corrected for selection bias,
leaving the correction of the distribution homogeneous and
inhomogeneous Malmquist biases as a task for the user.

As another test of differences between the SFI++A and SFI++B subsamples we
have also repeated the analysis of Fig.~\ref{H_bins}, but now to compare
the raw and corrected distances within each subsample. Fig.~\ref{H_bins2}
shows the comparison for the SFI++B subsample. In this case the difference in the Hubble
parameters in each bin vary from a minimum $0.01 \sigma$ to a maximum
$2.2\sigma$ in individual bins, with a weighted mean difference of $1\sigma$.
For SFI++A the difference in the Hubble parameters vary from $0.05\sigma$ to
$1.9\sigma$ in individual bins, with a weighted mean difference of $1.0\sigma$.
Thus again we do not see a significant difference between the subsamples.

It may appear surprising that the raw and uncorrected data only differ
by $1\sigma$ on average when binned by redshift. However,
once inhomogeneous Malmquist bias is accounted for the sign of
the correction is different at large redshifts as compared to low redshifts,
meaning that for an intermediate range the correction is small. The approach
by \citet{Tully13} of binning in redshift is not an appropriate one to use
when performing a parameter minimization that involves boosts to rest frames
in which the redshift is changed, as in \S\ref{LGframe}--\ref{min}.
Rather we followed \citet{WSMW} in binning by distance. This led to differences
between the raw and corrected data sets which are statistically much more
marked than is evident if one bins by redshift.

\section{Discussion}

We have investigated the extent to which it
is possible to define a standard of cosmological rest based on a frame
in which variation of the spherically averaged (monopole) nonlinear
Hubble expansion is a minimum. Such averages do not make any
assumptions about the geometry of space below the scale of statistical
homogeneity ($\lsim 100\hm$) of the sort implicit in the standard peculiar
velocities framework, which assumes a Euclidean spatial geometry.

We studied the systematic variation that arises when an arbitrary boost is
made from a frame of reference in which the spherically averaged Hubble
expansion is most uniform.
We found that such a systematic variation is indeed detected to a
statistically significant extent between the CMB and LG frames of reference,
using the \CS sample \citep{WFH1,WFH2}.

This supports the proposition made by \citet{WSMW} that the Local Group frame
may be a more suitable cosmic rest frame than the standard CMB frame, and
that consequently a significant fraction of the CMB dipole may be nonkinematic.
This proposition can be directly tested in numerical ray--tracing
simulations \citep{Bolejko}.
Given that the CMB frame is still the de facto choice for the cosmic rest
frame, this conclusion would have a far reaching impact for many areas of
cosmology \citep{WSMW}, including the question of large angle CMB anomalies
\citep{Piso}.

We extended the analysis to search for an improvement
on the Local Group as the standard of rest, using a variety of tests.
Very large boosts from the LG frame can be excluded if we simultaneously
demand that while the residual variation of the Hubble expansion should be
small in the regime of nonlinear expansion at small distances, a clear
signature of an emerging linear Hubble law should also be found at larger
distances. However, we found that in applying all possible tests there
is still freedom to perform quite large boosts close to the plane of
the galaxy, presumably because the lack of data in the Zone of Avoidance
leads to a lack of suitable constraints there. This hypothesis could
potentially be tested on simulated data using exact solutions of
Einstein's equations \citep{Bolejko}.

Since our conclusions depend on the \CS sample, we repeated
our analysis insofar as it was possible for the recently released
\textit{Cosmicflows}-2 sample \citep{Tully13}. This catalogue of 8162 galaxy,
groups and cluster distances is considerably larger than the \CS
sample, and potentially could provide more accurate results although it
is also of course limited in the Zone of Avoidance.

We have found very significant differences in the results for the CF2
and \CS samples, as can be seen by comparing Fig.~3 of \citet{WSMW}
and our Fig.~\ref{compare_vmod}. These
result from differences in the treatment of the Malmquist bias between
the SFI++ catalogue and CF2 catalogue, as previously noted
by \citet{Watkins14}. We also found apparent inconsistencies in the manner of
inclusion of subsamples of the SFI++ catalogue into the CF2 catalogue, with
respect to the treatment of Malmquist bias. More significantly,
since the reported CF2 distances do not include corrections for the
inhomogeneous distribution Malmquist bias they may be of limited use until
such corrections are applied.

The conclusions of \citet{WSMW} are dependent on the treatment of the
Malmquist bias in the SFI++ catalogue being accurate, as this constitutes the
largest part of the \CS sample. Naturally, one might question whether
any systematic procedure of \citet{springob07} could somehow spuriously
lead to an unusually uniform Hubble expansion in the LG frame through an
error in the Malmquist bias procedure.

We find no grounds for this. In particular, our analysis shows that the
difference between the CMB and LG frames has the distinctive signature of a
systematic boost offset (\ref{eq1}) noted by \citet{WSMW}. Nothing in the
Malmquist bias correction procedure of \citet{springob07} could obviously
introduce this signature through a systematic error. Their analysis does not
single out the LG, or LS, frame in any way; indeed all their redshifts are
referred to the CMB frame. Furthermore, although the remaining bias means
that the CF2 sample is currently unusable from the point of view of
determining a frame of minimum spherically averaged Hubble expansion variation
in the nonlinear regime, we observed in \S\ref{CF2boost} that the
{\em difference} of the CMB and LG frame spherical averages nonetheless still
shows the signature of the systematic boost offset in the CF2 data.

Since the boost offset is detectable in the independently reduced CF2 data,
it cannot be an artefact of the Malmquist bias treatment of \citet{springob07}.
Furthermore, the departure of the nonlinear expansion from the boost power
law (\ref{pl}) that is seen when comparing Figs.~\ref{Hdiff2} and
\ref{hdiff_CF2_gal} is precisely what is to be expected if there are
additional unaccounted uncertainties in individual CF2 distances as compared to
the SFI++ ones: the distance range of structures associated with the nonlinear
expansion is broadened.

If Fig.~\ref{compare_vmod} was based on accurate distances, it would
imply that the Hubble expansion in all frames of reference is far less
uniform than might reasonably be expected in any viable cosmological model;
in particular, there is a monopole or ``Hubble bubble'' variation of order
15--20\% in the range $20<r<60\hm$ in the CF2 sample, as compared to 4--5\% in
the \CS sample. The largest ``Hubble bubble'' variation that has ever been
claimed on such scales using more accurate Type Ia supernovae distances in
the CMB frame is $6.5\pm2.2$\% \citep{zrkd98}.

\citet{Tully13} chose not to correct for the distribution biases, as they
wished to separate ``the issues of distance measurements and velocity field
inferences''. Indeed, in the peculiar velocities approach the distribution
bias may be much less significant. In new work \citet*{HCT} use the CF2
catalogue to reconstruct large scale structure by means of the Wiener filter
and constrained realizations of Gaussian fields assuming a WMAP constrained
\LCDM\ model as a Bayesian prior. They observe that ``the Malmquist
bias introduces a spurious strong monopole term into the reconstructed
velocity field but is expected to hardly affect the bulk velocity which is
associated with the dipole of the velocity field'' \citep{HCT}. This would
appear to be the counterpart of the large monopole we observe in
Fig.~\ref{compare_vmod} in our analysis. \citet{HCT} corrected for the bias
but noted that the bulk velocity analysis is ``virtually unaffected by the
Malmquist bias''.

In our case, residual bulk flows on scales $\gsim100\hm$ may be an artefact of
using the CMB rest frame as the standard when it does not coincide with
the frame of minimum Hubble expansion variation. In particular, \citet{WSMW}
observed that a dipole fit to the Hubble expansion gave a dipole magnitude
which increased to a residual value in the outer shells as the shell radius
was increased in the CMB frame, but not in the LG frame. Thus large
scale bulk flows are not our primary interest. Rather,
we are interested in detecting the systematic monopole variation
(\ref{eq1}). In distinction to the peculiar velocity approach
our method by necessity is sensitive to a monopole bias.
In fact, our method of binning in radial shells by distance
with an anchoring to $\Ha$, is particularly sensitive to any distribution
bias which follows from a number density, $N(r)$, with strong gradients.
The bias effect in Fig.~\ref{compare_vmod} can be largely reproduced by
applying a uniform Hubble law to the CF2 redshifts, adding Gaussian scatter
to create a mock distance catalogue, and then applying our binning strategy
(R.~Watkins, private communication).

It may be possible to construct the $635\kms$ velocity attributed to
the LG within the \LCDM\ model, as has recently been claimed by \citet{Hess14}
who used constrained $N$-body simulations and nonlinear phase space
reconstructions to arrive at a value $v\Ns{LG}=685\pm137\kms$. However, this
itself does not constitute a proof of the standard kinematic interpretation,
but rather a verification within the 20\% uncertainty of a computer
simulation. The \LCDM\ model is certainly phenomenologically very successful,
and any competing model can only be viable insofar as many of its predictions
are close to the standard model, as is the case, for example, in the timescape
cosmology \citep{obs,SW,grb,nw}. What is important in testing the standard model
is to seek observations which are not expected in its framework. Although the
signature of the systematic monopole boost offset (\ref{eq1}) between the CMB
and LG frames should be checked in \LCDM\ simulations, it is not an observation
that should obviously arise if all cosmological redshifts arise purely from
a FLRW geometry plus local Lorentz boosts at any scale. Properly characterizing
and determining a frame of minimum nonlinear Hubble expansion variation is
therefore a fundamental question open to more precise observational tests in
future.

In conclusion, if we make no assumptions about our own standard of rest, but
make arbitrary boosts at our location, then analysis of the \CS sample
yields a degenerate set of frames for which the variation of the spherically
averaged nonlinear Hubble expansion is minimized. This set of candidate cosmic
rest frames includes the LG frame but excludes the CMB frame, as was already
established with very strong Bayesian evidence by \citet{WSMW}. The degeneracy
in defining the minimum variation frame is associated with a freedom to perform
boosts in the plane of the galaxy without changing the statistical significance
of the fit, probably due to the lack of constraining data in the Zone of
Avoidance.

The larger CF2 sample may potentially tighten the constraints on the definition
of the frame of minimum nonlinear expansion variation. However, it is first
necessary to reduce the data in the manner of \citet{springob07} to remove the
inhomogeneous distribution bias which appears to be the source of the large
discrepancies found in \S\ref{cf2}. The Universe is very inhomogeneous below
the scale of statistical homogeneity, and ironically it is only once the biases
associated with such inhomogeneities are removed that a picture of a
remarkably uniform average Hubble expansion actually emerges.

A careful treatment of inhomogeneous Malmquist bias is therefore key for the future progress of our understanding of the nature of cosmic expansion
as the surveys grow ever larger.

\section*{Acknowledgements}

We thank Krzysztof Bolejko, H\'el\`ene Courtois, Ahsan Nazer, Brent Tully,
Nezihe Uzun and Rick Watkins for helpful discussions and correspondence.

\appendix
%\section{Appendix A}\subsection*
\section{Linear regression via total least squares}

Consider a general linear model with errors in both the dependent and independent variables. We can express such a model as
\begin{align}
\begin{split}
y_i&=\beta_0+\beta_1x_i\\
(Y_i,X_i)&=(y_i,x_i)+(e_i,u_i)
\end{split}
\end{align}
where $(Y_i,X_i)$ are the observations, $(y_i,x_i)$ are the true values and $(e_i,u_i)$ are the measurement errors. We assume the measurement errors to be normally distributed with a covariance matrix
\beq
\Sigma=\left(\begin{array}{cc} \sigma_{ee} & \sigma_{eu}\\ \sigma_{ue} & \sigma_{uu} \end{array}\right).
\eeq
We can extend this analysis to allow for different errors at each point, which we will refer to as weights given by $\omega(X_i)=1/\sigma_{uu,i}$, $\omega(Y_i)=1/\sigma_{ee,i}$ and the correlation coefficient between the errors given by $\gamma_i=\sigma_{eu,i}\sqrt{\omega(X_i)\omega(Y_i)}$.

To carry out a least squares minimization we must calculate the statistical distance from an observation to the true value. In standard least squares this distance is the vertical distance from the data point to the model, since the values of the independent variable are assumed to be exact. Now, in the most simple case we would have the squared Euclidean distance from the observed data points to true value in the model as
\beq
\left[Y_i-(\beta_0+\beta_1x_i)\right]^2+(X_i-x_i)^2=e_i^2+u_i^2
\eeq
but if the variances of $e_i$ and $u_i$ are different from unity this statistical distance becomes $\sigma_{ee}^{-1}e_i^2+\sigma_{uu}^{-1}u_i^2$, and if these variances are correlated we must use the covariance matrix to give the ``statistical" distance
\beq\label{statdist}
\left(Y_i-\beta_0-\beta_1x_i,X_i-x_i\right)\Sigma^{-1}\left(Y_i-\beta_0-\beta_1x_i,X_i-x_i\right)^\intercal.
\eeq
In our case we must determine the values of $\beta_0$ and $\beta_1$ that
minimize (\ref{statdist}). That is, we must find the values $(\hat{x}_i,
\hat{y}_i)$ and $(\hat{\beta}_0,\hat{\beta}_1)$ that minimize this sum for the
given observations. First we fix the $x_i$ values by treating them as unknown
constants in a standard linear regression of the form
\beq
\left[\begin{array}{c} Y_i-\beta_0\\X_i\end{array}\right]=
\left[\begin{array}{c}\beta_1 \\ 1\end{array}
\right]x_i+\left[\begin{array}{c} e_i\\ u_i\end{array}\right]
\eeq
for which the generalized least squares estimator gives
\beq
\hat{x}_i=\left[(\beta_1,1)\Sigma^{-1}(\beta_1,1)^\intercal\right](\beta_1,1)
\Sigma^{-1}\left(Y_i-\beta_0,X_i\right)^\intercal.
\eeq
Substitution of $\hat{x_i}$ into (\ref{statdist}) gives
\beq
\frac{\left(Y_i-\beta_0-\beta_1X_i\right)^2}{\left(\sigma_{ee}-2\beta_1
\sigma_{eu}+\beta_1^2\sigma_{uu}\right)}
\eeq
so that after summing over all $N$ points we obtain
\beq\label{mindist}
S=\frac{\sum_{i=1}^N\left(Y_i-\beta_0-\beta_1X_i\right)^2}{\left(\sigma_{ee}
-2\beta_1\sigma_{eu}+\beta_1^2\sigma_{uu}\right)},
\eeq
which is the expression to be minimized. In \citep{York} the linear
equation that minimizes (\ref{mindist}) is given by
\beq\label{S}
\beta_1=\frac{\sum_{i=1}^NZ_i^2V_i\left[\frac{U_i}{\omega(Y_i)}+\frac{\beta_1V_i}{\omega(X_i)}-\frac{\gamma_iV_i}{\alpha_i}\right]}{\sum_{i=1}^NW_i^2U_i\left[\frac{U_i}{\omega(Y_i)}+\frac{\beta_1V_i}{\omega(X_i)}-\frac{\beta_1\gamma_iU_i}{\alpha_i}\right]}
\eeq
where
\begin{align*}
&\alpha_i^2=\omega(X_i)\omega(Y_i), \ \ \ U_i=X_i-\bar{X},\ \ \ V_i=Y_i-\bar{Y},\\
&\bar{X}=\left(\sum_{i=1}^NZ_iX_i\right)/\sum_{i=1}^NX_i \ \ \ \text{and} \ \ \ \bar{Y}=\left(\sum_{i=1}^NZ_iY_i\right)/\sum_{i=1}^NY_i,\\
& Z_i=\frac{\omega(X_i)\omega(Y_i)}{\omega(X_i)+b^2\omega(Y_i)-2b\gamma_i\alpha_i}.\\
\end{align*}
Clearly (\ref{S}) requires an iterative process to find $\beta_1$ which begins with an initial guess which may be found from performing a standard linear regression assuming the $X_i$ to be exact. After $\beta_1$ is obtained the value of $\beta_0$ is found from the fact that the mean must be on the best fit line and thus $\beta_0=\bar{Y}-\beta_1\bar{X}$.

The uncertainties in the parameter values, $\sigma_{\beta_0}$ and $\sigma_{\beta_1}$, are \citep{Titterington}
\begin{align}
\sigma_{\beta_0}^2&=\frac{\sum Z_ix_i^2}{\left(\sum Z_ix_i^2\right)\left(\sum Z_i\right)-\left(\sum Z_i x_i\right)^2},\\
\sigma_{\beta_1}^2&=\frac{\sum Z_i}{\left(\sum Z_ix_i^2\right)\left(\sum Z_i\right)-\left(\sum Z_i x_i\right)^2}.
\end{align}

We now return to the transformed model from (\ref{eq1}) which takes on the form
\beq
\log(\delta H_i)=p\log(\ave{r_i^2})+\log\left(\frac{v^2}{2\Ha}\right)\label{a12}
\eeq
such that we may identify $y_i=\log(\delta H_i)$, $x_i=\log(\ave{r_i^2})$, and $(\beta_1,\beta_0)=\left(p,\log\left(\frac{v^2}{2\Ha}\right)\right)$=$(p,\log A)$.

Since our distances are given in units $\hm$, independent of the overall
normalization of $\Ha$, rather than directly performing a fit to (\ref{pl}),
(\ref{a12}) we instead perform a fit to the relation
\beq
\Ha\left(H'_s-H_s\right)\approx\frac{v^2}{2}\left(\ave{r_i^2}_s\right)^p.
\eeq
This gives a direct estimate of $v$ in $\kms$, from which we obtain
$A\simeq v^2/(2\Ha)$.
%\end{document}
\newpage
% Erratum: Determing the frame of minimum nonlinear Hubble expansion variation
%-----------------------------------------------

\title{Erratum: Defining the frame of minimum nonlinear Hubble expansion variation}
\section*{\LARGE Erratum}

The paper ``Defining the frame of minimum nonlinear Hubble expansion
variation'' was originally published in\break MNRAS, 457, 3285 (2016).

There is a small numerical error\footnote{The error was first made in eq.\ (9) of \citet{WSMW}, and was later also repeated in eq.\ (2.8) of \citet{bnw}. However, as the value of $v$ inferred is not used in either of these papers, none of their results are affected in any way.} in the denominator of the final line of eq.~(12) of \citet{mw16} where the factor ``2'' should be ``3'', so that it
reads\footnote{This relation has now been extended by \citet{ks16} to include
the case that a bulk flow exists in the frame of minimum spherically averaged
Hubble expansion variation.}
$$\begin{matrix}
\Delta H_s=H'_s-H_s\hskip-8pt&\displaystyle\goesas\left(\sum_{i=1}^{N_s}{(v\cos\ph_i)^2\over\si_i^2}
\right)\left(\sum_{i=1}^{N_s}{cz_i r_i\over\si_i^2}\right)^{-1}\hfill\cr
&\displaystyle\approx\frac{v^2}{3\Ha\ave{r_i^2}_s}\hfill\cr
\end{matrix}\eqno(12)$$
Likewise three lines below eq.~(13), we should have $\langle (v \cos\ph_i)^2
\rangle \goesas\frn12 v^2 \int_0^\pi \sin\ph_i (\cos\ph_i)^2 d\ph_i =
\frn13 v^2$, while in the sentences below eq.~(15) and eq.~(A13) we should
have $A\approx v^2/(3\Ha)$, and eq.~(A13) itself should read
$$
\Ha\left(H'_s-H_s\right)\approx\frac{v^2}{3}\left(\ave{r_i^2}_s\right)^p.
\eqno({\rm A}13)
$$

Correction of this small error does not affect any of the conclusions of the
paper, but it does increase the value of the boost velocity, $v$,
inferred from the parameter $A$ in eq.~(12) or (A13) by 22.5\%. This has no
material consequence, however, since the uncertainties in $v$ are very large.
In particular, using the best-fitting value of $A$, in place of the values
given in section 3 we find
$v=946^{+1255}_{-539}\kms$ ($v=523^{+295}_{-188}\kms$) for the primed
(unprimed) shells respectively, both still being consistent with the
actual boost magnitude of $635\pm38\kms$.

The only instance in the paper where incorrect derived values of the boost
velocity were used in any further numerical analysis was in Section 4.3,
where these values were assigned to $\vd$ and compared to the true boost
velocity $\vt$, as parameters were varied in the $\{v,\ell,b\}$ parameter space.
The 22.5\% correction to $\vd$ results in slight changes to Fig.~9(a),
as shown, but no significant change to Fig.~9(b).

Using the corrected value to define the frame X of section 4.3, we find that
$\vd/\vt=1$ now occurs at $\vt=239\kms$ in a direction
$(\ell,b)=(59\deg,-3\deg)\pm(5\deg,4\deg)$, which is only changed by $1\deg$.
For this boost $p=-1.25\pm(0.92)\ns{stat}\pm(0.37)\ns{sys}$, which is still
consistent with $p=-1$ within the large uncertainties. (With only the primed
shells we now obtain $p=-1.17 \pm(1.06)\ns{stat}$.)

None of the physical conclusions
concerning frame X are changed from Section 4.3 of \citet{mw16}. In particular,
the changes to Fig.~10(b),(c) are so small that
they are indiscernible apart from a very small offset of
the first data point. Moreover,
as before frame $X$ is only weakly disfavoured compared to the global $\chN$
minimum, $\ln(P\Ns{MV}/P\Ns{X})= 1.7$, and is within $1\,\si$ of the global
minimum of $\chL$; similar to the LG frame on both counts.
\begin{figure}
{\centering
\includegraphics[width=0.486\textwidth]{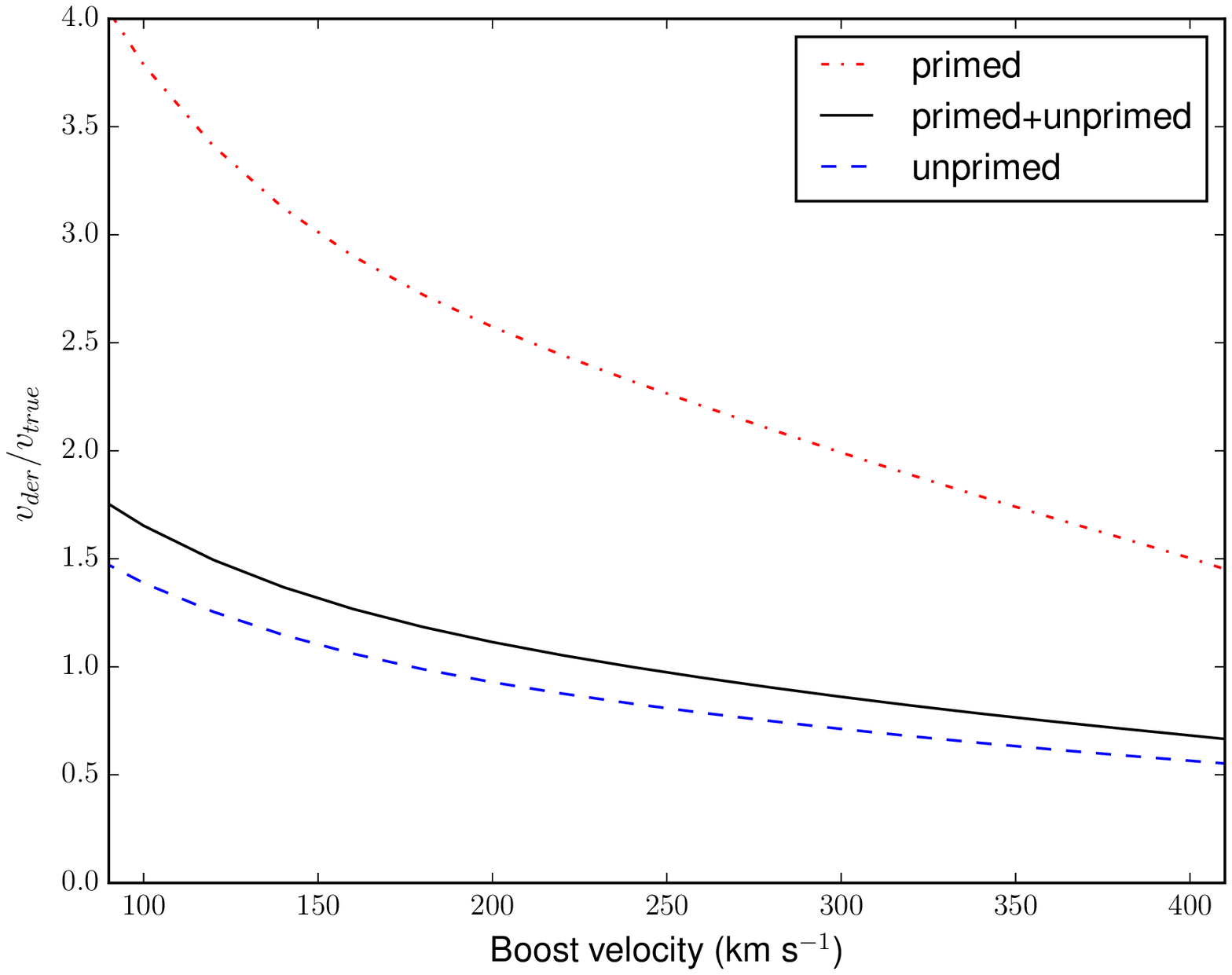}}

{\small{\bf Figure 9(a)} Corrected.
Ratio of the derived velocity from the best fit power
law to the true boost velocity. At each $v$ the direction is determined by
the best fit for which $p$ is closest to $-1$.}
\end{figure}

\section*{Acknowledgements}
We thank David Kraljic for pointing out the error.

\end{document}